\documentclass[12pt]{article}
\usepackage{amsmath}
\usepackage{graphicx}
\usepackage{enumerate}
\usepackage{natbib}
\setcitestyle{authoryear,round,semicolon,aysep={},yysep={,},notesep={, }}
\usepackage{url}
\usepackage{microtype}
\usepackage[utf8]{inputenc}
\usepackage{amssymb}
\usepackage{mathtools}
\usepackage{booktabs,longtable,needspace}
\usepackage{multirow}
\usepackage{threeparttable}
\usepackage{float}
\usepackage{caption}
\usepackage{placeins,pdflscape,rotating,afterpage}
\newtheorem{theorem}{Theorem}

\newtheorem{remark}{Remark}
\newcounter{example}

\newcommand{\blind}{0}

\usepackage{etoolbox}
\usepackage{xcolor}
\makeatletter
\patchcmd{\maketitle}{\parindent 1em}{\parindent 0pt\raggedright}{}{}
\patchcmd{\maketitle}{\hb@xt@1.8em}{\hb@xt@0pt}{}{}

\newcommand{\setcaptypefigure}{\def\@captype{figure}}
\makeatother
\usepackage{array}
\newcolumntype{R}[1]{>{\raggedleft\arraybackslash}p{#1}}
\newtheorem{lemma}{Lemma}
\providecommand{\bth}{\boldsymbol{\theta}}
\providecommand{\bX}{\boldsymbol{X}}

\providecommand{\bW}{\boldsymbol{W}}
\providecommand{\bx}{\boldsymbol{x}}
\providecommand{\bM}{\boldsymbol{M}}
\providecommand{\bV}{\boldsymbol{V}}
\providecommand{\bC}{\boldsymbol{C}}
\providecommand{\dM}{\dot{\boldsymbol{M}}}
\providecommand{\dm}{\dot{\boldsymbol{m}}}
\providecommand{\bSig}{\boldsymbol{\Sigma}}
\providecommand{\bPi}{\boldsymbol{\Pi}}
\providecommand{\bGam}{\boldsymbol{\Gamma}}
\providecommand{\bgam}{\boldsymbol{\gamma}}
\providecommand{\PX}{P_{\boldsymbol{X}}}

\newenvironment{proof}[1][Proof]{\par\smallskip\noindent\textit{#1.}\ }{\hfill$\square$\par\smallskip}
\begin{document}

\def\spacingset#1{\renewcommand{\baselinestretch}%
{#1}\small\normalsize} \spacingset{1}

\if0\blind
{
  \title{\bf Regression Model Checks Using Data-Dependent Partitions}
  \author{\begin{tabular}{@{}c@{\hspace{1.5em}}c@{}}
    Miguel A. Delgado\thanks{Corresponding author. Universidad Carlos III de Madrid, Calle Madrid 126--128, 28903 Getafe, Madrid, Spain. \protect\url{delgado@est-econ.uc3m.es}. Research funded by Ministerio de Ciencia, Innovación y Universidades
    (Spain), grant PID2021-125178NB-I00, and CEX2021-001181-M financed
    by MICIU/AEI/10.13039/501100011033.} &
    Antonio Raiola\thanks{University of Rome Tor Vergata, Via Columbia 2, 00133 Rome, Italy. \mbox{\texttt{antonio.raiola@uniroma2.it}}.}\\[0.3em]
    {\normalsize Universidad Carlos III de Madrid} & {\normalsize University of Rome Tor Vergata}
    \end{tabular}}
  \date{November 2026}
  \maketitle
} \fi

\if1\blind
{
  \bigskip
  \bigskip
  \bigskip
  \begin{center}
    {\LARGE\bf Regression Model Checks Using Data-Dependent Partitions\par}
\end{center}
  \medskip
} \fi

\bigskip
\begin{abstract}
We propose a Pearson-type chi-square test of regression model specification. The statistic compares observed and model-implied sample moments within a fixed number of cells and can be interpreted as a GMM $J$-test for overidentifying moment restrictions. Under suitable regularity conditions, its asymptotic chi-square distribution remains valid for data-dependent partitions, which can incorporate information about relevant alternatives. We derive the local power of the test and extend the construction to general conditional moment restrictions. In Monte Carlo experiments the test achieves size accuracy similar to that of bootstrap omnibus tests. Partitions that use information about the alternative can give large power gains, particularly as the number of regressors increases. In an application to production-function specification for Colombian manufacturing plants, the proposed specification tests often reject the Cobb--Douglas specification but find much weaker evidence against the constant-elasticity-of-substitution specification, and in the conditional-mean comparison omnibus tests reject neither. Tests of model specification by means of conditional moment restrictions give similar results.

\end{abstract}

\noindent%

\vfill

\newpage
\spacingset{1.45} 

\section{Introduction}

\label{sec:intro}

\looseness=-1 The identification of causal relations from data and the validity of subsequent inferences heavily rely on the correct specification
of models defined by conditional moment restrictions, usually
parametric regression models. This has led to the proposal of a variety
of model-checking methods that extend established goodness-of-fit tests for cumulative distribution functions (CDFs) to the regression
framework. We propose a Pearson-type specification test based on a partition
of the covariate space. It compares observed and model-implied sample moments
within the cells, and the test statistic has an asymptotic chi-square null distribution. With a fixed number of cells, the test detects only departures under which the regression errors have nonzero mean in some cell, so its power depends on the partition. Cells chosen with a given alternative in mind can increase power against that alternative.

\looseness=-1 The integrated regression function (IRF) fully characterizes the regression,
much as the CDF characterizes a distribution. Classical empirical-process
goodness-of-fit methods therefore have natural analogs for regression
specification testing, based on the integrated residual empirical
process (IREP). Early contributions
include \citet{bhattacharya1974,bhattacharya1976} on tests for
conditional mean independence. Extensions to composite
regression specifications include those of \citet{bierens1982consistent},
\citet{stute1997nonparametric}, \citet{bierens1997asymptotic}, and
\citet{andrews1997conditional}. With estimated parameters, the limiting
distributions of many IREP-based statistics depend on unknown features
of the data-generating process, and tests are implemented using
bootstrap methods. Asymptotically distribution-free alternatives
based on martingale transformations have been proposed by \citet{khmaladze1982martingale}
and \citet{10.1214/Khmaladze2004}. They can be hard
to put into practice when the number of covariates is large. Smoothing-based
procedures provide another alternative \citep{bickel1973some,hardle1993comparing}.
The standardized statistics are asymptotically normal when the amount
of smoothing converges to zero at a suitable rate as the sample size
diverges. However, these tests are sensitive to the amount of smoothing
chosen, and the normal approximation can be poor in finite samples.
For this reason, they are often implemented using the wild bootstrap
(see, for example, \citealp{hardle1993comparing}).

\looseness=-1 The \citet{pearson1900x} goodness-of-fit test remains a popular tool
for assessing distributional specifications. Related chi-square procedures
have also been proposed for conditional CDF models \citep{heckman1984chi,andrews1988chi}.
Recent contributions include the conditional-distribution test of
\citet{delgado2022conditional}, which uses the conditional probability integral transform to classify the data into a contingency table whose rows and columns are independent under the null hypothesis. The generalized Hosmer--Lemeshow
test of \citet{surjanovic2024generalized} derives critical values under
a specified conditional response distribution within a family of generalized
linear models. Our contribution concerns conditional-mean restrictions without further
assumptions on the conditional distribution of $Y$. Building on the results of
\citet{pollard1979general}, we show that the Pearson test for regression
specification using a fixed partition has the same asymptotic distribution as
its analog based on a data-dependent consistent estimator of that partition.

\looseness=-1 Correct specification requires the regression errors to have zero mean in every Borel region of the covariate space, but we only test this restriction on a partition into $L$ cells. The statistic is a weighted quadratic form in the differences between observed and fitted cell moments, which is in fact the GMM $J$-test of \citet{hansen1982large} for these restrictions. With finitely many cells, the restrictions do not characterize the conditional mean, and the partition determines which alternatives can be detected. When the partition is fixed, the test is a conditional moment test in the sense of \citet{newey1985conditional} and \citet{tauchen1985diagnostic}, with the cell indicators as instruments. When the cells are data-dependent, we extend the uniform-convergence argument of \citet{pollard1979general} to regression specification tests.

\looseness=-1 The paper makes three contributions. First, it proposes a chi-square test of regression specification that needs no bootstrap and remains valid when the cells are chosen from the data. Second, it shows how to build the partition from a guiding model, which need not be correctly specified, and uses the local power of the test to explain how the partition affects power. Third, it applies the test to general conditional moment restrictions, such as those of instrumental-variable models or the condition under which the slopes of the least-squares (LS) projection of log output on log inputs are elasticities of expected output. In an application to Colombian manufacturing plants, the tests often reject the Cobb--Douglas specification but rarely the constant-elasticity-of-substitution one, whereas the omnibus tests reject neither in the conditional-mean comparison.\enlargethispage{\baselineskip}

\looseness=-1 Section~\ref{sec:test} introduces the statistic, and Section~\ref{sec:ddp} establishes its asymptotic properties with data-dependent cells. Section~\ref{sec:new-contiguous} derives local power, and Section~\ref{sec:partition_choice} uses it to discuss partition choice. Section~\ref{sec:simulations} reports the Monte Carlo experiments. Section~\ref{sec:new-general} discusses testing general conditional moment restrictions, and Section~\ref{sec:applications} presents an application to the specification of production functions. Appendix A contains the proofs, and the supplementary Appendix B contains additional simulations, details of the applications, and technical remarks on partition choice. References prefixed A or B refer to the corresponding appendix.

\section{The Test Statistic}

\label{sec:test}

\looseness=-1 Let $\left\{ \boldsymbol{W}_{i}\right\} _{i=1}^{n}=\left\{ \left(Y_{i},\boldsymbol{X}_{i}^{\prime}\right)^{\prime}\right\} _{i=1}^{n}$
be i.i.d. observations of the $\mathbb{R}^{1+k}$-valued random vector
$\boldsymbol{W}=\left(Y,\boldsymbol{X}^{\prime}\right)^{\prime}$,
with induced probability distribution $P $ and $\mathbb{E}|Y|<\infty $, where $\boldsymbol{A}^{\prime}$ denotes the transpose of $\boldsymbol{A}$.
The null hypothesis is 
\begin{equation}
H_{0}:m\in\mathcal{M},\label{100}
\end{equation}
where $m(\boldsymbol{x})\coloneqq\mathbb{E}(Y\mid\boldsymbol{X}=\boldsymbol{x})$,
and, for a suitable parameter space $\Theta\subset\mathbb{R}^{p}$,
\[
\mathcal{M}\coloneqq\{m_{\boldsymbol{\theta}}:\boldsymbol{\theta}\in\Theta\} 
\]
is a class of measurable functions indexed by parameter vectors $\boldsymbol{\theta}\in\Theta $.

\looseness=-1 The regression function is fully characterized by the IRF 
\[
R(\boldsymbol{x})\coloneqq\mathbb{E}\!\left[Y\,1_{(-\boldsymbol{\infty},\boldsymbol{x}]}(\boldsymbol{X})\right]=\mathbb{E}\!\left[m(\boldsymbol{X})\,1_{(-\boldsymbol{\infty},\boldsymbol{x}]}(\boldsymbol{X})\right],
\]
where $1_{A}$ is the indicator of the set $A $, and $(-\boldsymbol{\infty},\boldsymbol{x}]=(-\infty,x_{1}]\times\cdots\times(-\infty,x_{k}]$.
Therefore, under $H_{0}$, there exists a $\boldsymbol{\theta}_{0}\in\Theta $
such that, for all $\boldsymbol{x}\in\mathbb{R}^{k}$, $R(\boldsymbol{x})=R_{\boldsymbol{\theta}_{0}}(\boldsymbol{x})$,
where $R_{\boldsymbol{\theta}}(\boldsymbol{x})\coloneqq\mathbb{E}\!\left[m_{\boldsymbol{\theta}}(\boldsymbol{X})1_{(-\boldsymbol{\infty},\boldsymbol{x}]}(\boldsymbol{X})\right]$
is the parametric version of $R(\boldsymbol{x})$ under $H_{0}$.

\looseness=-1 The sample versions of the unrestricted and model-implied IRFs, which
are analogous to the observed and expected frequencies
in the classical Pearson framework, are $\hat{R}(\boldsymbol{x})\coloneqq n^{-1}\sum_{i=1}^{n}Y_{i}1_{(-\boldsymbol{\infty},\boldsymbol{x}]}(\boldsymbol{X}_{i})$
and $\hat{R}_{\boldsymbol{\theta}}(\boldsymbol{x})\coloneqq n^{-1}\sum_{i=1}^{n}m_{\boldsymbol{\theta}}(\boldsymbol{X}_{i})1_{(-\boldsymbol{\infty},\boldsymbol{x}]}(\boldsymbol{X}_{i})$,
respectively. Existing tests are functionals of the IREP 
\[
\hat{V}_{\boldsymbol{\theta}}(\boldsymbol{x})\coloneqq\sqrt{n}\bigl(\hat{R}-\hat{R}_{\boldsymbol{\theta}}\bigr)(\boldsymbol{x})=\frac{1}{\sqrt{n}}\sum_{i=1}^{n}U_{\boldsymbol{\theta}}(\boldsymbol{W}_{i})1_{(-\boldsymbol{\infty},\boldsymbol{x}]}(\boldsymbol{X}_{i}),
\]
where $U_{\boldsymbol{\theta}}(y,\boldsymbol{x})\coloneqq y-m_{\boldsymbol{\theta}}(\boldsymbol{x})$
is the residual; we also write $U_{\boldsymbol{\theta}}(\boldsymbol{w})$
for $\boldsymbol{w}=(y,\boldsymbol{x}^{\prime})^{\prime}$ \citep[e.g.,][]{bhattacharya1974,bhattacharya1976,andrews1997conditional,stute1997nonparametric}.

\looseness=-1 Tests based on the IREP and/or its transformations are omnibus. They are
designed to detect any alternative to (\ref{100}), $H_{1}:m\notin\mathcal{M}$,
which can equivalently be expressed as 
\begin{equation*}
H_{1}:\quad\text{for every }\boldsymbol{\theta}\in\Theta,\quad R(\boldsymbol{x})\neq R_{\boldsymbol{\theta}}(\boldsymbol{x})\text{ for some }\boldsymbol{x}\in\mathbb{R}^{k}.
\end{equation*}
The limiting null distribution of $\hat{V}_{\boldsymbol{\tilde{\theta}}}(\boldsymbol{x})$,
with $\boldsymbol{\tilde{\theta}}=\boldsymbol{\theta}_{0}+O_{P}(n^{-1/2})$
under $H_{0}$, depends on unknown features of $P $ and the estimator
used (e.g., \citealt{stute1997nonparametric}, \citealt{bierens1997asymptotic}, or \citealt{andrews1997conditional}).
This is why the resulting tests are commonly implemented using bootstrap
approximations (e.g., \citealt{stute1998bootstrap}).

\looseness=-1 Given a partition of $\mathbb{R}^{k}$ into $L>p $ disjoint cells,
$\boldsymbol{C}=\{C_{1},\ldots,C_{L}\}$, say, consider tests designed
to detect alternatives of the form 
\[
H_{1\boldsymbol{C}}:\boldsymbol{M}\neq\boldsymbol{M}_{\boldsymbol{\theta}}\text{ for all }\boldsymbol{\theta}\in\Theta,
\]
where $\boldsymbol{M}=\mathbb{E}[Y\mathbf{1}_{\boldsymbol{C}}(\boldsymbol{X})]$
and $\boldsymbol{M}_{\boldsymbol{\theta}}=\mathbb{E}[m_{\boldsymbol{\theta}}(\boldsymbol{X})\mathbf{1}_{\boldsymbol{C}}(\boldsymbol{X})]$,
with $\mathbf{1}_{\boldsymbol{C}}(\boldsymbol{x})=(1_{C_{1}}(\boldsymbol{x}),\ldots,1_{C_{L}}(\boldsymbol{x}))^{\prime}$.
For any Borel set $A\subset\mathbb{R}^{k}$, define the signed measures
$M(A)\coloneqq\mathbb{E}[Y1_{A}(\boldsymbol{X})]$ and $M_{\boldsymbol{\theta}}(A)\coloneqq\mathbb{E}[m_{\boldsymbol{\theta}}(\boldsymbol{X})1_{A}(\boldsymbol{X})]$.
Then $\boldsymbol{M}$ and $\boldsymbol{M}_{\boldsymbol{\theta}}$
are the corresponding $L\times1$ vectors of cell moments, with $\ell $-th
components $M(C_{\ell})$ and $M_{\boldsymbol{\theta}}(C_{\ell})$,
respectively, $\ell=1,\ldots,L $. Thus, $H_{1\boldsymbol{C}}$ restricts
$H_{1}$ to alternatives such that, for every $\boldsymbol{\theta}\in\Theta $,
the residual $U_{\boldsymbol{\theta}}(\boldsymbol{W})$ has a nonzero
moment $\mathbb{E}[U_{\boldsymbol{\theta}}(\boldsymbol{W})1_{C_{\ell}}(\boldsymbol{X})]$
in at least one cell $C_{\ell}$.
The sample versions are $\boldsymbol{\hat{M}}\coloneqq n^{-1}\sum_{i=1}^{n}Y_{i}\mathbf{1}_{\boldsymbol{C}}(\boldsymbol{X}_{i})$
and $\boldsymbol{\hat{M}}_{\boldsymbol{\theta}}\coloneqq n^{-1}\sum_{i=1}^{n}m_{\boldsymbol{\theta}}(\boldsymbol{X}_{i})\mathbf{1}_{\boldsymbol{C}}(\boldsymbol{X}_{i})$.
Write $\hat{M}_{\ell}$ and $\hat{M}_{\boldsymbol{\theta}\ell}$ for
their $\ell $-th components, that is, the sample versions of $M(C_{\ell})$
and $M_{\boldsymbol{\theta}}(C_{\ell})$, respectively. These are
the observed and model-implied sample cell moments. Their scaled difference
is 
\[
\boldsymbol{\hat{V}}_{\boldsymbol{\theta}}\coloneqq\sqrt{n}\left(\boldsymbol{\hat{M}}-\boldsymbol{\hat{M}}_{\boldsymbol{\theta}}\right)=\frac{1}{\sqrt{n}}\sum_{i=1}^{n}U_{\boldsymbol{\theta}}(\boldsymbol{W}_{i})\mathbf{1}_{\boldsymbol{C}}(\boldsymbol{X}_{i}).
\]
Under $H_{0}$, there is a $\boldsymbol{\theta}_{0}\in\Theta $ such
that $\boldsymbol{\hat{V}}_{\boldsymbol{\theta}_{0}}=\boldsymbol{\hat{V}}_{0}$
a.s.-$P $, where 
\[
\boldsymbol{\hat{V}}_{0}\coloneqq\frac{1}{\sqrt{n}}\sum_{i=1}^{n}U_{0}(\boldsymbol{W}_{i})\mathbf{1}_{\boldsymbol{C}}(\boldsymbol{X}_{i}).
\]
Here $U_{0}(\boldsymbol{w})\coloneqq y-m(\boldsymbol{x})$ is the
true regression error, so $\mathbb{E}[U_{0}(\boldsymbol{W})\mid\boldsymbol{X}]=0$.
Under $H_{0}$, $m_{\boldsymbol\theta_0}(\boldsymbol X)=m(\boldsymbol X)$ a.s.-$P_{\boldsymbol X}$, where $P_{\boldsymbol\xi}$ denotes the probability distribution induced by a random vector $\boldsymbol\xi$, and hence
$U_{\boldsymbol{\theta}_{0}}(\boldsymbol{W})=U_{0}(\boldsymbol{W})$
a.s.-$P $. Thus, $\boldsymbol{\hat{V}}_{0}$ is centered at zero with
covariance matrix 
\begin{equation*}
\operatorname{Var}(\boldsymbol{\hat{V}}_{0})=\operatorname{diag}\left\{ \mathbb{E}\!\left[\operatorname{Var}(Y\mid\boldsymbol{X})\mathbf{1}_{\boldsymbol{C}}(\boldsymbol{X})\right]\right\} =:\boldsymbol{\Sigma}_{0}.
\end{equation*}
Its $\ell $-th diagonal component is $\sigma_{0\ell}^{2}=\mathbb{E}[U_{0}^{2}(\boldsymbol{W})1_{C_{\ell}}(\boldsymbol{X})]$.

\looseness=-1 When $\boldsymbol{\theta}_{0}$ is known, the test statistic is $\hat{\mathsf{X}}_{\boldsymbol{\theta}_{0}}^{2}(\boldsymbol{\theta}_{0})$,
where 
\begin{equation}
\hat{\mathsf{X}}_{\boldsymbol{\theta}_{2}}^{2}(\boldsymbol{\theta}_{1})\coloneqq n\sum_{\ell=1}^{L}\frac{(\hat{M}_{\ell}-\hat{M}_{\boldsymbol{\theta}_{1}\ell})^{2}}{\hat{\sigma}_{\boldsymbol{\theta}_{2}\ell}^{2}}=\boldsymbol{\hat{V}}_{\boldsymbol{\theta}_{1}}^{\prime}\boldsymbol{\hat{\Sigma}}_{\boldsymbol{\theta}_{2}}^{-1}\boldsymbol{\hat{V}}_{\boldsymbol{\theta}_{1}},\label{5}
\end{equation}
with $\boldsymbol{\hat{\Sigma}}_{\boldsymbol{\theta}}\coloneqq\operatorname{diag}\{\hat{\sigma}_{\boldsymbol{\theta}1}^{2},\ldots,\hat{\sigma}_{\boldsymbol{\theta}L}^{2}\}$
and $\hat{\sigma}_{\boldsymbol{\theta}\ell}^{2}\coloneqq n^{-1}\sum_{i=1}^{n}U_{\boldsymbol{\theta}}^{2}(\boldsymbol{W}_{i})1_{C_{\ell}}(\boldsymbol{X}_{i})$.
The statistic $\hat{\mathsf{X}}_{\boldsymbol{\theta}_{0}}^{2}(\boldsymbol{\theta}_{0})$
extends the Pearson statistic to regression model specification under
simple hypotheses. As in the classical case, it is the sum of the squared
differences between the unrestricted (``observed'') and model-implied
(``expected'') sample cell moments, $\hat M_\ell$ and
$\hat M_{\boldsymbol\theta_0\ell}$ respectively, weighted by the inverse of their estimated variances. If $\mathbb{E}|Y|^{2}<\infty $, then, under $H_{0}$,
$\boldsymbol{\hat{\Sigma}}_{\boldsymbol{\theta}_{0}}\rightarrow_{P}\boldsymbol{\Sigma}_{0}$
and $\boldsymbol{\hat{V}}_{0}\rightarrow_{d}\mathcal{N}_{L}(\boldsymbol{0},\boldsymbol{\Sigma}_{0})$.
Hence, if $\sigma_{0\ell}^{2}>0$ for all $\ell=1,\ldots,L $, $\hat{\mathsf{X}}_{\boldsymbol{\theta}_{0}}^{2}(\boldsymbol{\theta}_{0})\longrightarrow_{d}\chi_{L}^{2}$
by the continuous mapping theorem.

\looseness=-1 Under composite hypotheses, we minimize the Pearson criterion function using the GMM estimator constructed from the cell moment restrictions. For any preliminary
estimator $\boldsymbol{\tilde{\theta}}=\boldsymbol{\theta}_{0}+o_{P}(1)$
under $H_{0}$, take a measurable $\boldsymbol{\hat{\theta}}\in\Theta $
satisfying 
\[
\hat{\mathsf{X}}_{\boldsymbol{\tilde\theta}}^2(\boldsymbol{\hat\theta})
\leq\inf_{\boldsymbol\theta\in\Theta}
\hat{\mathsf{X}}_{\boldsymbol{\tilde\theta}}^2(\boldsymbol\theta)+o_P(1). \]
The proposed test statistic is 
\[
\hat{\mathsf{X}}_{\boldsymbol{\tilde{\theta}}}^{2}(\boldsymbol{\hat{\theta}})=\boldsymbol{\hat{V}}_{\boldsymbol{\hat{\theta}}}^{\prime}\boldsymbol{\hat{\Sigma}}_{\boldsymbol{\tilde{\theta}}}^{-1}\boldsymbol{\hat{V}}_{\boldsymbol{\hat{\theta}}}.
\]

Under $H_{1\boldsymbol C}$, the test is consistent if $n^{-1}\hat{\mathsf X}^2_{\tilde{\boldsymbol\theta}}(\cdot)$ converges uniformly over $\Theta$ to a function with a strictly positive infimum and the limiting weighting matrix is nonsingular.

\looseness=-1 The following assumptions are regression counterparts of those used
by \citet{birch1964detection} for grouped-data goodness-of-fit tests,
following the presentation in \citet{pollard1979general}. 
Besides the usual identification, moment, differentiability and rank
conditions, we need Glivenko--Cantelli and Donsker conditions on the
regression class, so that sample averages and centered empirical processes
can be controlled uniformly; see \citet[chap.~2]{van1996weak}. We assume
that the classes involved are pointwise measurable, or more generally that
they satisfy the measurability conditions of the results we cite. For $\epsilon>0$, let
$\mathcal{M}_{\epsilon}\coloneqq\{m_{\boldsymbol{\theta}}:\boldsymbol{\theta}\in\Theta,\ \|\boldsymbol{\theta}-\boldsymbol{\theta}_{0}\|<\epsilon\}$.
Throughout, $\|\cdot\|$ denotes the Euclidean norm, and
$\|\boldsymbol{a}\|_{Q,2}\coloneqq[\int\boldsymbol{a}^{\prime}(\boldsymbol{x})\boldsymbol{a}(\boldsymbol{x})\,Q(d\boldsymbol{x})]^{1/2}$
denotes the $L^2(Q)$ norm of a vector-valued function $\boldsymbol a$.
\Needspace{8\baselineskip}
\begin{description}
\item [{A1}] ~ \vspace{-0.35\baselineskip}
\begin{description}\setlength{\topsep}{0pt}\setlength{\partopsep}{0pt}
\item [{(i)}] $\boldsymbol{\theta}_{0}$ is an interior point of $\Theta $, 
\item [{(ii)}] for every $\epsilon>0$, there is an $\eta(\epsilon)>0$
such that $\inf_{\|\boldsymbol{\theta}-\boldsymbol{\theta}_{0}\|>\epsilon}\|\boldsymbol{M}_{\boldsymbol{\theta}}-\boldsymbol{M}_{\boldsymbol{\theta}_{0}}\|>\eta(\epsilon)$,
\item [{(iii)}] $\mathbb{E}|Y|^{2}<\infty $, $\sigma_{0\ell}^{2}>0$ for
each $\ell=1,\ldots,L $, and $L>p $, 
\item [{(iv)}] $\boldsymbol{\theta}\mapsto m_{\boldsymbol{\theta}}$ is
$L^{2}(P_{\boldsymbol{X}})$-differentiable at $\boldsymbol{\theta}_{0}$,
i.e., there is a $p\times1$ derivative vector $\boldsymbol{\dot{m}}\in L^{2}(P_{\boldsymbol{X}})$
such that $\|m_{\boldsymbol{\theta}}-m_{\boldsymbol{\theta}_{0}}-\boldsymbol{\dot{m}}^{\prime}(\boldsymbol{\theta}-\boldsymbol{\theta}_{0})\|_{P_{\boldsymbol{X}},2}=o(\|\boldsymbol{\theta}-\boldsymbol{\theta}_{0}\|)\quad\text{as }\boldsymbol{\theta}\to\boldsymbol{\theta}_{0},$
\item [{(v)}] the $L\times p $ matrix $\dot{\boldsymbol{M}}=\mathbb{E}[\mathbf{1}_{\boldsymbol{C}}(\boldsymbol{X})\boldsymbol{\dot{m}}(\boldsymbol{X})^{\prime}]$
has full column rank $p $, 
\item [{(vi)}] the global class $\mathcal{M}$ is $P_{\boldsymbol{X}}$-Glivenko--Cantelli
with an envelope in $L^{2}(P_{\boldsymbol{X}})$; for some $\epsilon>0$,
the local class $\mathcal{M}_{\epsilon}$ is $P_{\boldsymbol{X}}$-Donsker. 
\end{description}
\end{description}
Assumptions A1(i) and A1(ii) are needed
for consistency, while A1(iii) ensures that the residual variance within each cell is finite
and positive. Assumption A1(iv) implies
differentiability of the cell moment vector. Indeed, by the Cauchy--Schwarz
inequality and the fact that the cells form a partition, 
\begin{align}
 & \left\Vert \boldsymbol{M}_{\boldsymbol{\theta}}-\boldsymbol{M}_{\boldsymbol{\theta}_{0}}-\dot{\boldsymbol{M}}(\boldsymbol{\theta}-\boldsymbol{\theta}_{0})\right\Vert \nonumber \\
 & \qquad\leq\left\Vert m_{\boldsymbol{\theta}}-m_{\boldsymbol{\theta}_{0}}-\boldsymbol{\dot{m}}^{\prime}(\boldsymbol{\theta}-\boldsymbol{\theta}_{0})\right\Vert _{P_{\boldsymbol{X}},2}=o(\|\boldsymbol{\theta}-\boldsymbol{\theta}_{0}\|)\quad\text{as }\boldsymbol{\theta}\to\boldsymbol{\theta}_{0}.\label{eq:03}
\end{align}
The implied $L^{2}(P_{\boldsymbol{X}})$-continuity also provides
the continuity required for consistency of the variance estimators.
Finally, A1(vi) supplies the uniform convergence and stochastic equicontinuity
needed for consistency and the asymptotic expansion, respectively.

\begin{remark}\label{rem:primitive} Primitive sufficient conditions
for A1(vi) are standard. The global Glivenko--Cantelli and envelope
requirements hold if $\Theta $ is compact, $\boldsymbol{\theta}\mapsto m_{\boldsymbol{\theta}}(\boldsymbol{x})$
is continuous on $\Theta $ for $P_{\boldsymbol{X}}$-almost every
$\boldsymbol{x}$, and $\mathcal{M}$ has an $L^{2}(P_{\boldsymbol{X}})$
envelope; see \citet[Lemma~2.4]{NeweyMcFadden1994}. In linear regression
with a compact parameter space, this envelope exists if the regressors
have finite second moments. If the parameter space is not bounded,
we first prove consistency separately. The rest of the asymptotic
analysis can then be carried out on a bounded neighborhood of the true
parameter, where a square-integrable envelope is easier to obtain.
In linear regression the minimizer has a closed form, so consistency follows directly from the convergence of the cell averages. For the local Donsker
requirement, it suffices that, for some $\epsilon>0$, 
\[
|m_{\boldsymbol{\theta}_{1}}(\boldsymbol{x})-m_{\boldsymbol{\theta}_{2}}(\boldsymbol{x})|\leq G(\boldsymbol{x})\|\boldsymbol{\theta}_{1}-\boldsymbol{\theta}_{2}\|,\qquad G\in L^{2}(P_{\boldsymbol{X}}),
\]
for $\|\boldsymbol{\theta}_{j}-\boldsymbol{\theta}_{0}\|<\epsilon $,
$j=1,2$. Under this condition the local class has polynomial covering numbers
and an envelope in $L^{2}(P_{\boldsymbol{X}})$, so it is
$P_{\boldsymbol{X}}$-Donsker, given the usual measurability conditions.
The Lipschitz bound holds, for instance, if $m_{\boldsymbol{\theta}}(\boldsymbol{x})$
is continuously differentiable in $\boldsymbol{\theta}$ on a small
convex neighborhood of $\boldsymbol{\theta}_{0}$ and its derivative
has an envelope in $L^{2}(P_{\boldsymbol{X}})$.
\end{remark}

\begin{theorem}\label{thm:fixed} Under $H_{0}$, if A1 holds, $\hat{\mathsf{X}}_{\boldsymbol{\tilde{\theta}}}^{2}(\boldsymbol{\hat{\theta}})\longrightarrow_{d}\chi_{L-p}^{2}$
for any $\boldsymbol{\tilde{\theta}}=\boldsymbol{\theta}_{0}+o_{P}(1)$.
\end{theorem}

All proofs are in Appendix~A. The proof of Theorem~\ref{thm:fixed} combines~\eqref{eq:03} with uniform empirical-process approximations to obtain a quadratic expansion of the criterion.

\section{Data-Dependent Partitions}

\label{sec:ddp}

The cells may depend on characteristics of the data or the model and,
in particular, may use information on alternatives of interest.
Following \citet{pollard1979general}, we account for the randomness
of the partition by restricting the class of admissible cells and
imposing convergence of the estimated partition.

\looseness=-1 Let $\mathcal{C}\subset\mathcal{B}^{k}$ be a class of Borel subsets
of $\mathbb{R}^{k}$ and define the class of ordered partitions 
\[
\mathcal{G}\coloneqq\left\{ \boldsymbol{\gamma}=(\gamma_{1},\ldots,\gamma_{L})\in\mathcal{C}^{L}:\gamma_{1},\ldots,\gamma_{L}\text{ are disjoint and }\bigcup_{j=1}^{L}\gamma_{j}=\mathbb{R}^{k}\right\} .
\]
For $A,B\in\mathcal{C}$, let $d_{P_{\boldsymbol{X}}}(A,B)\coloneqq\|1_{A}-1_{B}\|_{P_{\boldsymbol{X}},2}=P_{\boldsymbol{X}}(A\triangle B)^{1/2}$,
where $\triangle $ denotes symmetric difference. We equip $\mathcal{G}$
with the product semimetric 
\[
d_{P_{\boldsymbol{X}}}(\boldsymbol{\gamma},\widetilde{\boldsymbol{\gamma}})\coloneqq\left\{ \sum_{\ell=1}^{L}P_{\boldsymbol{X}}(\gamma_{\ell}\triangle\widetilde{\gamma}_{\ell})\right\} ^{1/2}.
\]
For a random partition $\hat{\boldsymbol{\Gamma}}\in\mathcal{G}$,
we write $\hat{\boldsymbol{\Gamma}}\to_{P}\boldsymbol{\Gamma}$ if
$d_{P_{\boldsymbol{X}}}(\hat{\boldsymbol{\Gamma}},\boldsymbol{\Gamma})\to_{P}0$.
Since $L $ is fixed, this is equivalent to $P_{\boldsymbol{X}}(\hat{\Gamma}_{\ell}\triangle\Gamma_{\ell})\to_{P}0$
for every $\ell=1,\ldots,L $. The statistic is invariant to a common permutation of the cells -- labels are used only to align the components of the moment vector and to define convergence in the product semimetric. Since the statistic does not depend on the order of the cells, it is enough that the cells converge after a suitable relabeling.

\looseness=-1 We now introduce the notation that depends on the partition. For $\boldsymbol{\gamma}\in\mathcal{G}$,
let $\mathbf{1}_{\boldsymbol{\gamma}}(\boldsymbol{x})=(1_{\gamma_{1}}(\boldsymbol{x}),\ldots,1_{\gamma_{L}}(\boldsymbol{x}))^{\prime}$,
and define $\boldsymbol{M}(\boldsymbol{\gamma})=\mathbb{E}[Y\mathbf{1}_{\boldsymbol{\gamma}}(\boldsymbol{X})]$
and $\boldsymbol{M}_{\boldsymbol{\theta}}(\boldsymbol{\gamma})=\mathbb{E}[m_{\boldsymbol{\theta}}(\boldsymbol{X})\mathbf{1}_{\boldsymbol{\gamma}}(\boldsymbol{X})]$.
The residual process and the derivative of the cell moment vector are
\[
\hat{\boldsymbol{V}}_{\boldsymbol{\theta}}(\boldsymbol{\gamma})=\sqrt{n}(\hat{\boldsymbol{M}}-\hat{\boldsymbol{M}}_{\boldsymbol{\theta}})(\boldsymbol{\gamma})=\frac{1}{\sqrt{n}}\sum_{i=1}^{n}U_{\boldsymbol{\theta}}(\boldsymbol{W}_{i})\mathbf{1}_{\boldsymbol{\gamma}}(\boldsymbol{X}_{i}),\quad\text{and}\quad\dot{\boldsymbol{M}}(\boldsymbol{\gamma})=\mathbb{E}[\mathbf{1}_{\boldsymbol{\gamma}}(\boldsymbol{X})\boldsymbol{\dot{m}}(\boldsymbol{X})^{\prime}],
\]
respectively.
Write $\hat{\boldsymbol{V}}_{0}(\boldsymbol{\gamma})$ for the same
process with $U_{0}$ in place of $U_{\boldsymbol{\theta}}$. For
a cell $A $, let 
\[
\hat{\sigma}_{\boldsymbol{\theta}}^{2}(A)=\frac{1}{n}\sum_{i=1}^{n}U_{\boldsymbol{\theta}}^{2}(\boldsymbol{W}_{i})1_{A}(\boldsymbol{X}_{i})\quad\text{and}\quad\sigma_{0}^{2}(A)=\mathbb{E}[U_{0}^{2}(\boldsymbol{W})1_{A}(\boldsymbol{X})],
\]
and set $\hat{\boldsymbol{\Sigma}}_{\boldsymbol{\theta}}(\boldsymbol{\gamma})=\operatorname{diag}\{\hat{\sigma}_{\boldsymbol{\theta}}^{2}(\gamma_{1}),\ldots,\hat{\sigma}_{\boldsymbol{\theta}}^{2}(\gamma_{L})\}$
and $\boldsymbol{\Sigma}_{0}(\boldsymbol{\gamma})=\operatorname{diag}\{\sigma_{0}^{2}(\gamma_{1}),\ldots,\sigma_{0}^{2}(\gamma_{L})\}$.
The analog of \eqref{5}, whenever $\hat\sigma_{\boldsymbol\theta_2}^{2}(\gamma_\ell)>0$ for every $\ell$, is 
\[
\hat{\mathsf{X}}_{\boldsymbol{\theta}_{2}}^{2}(\boldsymbol{\gamma},\boldsymbol{\theta}_{1})\coloneqq\hat{\boldsymbol{V}}_{\boldsymbol{\theta}_{1}}(\boldsymbol{\gamma})^{\prime}\hat{\boldsymbol{\Sigma}}_{\boldsymbol{\theta}_{2}}^{-1}(\boldsymbol{\gamma})\hat{\boldsymbol{V}}_{\boldsymbol{\theta}_{1}}(\boldsymbol{\gamma}).
\]
For a preliminary estimator $\tilde{\boldsymbol{\theta}}=\boldsymbol{\theta}_{0}+o_{P}(1)$
under $H_{0}$, use approximate minimizers satisfying
\[
\hat{\mathsf{X}}_{\tilde{\boldsymbol\theta}}^2
(\boldsymbol\gamma,\hat{\boldsymbol\theta}(\boldsymbol\gamma))
\leq\inf_{\boldsymbol\theta\in\Theta}
\hat{\mathsf{X}}_{\tilde{\boldsymbol\theta}}^2(\boldsymbol\gamma,\boldsymbol\theta)
+\hat r(\boldsymbol\gamma),\qquad
\sup_{\boldsymbol\gamma\in\mathcal G}\hat r(\boldsymbol\gamma)=o_P(1).
\]
The optimization error $\hat r$ is zero for exact minimizers. The proposed statistic is $\hat{\mathsf{X}}_{\tilde{\boldsymbol{\theta}}}^{2}(\hat{\boldsymbol{\Gamma}},\hat{\boldsymbol{\theta}})$, where $\hat{\boldsymbol{\theta}}=\hat{\boldsymbol{\theta}}(\hat{\boldsymbol{\Gamma}})$.

The cell class $\mathcal{C}$ is assumed to be Vapnik--Chervonenkis
(VC). Its indicator class $\mathcal{I}_{\mathcal{C}}=\{1_{C}:C\in\mathcal{C}\}$
therefore has a bounded uniform entropy integral (BUEI), with envelope
one \citep[sec.~2.6]{van1996weak}.

\looseness=-1 Two distinct uniform empirical-process arguments are needed. First,
with the regression error $U_{0}$ held fixed, the relevant marked
class is 
\[
\mathcal{F}_{0}\coloneqq\{(y,\boldsymbol{x})\mapsto U_{0}(y,\boldsymbol{x})1_{C}(\boldsymbol{x}):C\in\mathcal{C}\}.
\]
Under $H_{0}$ and $\mathbb{E}|Y|^{2}<\infty $, this class has the
square-integrable envelope $|U_{0}|$. Since $\mathcal{I}_{\mathcal{C}}$
has BUEI and $U_{0}$ is a fixed square-integrable mark, the marked
class $\mathcal{F}_{0}$ is $P $-Donsker under the conventions stated above; see \citet[Theorems~2.5.2 and~2.6.7]{van1996weak}. Since $\mathbb{E}[U_{0}(\boldsymbol{W})\mid\boldsymbol{X}]=0$,
this gives 
\[
\hat{\boldsymbol{V}}_{0}(\cdot)\longrightarrow_{d}\boldsymbol{V}_{0}(\cdot)\quad\text{in }\ell^{\infty}(\mathcal{G};\mathbb{R}^{L}),
\]
where $\ell^{\infty}(\mathcal{G};\mathbb{R}^{L})$ is the space of
bounded functions from $\mathcal{G}$ to $\mathbb{R}^{L}$, endowed
with the supremum norm, and $\boldsymbol{V}_{0}$ is a tight, centered
Gaussian process. Notice that $\operatorname{Var}[\boldsymbol{V}_{0}(\boldsymbol{\gamma})]=\boldsymbol{\Sigma}_{0}(\boldsymbol{\gamma})$.

\looseness=-1 Second, replacing the fitted parameters while allowing the cells to vary requires stochastic equicontinuity over the local product class
$\{\boldsymbol{x}\mapsto m_{\boldsymbol{\theta}}(\boldsymbol{x})1_C(\boldsymbol{x}):m_{\boldsymbol{\theta}}\in\mathcal M_\epsilon,C\in\mathcal C\}$.
The BUEI assumption on $\mathcal M_\epsilon$ below and the indicator entropy bound make this class Donsker. Together with $L^2(P_{\boldsymbol X})$-continuity at $\boldsymbol\theta_0$, this gives the following centered-increment bound for every deterministic $\delta_n\downarrow0$:
\[
\sup_{\substack{\boldsymbol{\gamma}\in\mathcal{G}\\
\boldsymbol{\theta}\in\Theta:\,\|\boldsymbol{\theta}-\boldsymbol{\theta}_{0}\|\leq\delta_{n}
}
}\left\Vert \hat{\boldsymbol{V}}_{\boldsymbol{\theta}}(\boldsymbol{\gamma})-\hat{\boldsymbol{V}}_{0}(\boldsymbol{\gamma})+\sqrt{n}\{\boldsymbol{M}_{\boldsymbol{\theta}}(\boldsymbol{\gamma})-\boldsymbol{M}_{\boldsymbol{\theta}_{0}}(\boldsymbol{\gamma})\}\right\Vert =o_{P}(1).
\]
See the argument in Appendix A. The $L^{2}(P_{\boldsymbol{X}})$-differentiability assumption
also yields, by the same bound as in (\ref{eq:03}), 
\[
\sup_{\boldsymbol{\gamma}\in\mathcal{G}}\left\Vert \boldsymbol{M}_{\boldsymbol{\theta}}(\boldsymbol{\gamma})-\boldsymbol{M}_{\boldsymbol{\theta}_{0}}(\boldsymbol{\gamma})-\dot{\boldsymbol{M}}(\boldsymbol{\gamma})(\boldsymbol{\theta}-\boldsymbol{\theta}_{0})\right\Vert =o(\|\boldsymbol{\theta}-\boldsymbol{\theta}_{0}\|)\quad\text{as }\boldsymbol{\theta}\to\boldsymbol{\theta}_{0}.
\]
The Glivenko--Cantelli and envelope conditions below are global and give uniform laws of large numbers for the cell moments, which yield consistency. A local law for the squared residuals gives consistent weights at the preliminary estimate, and local differentiability and Donsker conditions give the quadratic expansion uniformly over partitions. We assume the following.

\begin{description}
\item [{A2}] ~ \vspace{-0.35\baselineskip}
\begin{description}\setlength{\topsep}{0pt}\setlength{\partopsep}{0pt}
\item [{(i)}] $\hat{\boldsymbol{\Gamma}}\rightarrow_{P}\boldsymbol{\Gamma}\in\mathcal{G}$
for a fixed, deterministic partition $\boldsymbol{\Gamma}$; 
\item [{(ii)}] $\boldsymbol{\theta}_{0}$ is an interior point of $\Theta $; 
\item [{(iii)}] for each $\epsilon>0$, there is an $\eta(\epsilon)>0$
such that 
\[
\inf_{\|\boldsymbol{\theta}-\boldsymbol{\theta}_{0}\|>\epsilon}\|\boldsymbol{M}_{\boldsymbol{\theta}}(\boldsymbol{\Gamma})-\boldsymbol{M}_{\boldsymbol{\theta}_{0}}(\boldsymbol{\Gamma})\|>\eta(\epsilon);
\]
\item [{(iv)}] $\mathbb{E}|Y|^{2}<\infty $ and $\sigma_{0}^{2}(\Gamma_{\ell})>0$
for each $\ell=1,\ldots,L $, with $L>p $; 
\item [{(v)}] $\boldsymbol{\theta}\mapsto m_{\boldsymbol{\theta}}(\cdot)$
is differentiable in $L^{2}(P_{\boldsymbol{X}})$ at $\boldsymbol{\theta}_{0}$
with derivative $\dot{\boldsymbol{m}}\in L^{2}(P_{\boldsymbol{X}})$,
i.e., 
\[
\|m_{\boldsymbol{\theta}}-m_{\boldsymbol{\theta}_{0}}-\dot{\boldsymbol{m}}'(\boldsymbol{\theta}-\boldsymbol{\theta}_{0})\|_{P_{\boldsymbol{X}},2}=o(\|\boldsymbol{\theta}-\boldsymbol{\theta}_{0}\|)\quad\text{as }\boldsymbol{\theta}\to\boldsymbol{\theta}_{0};
\]
\item [{(vi)}] the $L\times p $ matrix $\dot{\boldsymbol{M}}(\boldsymbol{\Gamma})$
has rank $p $; 
\item [{(vii)}] $\mathcal{M}$ is $P_{\boldsymbol{X}}$-Glivenko--Cantelli
with envelope in $L^{2}(P_{\boldsymbol{X}})$, and the local parametric
class $\mathcal{M}_{\epsilon}$ is BUEI with an envelope in $L^{2}(P_{\boldsymbol{X}})$
for some $\epsilon>0$; 
\item [{(viii)}] the class of sets $\mathcal{C}$ from which the cells
are chosen is a VC class of sets. 
\end{description}
\end{description}
A2(iv) and A2(viii)
give the Donsker property of the fixed-error process. Its natural
increment semimetric is 
\[
\rho_{0}^{2}(\boldsymbol{\gamma},\widetilde{\boldsymbol{\gamma}})\coloneqq\sum_{\ell=1}^{L}\mathbb{E}[U_{0}^{2}(\boldsymbol{W})1_{\gamma_{\ell}\triangle\widetilde{\gamma}_{\ell}}(\boldsymbol{X})].
\]
Since $\mathbb{E}[U_{0}^{2}(\boldsymbol{W})]<\infty $, absolute continuity
of the integral yields $\rho_{0}(\boldsymbol{\gamma}_{n},\boldsymbol{\gamma})\to0$
whenever $d_{P_{\boldsymbol{X}}}(\boldsymbol{\gamma}_{n},\boldsymbol{\gamma})\to0$.
Therefore, A2(i) and stochastic equicontinuity imply 
\[
\hat{\boldsymbol{V}}_{0}(\hat{\boldsymbol{\Gamma}})=\hat{\boldsymbol{V}}_{0}(\boldsymbol{\Gamma})+o_{P}(1).
\]
No independence between the estimated partition and the observations
is required. Primitive conditions for A2(i) for the procedures considered
below are discussed in Section~\ref{sec:partition_choice}.

\looseness=-1 Assumptions A2(ii)--A2(vi) are the counterparts of A1(i)--A1(v)
at the limiting partition. A2(vii) supplies the additional uniformity
needed for parameter estimation, and the VC assumption in A2(viii)
controls variation across cells. More generally, the latter can be
replaced by BUEI of the indicator class. Uniform laws of large numbers over $\mathcal G$ and consistency of the preliminary estimator give $\sup_{\boldsymbol\gamma\in\mathcal G}\|\hat{\boldsymbol\Sigma}_{\tilde{\boldsymbol\theta}}(\boldsymbol\gamma)-\boldsymbol\Sigma_0(\boldsymbol\gamma)\|\to_P0$. Since A2(i) gives $\hat{\boldsymbol\Gamma}\to_P\boldsymbol\Gamma$, continuity of the population cell moments then yields
$\hat{\boldsymbol\Sigma}_{\tilde{\boldsymbol\theta}}(\hat{\boldsymbol\Gamma})\to_P\boldsymbol\Sigma_0(\boldsymbol\Gamma)$.
Positivity in A2(iv) makes the inverse well defined with probability tending to one. Continuity also implies that the population derivative matrix evaluated at the estimated partition satisfies
$\dot{\boldsymbol M}(\hat{\boldsymbol\Gamma})\to_P\dot{\boldsymbol M}(\boldsymbol\Gamma)$.
Combining
these facts with the uniform local criterion expansion and approximate minimization gives
\begin{equation}\label{eq:partition-equivalence}
\hat{\mathsf{X}}_{\tilde{\boldsymbol{\theta}}}^{2}(\hat{\boldsymbol{\Gamma}},\hat{\boldsymbol{\theta}}(\hat{\boldsymbol{\Gamma}}))-\hat{\mathsf{X}}_{\tilde{\boldsymbol{\theta}}}^{2}(\boldsymbol{\Gamma},\hat{\boldsymbol{\theta}}(\boldsymbol{\Gamma}))=o_{P}(1).
\end{equation}

\begin{theorem}\label{thm:ddp} Under $H_{0}$, if A2 holds, $\hat{\mathsf{X}}_{\tilde{\boldsymbol{\theta}}}^{2}(\hat{\boldsymbol{\Gamma}},\hat{\boldsymbol{\theta}})\longrightarrow_{d}\chi_{L-p}^{2}$
for any $\tilde{\boldsymbol{\theta}}=\boldsymbol{\theta}_{0}+o_{P}(1)$.
\end{theorem}

The result follows from \eqref{eq:partition-equivalence} and the
fixed-partition limit in Theorem~\ref{thm:fixed}. Section~\ref{app:S.1.2}
provides the uniform convergence over partitions and the quadratic-expansion arguments.

\looseness=-1 Convergence of $\hat{\boldsymbol\Gamma}$ to a fixed partition is sufficient, but not necessary, for the chi-square limit. In the proof it plays two roles. It keeps $\hat{\boldsymbol\Sigma}_{\tilde{\boldsymbol\theta}}(\hat{\boldsymbol\Gamma})$ and $\dot{\boldsymbol M}(\hat{\boldsymbol\Gamma})$ away from singularity, and, through $\hat{\boldsymbol V}_0(\hat{\boldsymbol\Gamma})=\hat{\boldsymbol V}_0(\boldsymbol\Gamma)+o_P(1)$, it reduces the problem to a fixed partition. As \citet[sec.~6]{pollard1979general} observes, the first role only requires that, with probability arbitrarily close to one, $\hat{\boldsymbol\Gamma}$ lies in a compact subset of $(\mathcal G,d_{P_{\boldsymbol X}})$ on which A2(iv) and A2(vi) hold. For the second role, it suffices that the pair $\{\hat{\boldsymbol V}_0(\cdot),\hat{\boldsymbol\Gamma}\}$ converges in distribution, as a random element of $\ell^\infty(\mathcal G;\mathbb R^L)\times\mathcal G$, to $\{\boldsymbol V_0(\cdot),\boldsymbol\Gamma^*\}$, where $\boldsymbol\Gamma^*$ is a random partition independent of $\boldsymbol V_0(\cdot)$. Then $\hat{\boldsymbol V}_0(\hat{\boldsymbol\Gamma})\to_d\boldsymbol V_0(\boldsymbol\Gamma^*)$, which given $\boldsymbol\Gamma^*=\boldsymbol\gamma$ is distributed as $\mathcal N_L(\boldsymbol 0,\boldsymbol\Sigma_0(\boldsymbol\gamma))$. The statistic therefore has a $\chi^2_{L-p}$ limit given each $\boldsymbol\gamma$, and hence also unconditionally. Pollard sketches, without a formal proof, why this independence would hold when $\hat{\boldsymbol\Gamma}$ converges in probability to a possibly random partition $\boldsymbol\Gamma^*$. Under A2(i), $\boldsymbol\Gamma^*=\boldsymbol\Gamma$ is fixed and independence is trivial.

\section{Local Power}

\label{sec:new-contiguous}

\looseness=-1 With data-dependent cells, the consistency argument of Section~\ref{sec:test} applies with $\boldsymbol{C}$ replaced by the limit $\boldsymbol\Gamma$ of $\hat{\boldsymbol\Gamma}$ under the alternative. To compare partitions, we consider alternatives that approach the null at rate $n^{-1/2}$. Under these alternatives the statistic converges to a noncentral chi-square distribution whose noncentrality parameter depends on the partition, and Section~\ref{sec:partition_choice} uses this parameter to compare partitions. \citet[sec.~3]{stute1997nonparametric} and \citet{delgado2008distribution} also study local alternatives of this kind, and our derivation of the noncentrality parameter follows \citet[sec.~6]{pollard1980minimum}.

\looseness=-1 Let $\{(\varepsilon_{i},\boldsymbol{X}_{i})\}_{i=1}^{n}$ be i.i.d.
with a fixed joint law, $\mathbb{E}(\varepsilon\mid\boldsymbol{X})=0$
and $\mathbb{E}(\varepsilon^{2})<\infty$. Consider the local regression
sequence
\[
H_{1n}:\quad Y_{i}=m_{\boldsymbol{\theta}_{0}}(\boldsymbol{X}_{i})+\frac{\delta(\boldsymbol{X}_{i})}{\sqrt{n}}+\varepsilon_{i},\qquad\delta\in L^{2}(P_{\boldsymbol{X}}),\qquad\mathbb{E}\{\delta(\boldsymbol{X})\}=0.
\]
The errors may be heteroskedastic. We suppress the dependence of $Y_{i}$ on $n$ and take expectations under $H_{1n}$. Only Remark~\ref{rem:constant-null-power} uses $\mathbb{E}\{\delta(\boldsymbol{X})\}=0$, which keeps $\mathbb{E}(Y)$ as under $H_{0}$.
\begin{description}
\item [{A3 (Local regularity)}] \label{ass:new-local} Under $H_{1n}$, $\hat{\boldsymbol{\Gamma}}\to_{P}\boldsymbol{\Gamma}$,
fixed, in the metric of Section~\ref{sec:ddp}, with $\hat{\boldsymbol{\Gamma}}\in\mathcal{G}$, and A2(ii)--A2(viii) hold at $\delta=0$, with the partition-dependent conditions evaluated at $\boldsymbol{\Gamma}$.
The objective function $\hat{\mathsf{X}}_{\tilde{\boldsymbol{\theta}}}^{2}(\hat{\boldsymbol{\Gamma}},\boldsymbol{\theta})$
is minimized up to a nonnegative error that is $o_{P}(1)$.
For a deterministic partition, A1 may replace A2.
\end{description}
Let $\boldsymbol{D}(\boldsymbol{\Gamma})=\mathbb{E}\{\delta(\boldsymbol{X})\mathbf{1}_{\boldsymbol{\Gamma}}(\boldsymbol{X})\}$
be the cell drift, let $\boldsymbol{\Sigma}_{0}(\boldsymbol{\Gamma})=\operatorname{diag}\{\mathbb{E}[\varepsilon^{2}\mathbf{1}_{\boldsymbol{\Gamma}}(\boldsymbol{X})]\}$,
and let $\boldsymbol{\dot{M}}(\boldsymbol{\Gamma})$ be the matrix of Section~\ref{sec:ddp}. Define
$\lambda(\boldsymbol{\Gamma})=\boldsymbol{d}(\boldsymbol{\Gamma})'\boldsymbol{\Pi}(\boldsymbol{\Gamma})\boldsymbol{d}(\boldsymbol{\Gamma})$,
where $\boldsymbol{d}(\boldsymbol{\Gamma})=\boldsymbol{\Sigma}_{0}^{-1/2}(\boldsymbol{\Gamma})\boldsymbol{D}(\boldsymbol{\Gamma})$
and
\begin{equation}
\boldsymbol{\Pi}(\boldsymbol{\Gamma})=\boldsymbol{I}_{L}-\boldsymbol{\Sigma}_{0}^{-1/2}(\boldsymbol{\Gamma})\boldsymbol{\dot{M}}(\boldsymbol{\Gamma})\{\boldsymbol{\dot{M}}(\boldsymbol{\Gamma})'\boldsymbol{\Sigma}_{0}^{-1}(\boldsymbol{\Gamma})\boldsymbol{\dot{M}}(\boldsymbol{\Gamma})\}^{-1}\boldsymbol{\dot{M}}(\boldsymbol{\Gamma})'\boldsymbol{\Sigma}_{0}^{-1/2}(\boldsymbol{\Gamma})\label{eq:Pi}
\end{equation}
is the orthogonal projection of rank $L-p$ that annihilates the column
space of $\boldsymbol{\Sigma}_{0}^{-1/2}(\boldsymbol{\Gamma})\boldsymbol{\dot{M}}(\boldsymbol{\Gamma})$.
To lighten notation, we drop the argument $\boldsymbol{\Gamma}$ of $\boldsymbol{D}$, $\boldsymbol{\Sigma}_{0}$, $\boldsymbol{\dot{M}}$, $\boldsymbol{d}$, and $\boldsymbol{\Pi}$ in what follows, and write $\hat{\boldsymbol{\theta}}$ for $\hat{\boldsymbol{\theta}}(\hat{\boldsymbol{\Gamma}})$. We keep it in $\hat{\boldsymbol{V}}_{\boldsymbol{\theta}_{0}}(\boldsymbol{\Gamma})$ and $\lambda(\boldsymbol{\Gamma})$.

\begin{theorem}[Local alternatives] \label{thm:new-local} Under
$H_{1n}$ and A3, with a preliminary estimator satisfying $\tilde{\boldsymbol{\theta}}\to_{P}\boldsymbol{\theta}_{0}$, $\hat{\mathsf{X}}_{\tilde{\boldsymbol{\theta}}}^{2}(\hat{\boldsymbol{\Gamma}},\hat{\boldsymbol{\theta}})\ \to_{d}\ \chi_{L-p}^{2}\{\lambda(\boldsymbol{\Gamma})\}$.
\end{theorem}

\looseness=-1 Section~\ref{app:S.1.3} proves Theorem~\ref{thm:new-local},
adapting the proof of Theorem~\ref{thm:ddp} to replace $\hat{\boldsymbol{\Gamma}}$
by $\boldsymbol{\Gamma}$. We outline where $\lambda(\boldsymbol{\Gamma})$
comes from. The departure enters through the cell moments at the true
parameter. Since, under $H_{1n}$, $U_{\boldsymbol{\theta}_{0}}(\boldsymbol{W}_{i})=\varepsilon_{i}+n^{-1/2}\delta(\boldsymbol{X}_{i})$,
\begin{equation}
\hat{\boldsymbol{V}}_{\boldsymbol{\theta}_{0}}(\boldsymbol{\Gamma})=\frac{1}{\sqrt{n}}\sum_{i=1}^{n}\varepsilon_{i}\mathbf{1}_{\boldsymbol{\Gamma}}(\boldsymbol{X}_{i})+\frac{1}{n}\sum_{i=1}^{n}\delta(\boldsymbol{X}_{i})\mathbf{1}_{\boldsymbol{\Gamma}}(\boldsymbol{X}_{i})\ \to_{d}\ \boldsymbol{D}+\boldsymbol{\Sigma}_{0}^{1/2}\boldsymbol{G},\label{eq:V0-local}
\end{equation}
with $\boldsymbol{G}\sim\mathcal{N}_{L}(\boldsymbol{0},\boldsymbol{I}_{L})$.
If $\boldsymbol{\theta}_{0}$ were known, the statistic evaluated at $\boldsymbol{\theta}_{0}$, $\hat{\mathsf{X}}_{\tilde{\boldsymbol{\theta}}}^{2}(\hat{\boldsymbol{\Gamma}},\boldsymbol{\theta}_{0})$, would therefore converge to $\|\boldsymbol{G}+\boldsymbol{d}\|^{2}$, which is distributed
as $\chi_{L}^{2}\{\lambda_{1}(\boldsymbol{\Gamma})\}$ with $\lambda_{1}(\boldsymbol{\Gamma})=\boldsymbol{d}'\boldsymbol{d}$.
Estimating $\boldsymbol{\theta}_{0}$ changes this limit, because
\begin{equation}
\sqrt{n}(\hat{\boldsymbol{\theta}}-\boldsymbol{\theta}_{0})=(\boldsymbol{\dot{M}}'\boldsymbol{\Sigma}_{0}^{-1}\boldsymbol{\dot{M}})^{-1}\boldsymbol{\dot{M}}'\boldsymbol{\Sigma}_{0}^{-1}\hat{\boldsymbol{V}}_{\boldsymbol{\theta}_{0}}(\boldsymbol{\Gamma})+o_{P}(1)\ \to_{d}\ \boldsymbol{b}+(\boldsymbol{\dot{M}}'\boldsymbol{\Sigma}_{0}^{-1}\boldsymbol{\dot{M}})^{-1}\boldsymbol{\dot{M}}'\boldsymbol{\Sigma}_{0}^{-1/2}\boldsymbol{G},\label{eq:theta-local}
\end{equation}
with asymptotic bias $\boldsymbol{b}=(\boldsymbol{\dot{M}}'\boldsymbol{\Sigma}_{0}^{-1}\boldsymbol{\dot{M}})^{-1}\boldsymbol{\dot{M}}'\boldsymbol{\Sigma}_{0}^{-1}\boldsymbol{D}$.
The estimator is not centered at $\boldsymbol{\theta}_{0}$, because
it offsets the part of the drift that a change in the parameter can
reproduce. Using arguments in the proof of Theorem~\ref{thm:ddp}
and~\eqref{eq:theta-local},
\[
\boldsymbol{\Sigma}_{0}^{-1/2}\hat{\boldsymbol{V}}_{\hat{\boldsymbol{\theta}}}(\hat{\boldsymbol{\Gamma}})=\boldsymbol{\Sigma}_{0}^{-1/2}\hat{\boldsymbol{V}}_{\boldsymbol{\theta}_{0}}(\boldsymbol{\Gamma})-\boldsymbol{\Sigma}_{0}^{-1/2}\boldsymbol{\dot{M}}\sqrt{n}(\hat{\boldsymbol{\theta}}-\boldsymbol{\theta}_{0})+o_{P}(1)=\boldsymbol{\Pi}\boldsymbol{\Sigma}_{0}^{-1/2}\hat{\boldsymbol{V}}_{\boldsymbol{\theta}_{0}}(\boldsymbol{\Gamma})+o_{P}(1),
\]
so, by~\eqref{eq:V0-local}, the statistic converges to $\|\boldsymbol{\Pi}(\boldsymbol{G}+\boldsymbol{d})\|^{2}$,
which is distributed as $\chi_{L-p}^{2}(\boldsymbol{d}'\boldsymbol{\Pi}\boldsymbol{d})$, because $\boldsymbol{\Pi}$ is idempotent and $\operatorname{rank}(\boldsymbol{\Pi})=\operatorname{trace}(\boldsymbol{\Pi})=L-p$.

\looseness=-1 The noncentrality parameter splits into the two sources of the limit,
\[
\lambda(\boldsymbol{\Gamma})=\boldsymbol{d}'\boldsymbol{d}-\boldsymbol{d}'(\boldsymbol{I}_{L}-\boldsymbol{\Pi})\boldsymbol{d}=\lambda_{1}(\boldsymbol{\Gamma})-\lambda_{2}(\boldsymbol{\Gamma}),\qquad\lambda_{2}(\boldsymbol{\Gamma})=\boldsymbol{b}'\boldsymbol{\dot{M}}'\boldsymbol{\Sigma}_{0}^{-1}\boldsymbol{\dot{M}}\boldsymbol{b}.
\]
The first term comes from the limit of $\hat{\boldsymbol{V}}_{\boldsymbol{\theta}_{0}}(\boldsymbol{\Gamma})$
alone and depends only on the cell drift. The second is due to the
asymptotic bias of the estimator, which shifts the fitted cell moments
by $\boldsymbol{\dot{M}}\boldsymbol{b}$ and offsets the component
of the drift in the column space of $\boldsymbol{\dot{M}}$. Hence
$\lambda(\boldsymbol{\Gamma})\le\lambda_{1}(\boldsymbol{\Gamma})$,
with equality if and only if $\boldsymbol{\dot{M}}'\boldsymbol{\Sigma}_{0}^{-1}\boldsymbol{D}=\boldsymbol{0}$.
Equivalently, $\lambda(\boldsymbol{\Gamma})$ is the squared $\boldsymbol{\Sigma}_{0}^{-1}$-weighted
distance from $\boldsymbol{D}$ to the column space of $\boldsymbol{\dot{M}}$,
attained at $\boldsymbol{\dot{M}}\boldsymbol{b}$.

\looseness=-1 A nonzero departure need not yield positive noncentrality, because
its cell averages may vanish or its cell drift may lie in the column
space of $\boldsymbol{\dot{M}}$. In particular, a parametric tangent
direction $\delta(\boldsymbol{x})=\boldsymbol{\dot{m}}(\boldsymbol{x})^{\prime}\boldsymbol{t}$
has $\boldsymbol{D}=\boldsymbol{\dot{M}}\boldsymbol{t}$ and $\lambda(\boldsymbol{\Gamma})=0$.
The condition $\mathbb{E}\{\delta(\boldsymbol{X})\}=0$, which
preserves the unconditional mean, implies $\boldsymbol{1}'\boldsymbol{D}=0$,
but not $\boldsymbol{\dot{M}}^{\prime}\boldsymbol{\Sigma}_{0}^{-1}\boldsymbol{D}=\boldsymbol{0}$,
so it does not make $\lambda_{2}(\boldsymbol{\Gamma})$ vanish.

\section{Partition Choice}

\label{sec:partition_choice}

\looseness=-1 The partition determines the departures from the null that the test can detect, through the noncentrality $\lambda(\boldsymbol\Gamma)$ of Theorem~\ref{thm:new-local}. Generic partitions are built from the covariates alone, and guided
partitions use the fitted values of a parametric model. In either
case, every cell of the sample partition must contain at least one
observation, and Theorem~\ref{thm:ddp} requires the final cells
to satisfy the convergence, variance and rank conditions in A2(i), A2(iv) and A2(vi). Because the cells select the moment conditions,
the rank requirement is an identification condition, as in other GMM
procedures. Specifically, $\boldsymbol{\dot{M}}(\boldsymbol{\Gamma})$ has rank
smaller than $p $ if and only if some combination $\boldsymbol{c}'\boldsymbol{\dot{m}}(\boldsymbol{X})$,
$\boldsymbol{c}\neq\boldsymbol{0}$, has zero mean within every cell of $\boldsymbol\Gamma$. Appendix~B justifies the rank and convergence conditions for the data-dependent partitions considered below.

\paragraph{Generic partitions.}

We consider two data-dependent partitions. Balanced coordinate (BC) partitions place approximately $n/L$ observations with the smallest values of $X_1$ in the first cell, then approximately $n/L$ remaining observations with the smallest values of $X_2$ in the second, cycling through the coordinates until $L-1$ cells have been formed. The last cell contains the remaining observations.

With fixed $k$, $L$ and coordinate order, the BC cells converge when the successive conditional cutoffs are unique and have probability-zero boundaries. Appendix~B gives the argument, by induction over the cuts.

$K $-means (KM) partitions group observations that are close in Euclidean distance, and their cells may contain quite different numbers of observations. If the population KM problem has
a unique solution, up to relabeling, the sample centers that minimize the $K$-means objective converge to the population centers \citep{pollard1981strong}, and the KM cells converge
to the corresponding Voronoi partition. Uniqueness may fail, however.
For instance, if $\boldsymbol{X}$ is isotropic Gaussian and $k>1$,
the problem is invariant to rotations, so any rotation of an optimal
partition is also optimal. In that case the cells produced by a rotation-equivariant
implementation need not converge to a fixed partition.

\paragraph{Partitions based on fitted guiding models.}

\looseness=-1 By a guiding model we mean a parametric regression that reflects the
departures we want to detect. It can be the null model itself, and
it does not have to be correctly specified. Let $\boldsymbol{x}\mapsto m_{\boldsymbol{\alpha}}^{*}(\boldsymbol{x})$
be a guiding model indexed by $\boldsymbol{\alpha}$, with $\hat{\boldsymbol{\alpha}}=\boldsymbol{\alpha}^{*}+o_{P}(1)$.
The fitted-value (FV) partition sorts observations by
$m_{\hat{\boldsymbol\alpha}}^*(\boldsymbol x)$ and cuts at empirical
quantiles. When the null supplies the fitted values, this is the grouping principle of the Hosmer--Lemeshow test and its extensions
\citep{hosmer1980goodness,surjanovic2024generalized}. The partitions are constructed from the preliminary estimates and held fixed while the Pearson criterion is minimized. Convergence of the guide in $L^2(P_{\boldsymbol X})$, unique cutoffs,
and zero probability at their boundaries imply A2(i).

\looseness=-1 We also adapt to regression the Neyman--Pearson (NP) classes proposed by
\citet[p.~40]{greenwood1996guide} for the classical Pearson test. Let
$\hat{\Delta}(\boldsymbol{x})=m_{\hat{\boldsymbol{\alpha}}}^{*}(\boldsymbol{x})-m_{\tilde{\boldsymbol{\theta}}}(\boldsymbol{x})$,
where $\tilde{\boldsymbol{\theta}}\rightarrow_{P}\boldsymbol{\theta}^{*}$ and
$\boldsymbol{\theta}^{*}=\boldsymbol{\theta}_{0}$ under $H_{0}$. Observations
with $\hat{\Delta}(\boldsymbol{x})\le0$ form the class $\hat{\Gamma}^{-}$ and
the rest form $\hat{\Gamma}^{+}$; each class is then divided into FV, BC or KM
cells, giving $L$ cells in total. Splitting positive and negative deviations can reduce cancellation within cells and improve power. However, the gain also depends on how the sign classes are subdivided and on which departures can be offset by fitting the null model. Thus, NP partitions need not outperform direct FV partitions. In Theorem~\ref{thm:new-local}, cancellation within cells reduces $\lambda_1(\boldsymbol\Gamma)$, and the estimation effect, by which the fitted null offsets part of the departure, enters through $\lambda_2(\boldsymbol\Gamma)$.

The sign classes converge if the limiting discrepancy $\Delta(\boldsymbol x)$ has no mass at
zero under $P_{\boldsymbol X}$. When the guide nests the null, however, $\Delta(\boldsymbol x)=0$ a.s.-$P_{\boldsymbol X}$ under $H_0$ and
estimation error determines the split. Example~3 states the additional
conditions needed in that case.

\begin{figure}[!t]
\centering \renewcommand{\baselinestretch}{1}\selectfont \captionsetup{font=small,skip=3pt} \includegraphics[width=0.80\textwidth]{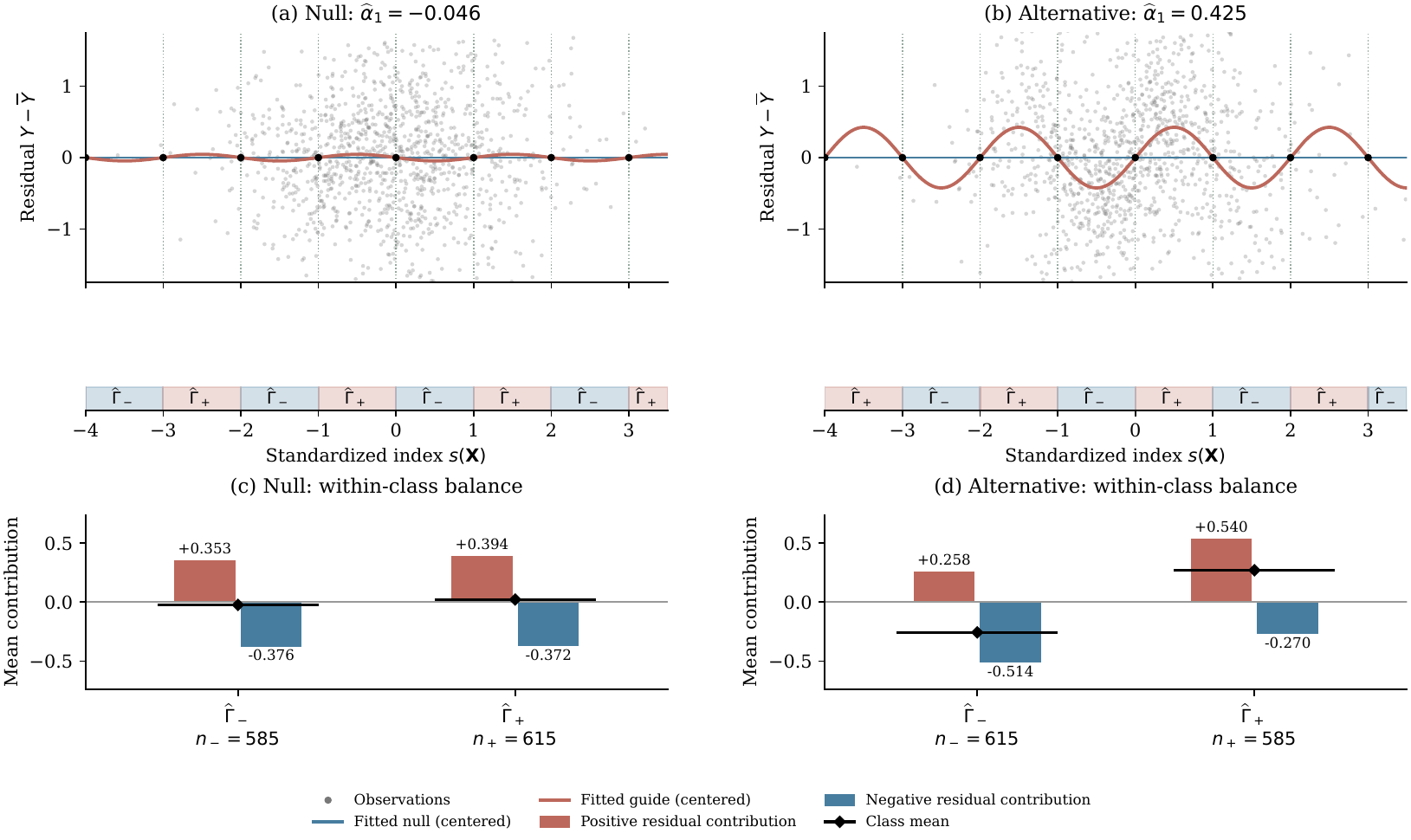}
\caption{Example~1: $n=1200$, $k=10$, $\rho=0$, and departure coefficient $d=0.40$ in the conditional mean $1+d\sin\{\pi s(\boldsymbol x)\}$, with shared
covariates and independent disturbances. Panels (a)--(b) show centered LS
fits, crossings and preliminary sign classes (the plotted index is $s(\boldsymbol X)$); (c)--(d) show within-class
residual contributions. The counts $n_-$ and $n_+$ are the numbers of observations in the two preliminary sign classes.}
\label{fig:figures/Figure1} 
\end{figure}

\refstepcounter{example}\label{ex:mean-independence}

\paragraph{Example~\theexample\ (mean independence).}

\looseness=-1 To test $H_0:\mathbb E(Y\mid\boldsymbol X)=\mathbb E(Y)$, consider the LS guide
$m^*_{\boldsymbol\alpha}(\boldsymbol x)=\alpha_0+\alpha_1\sin\{\pi s(\boldsymbol x)\}$.
Let $\boldsymbol X\sim\mathcal N(\boldsymbol0,\boldsymbol\Sigma_\rho)$, where $(\boldsymbol\Sigma_\rho)_{j\ell}=\rho^{|j-\ell|}$, and set
\begin{equation}
s(\boldsymbol x)=\frac{\boldsymbol 1'\boldsymbol x}{(\boldsymbol 1'\boldsymbol\Sigma_\rho\boldsymbol 1)^{1/2}},\label{eq:10}
\end{equation}
where $\boldsymbol 1$ is a vector of ones. The sine guide is nearly orthogonal to the linear covariate directions. The FV partition cuts $\sin\{\pi s(\boldsymbol X)\}$ at its quantiles. The NP partitions split the observations by the sign of the fitted guide minus the fitted constant, and then subdivide each class. This discrepancy is a multiple of the centered sine regressor, so the two classes have a fixed limit, up to relabeling. Figure~\ref{fig:figures/Figure1} illustrates the fits and sign classes. Design~1 in Section~\ref{sec:simulations} reports simulation results for this design.

\begin{remark}[Partition choice in Example~\ref{ex:mean-independence}]\label{rem:constant-null-power}
To isolate the role of the partition, consider Example~\ref{ex:mean-independence}, where the null is a
constant, and write $\pi_{\ell}=P(\boldsymbol{X}\in\Gamma_{\ell})$,
$\mu_{\ell}=\mathbb{E}\{\delta(\boldsymbol{X})\mid\boldsymbol{X}\in\Gamma_{\ell}\}$,
and $\omega_{\ell}=\mathbb{E}\{\varepsilon^{2}1_{\Gamma_{\ell}}(\boldsymbol{X})\}$.
The noncentrality of Theorem~\ref{thm:new-local} gives
\[
\lambda(\boldsymbol{\Gamma})=\sum_{\ell=1}^{L}\frac{\pi_{\ell}^{2}(\mu_{\ell}-t^{*})^{2}}{\omega_{\ell}},\qquad t^{*}=\frac{\sum_{\ell=1}^{L}(\pi_{\ell}^{2}/\omega_{\ell})\mu_{\ell}}{\sum_{\ell=1}^{L}\pi_{\ell}^{2}/\omega_{\ell}},
\]
\looseness=-1 where $t^{*}$ is the mean of the limiting distribution of $\sqrt n(\hat\theta-\theta_0)$. Under homoskedasticity, $\omega_{\ell}=\sigma^{2}\pi_{\ell}$
and $\mathbb{E}\{\delta(\boldsymbol{X})\}=0$ give $t^{*}=0$, so estimation does not reduce the noncentrality parameter, and
$\lambda(\boldsymbol{\Gamma})=\sigma^{-2}\sum_{\ell}\pi_{\ell}\mu_{\ell}^{2}$, the between-cell variation of the departure.
When the two preliminary classes separate the positive and negative values of the departure, no cell mixes values of opposite sign, and the split of each class into cells determines how much of the remaining variation the partition retains.
Under heteroskedasticity, $t^{*}\neq0$ in general, and the cell weights $\pi_{\ell}^{2}/\omega_{\ell}$ also matter.
\end{remark}

\refstepcounter{example}\label{ex:nonnested-guides}
\paragraph{Example~\theexample\ (non-nested guides).}

\looseness=-1 The Hermite guide describes an interaction that is not a function
of a single linear index. Let $\ln X_{1},\ln X_{2}$ be independent
$\mathcal{N}(0,1/4)$ variables, and fit by LS 
\begin{equation}
m_{\boldsymbol{\theta}}(\boldsymbol{x})=\theta_{0}+\theta_{1}x_{1}+\theta_{2}x_{2},\qquad m_{\boldsymbol{\alpha}}^{*}(\boldsymbol{x})=\alpha_{0}+\alpha_{H}q_{H}(\boldsymbol{x}),\label{eq:hermite_null_guide}
\end{equation}
where 
\begin{equation}
q_{H}(\boldsymbol{x})=[(2\ln x_{1})^{2}-1][(2\ln x_{2})^{2}-1].\label{eq:hermite_interaction}
\end{equation}
The two models in~\eqref{eq:hermite_null_guide} are non-nested.

\looseness=-1 The covariate distribution and the construction of $q_H$ imply $\mathbb{E}(q_H)=\mathbb{E}(X_jq_H)=0$ for $j=1,2$, so $q_H$ is uncorrelated with a linear null. Under a nonconstant linear null and under local alternatives, the fitted Hermite guide therefore converges to the constant $\mathbb{E}(Y)$. The limiting discrepancy $\Delta(\boldsymbol x)=\mathbb{E}(Y)-m_{\boldsymbol{\theta}_{0}}(\boldsymbol{x})$ is then a nonconstant linear function that vanishes only on a line, a set of probability zero, so the sign classes have fixed limits.

Design~2 in Section~\ref{sec:simulations} uses this guide and a non-nested linear--log guide.

\refstepcounter{example}\label{ex:nested-guides}
\paragraph{Example~\theexample\ (nested guiding models).}

\looseness=-1 Nested guides target departures that add nonlinear components while retaining the regressors of the null. The LS fitted models under the null and the guide are $m_{\hat{\boldsymbol\theta}}(\boldsymbol x)=\boldsymbol z'(\boldsymbol x)\hat{\boldsymbol\theta}$ and $m^*_{\hat{\boldsymbol\alpha}}(\boldsymbol x)=\boldsymbol z'(\boldsymbol x)\hat{\boldsymbol\alpha}_1+\boldsymbol h'(\boldsymbol x)\hat{\boldsymbol\alpha}_2$, respectively, where $\boldsymbol z(\boldsymbol x)=(1,\boldsymbol x')'$ and $\boldsymbol x\mapsto\boldsymbol h(\boldsymbol x)\in\mathbb R^q$. Let $\hat{\boldsymbol B}$ be the $(k+1)\times q$ matrix whose $j$th column contains the LS coefficients from projecting $h_j(\boldsymbol X)$ on $\boldsymbol z(\boldsymbol X)$. The LS normal equations imply $\hat{\boldsymbol\theta}=\hat{\boldsymbol\alpha}_1+\hat{\boldsymbol B}\hat{\boldsymbol\alpha}_2$.
Consequently,
\[
m^*_{\hat{\boldsymbol\alpha}}(\boldsymbol x)-m_{\hat{\boldsymbol\theta}}(\boldsymbol x)
=\tilde{\boldsymbol h}'(\boldsymbol x)\hat{\boldsymbol\alpha}_2,
\qquad
\tilde{\boldsymbol h}(\boldsymbol x)
:=\boldsymbol h(\boldsymbol x)-\hat{\boldsymbol B}'\boldsymbol z(\boldsymbol x).
\]
\looseness=-1 When $q=1$, the NP sign classes are determined by $\tilde h(\boldsymbol x)\hat\alpha_2$. If $\hat{\boldsymbol B}\to_p\boldsymbol B_0$, $P_{\boldsymbol X}\{\tilde h_0(\boldsymbol X)=0\}=0$, and $P(\hat\alpha_2=0)\to0$, where $\tilde h_0(\boldsymbol x)=h(\boldsymbol x)-\boldsymbol B_0'\boldsymbol z(\boldsymbol x)$, they converge up to relabeling. This covers the Hermite guide and the single-index guide of Design~3. When $L$ is odd, however, the class that receives the extra cell depends on the sign of $\hat\alpha_2$, which is random under the null, so A2(i) can fail. When $q>1$, the boundary $\tilde{\boldsymbol h}(\boldsymbol x)'\hat{\boldsymbol\alpha}_2=0$ depends on the direction $\hat{\boldsymbol\alpha}_2/\|\hat{\boldsymbol\alpha}_2\|$, which is also random under the null, so even the two classes need not converge. In Design~1 the cell counts and processing order are fixed in advance, so the BC and Index refinements converge, whereas guided KM needs separate conditions. In simulations, null rejection rates remain close to the nominal level. See Appendix~B for details.

\section{Monte Carlo Experiments}

\label{sec:simulations}
\looseness=-1 The simulations examine the partition rules of Section~\ref{sec:partition_choice} under fixed null and alternative models. All three designs use
\[
Y=m(\boldsymbol X)+\sigma(\boldsymbol X)\varepsilon,
\qquad \varepsilon\sim\mathcal N(0,1),\qquad \varepsilon\perp\boldsymbol X.
\]
The covariate distribution and scale vary across designs but remain fixed within each null--alternative comparison, with $\mathbb E\{\sigma^2(\boldsymbol X)\}=1$.
Unless otherwise indicated, results use 1,000 replications at the five percent significance level, and power is not size-adjusted.

\looseness=-1 The test behaves similarly across designs. At the configurations reported in the main tables, null rejection frequencies are generally close to the nominal level. The cell-count experiments show increasing conservatism when cells contain few observations. In the reported experiments, the most powerful partition in every design is one built with a guide that contains the true regression function. In the configurations of the main tables, such a partition is at least as powerful as the omnibus tests, and much more powerful when the number of covariates is large, where the omnibus tests have little power or a size far from the nominal level. This comparison favors the guided partitions, which use information on the alternative that the omnibus tests ignore. In practice, however, only a few parametric models are serious candidates, and Section~\ref{sec:applications} shows that the procedure discriminates among them in a sensible way. Generic partitions, which use no information about the departure, have good power against some departures, but their power may be close to the nominal level when the departure averages out within their cells. Partitions that use partial information on the alternative, such as a single-index model with an unknown link function, can improve power substantially. Even a guide from a misspecified model can raise power above that of the generic partitions through its sign classes.

Table~\ref{tab:partition-labels} summarizes the partition labels used below. A prefix $g$ identifies the guide used for the preliminary sign split. FV--$g$ uses its fitted values directly, whereas $g$--FV subdivides each sign class by fitted-value quantiles. Appendix~B explains odd-$L$ allocations.

\begin{table}[!t]
\centering
\begin{threeparttable}
\caption{Partition rules.}\label{tab:partition-labels}
\footnotesize\renewcommand{\baselinestretch}{1}\selectfont\setlength{\tabcolsep}{5pt}
\begin{tabular}{@{}lll@{}}
\toprule
Label & Preliminary split & Cells \\
\midrule
BC & none & balanced cuts, cycling through the coordinates \\
KM & none & $K$-means (Voronoi) cells of the covariates \\
Index & none & quantiles of a given index $s(\boldsymbol X)$ \\
FV--$g$ & none & quantiles of the fitted values of guide $g$ \\
$g$--FV & sign of $\hat\Delta$ & quantiles of the fitted values of $g$ within each class \\
$g$--BC, $g$--KM, $g$--Index & sign of $\hat\Delta$ & BC, KM, or Index cells within each class \\
\bottomrule
\end{tabular}
\begin{tablenotes}[flushleft]\scriptsize
\item \textit{Notes:} $\hat\Delta=m^*_{\hat{\boldsymbol\alpha}}-m_{\tilde{\boldsymbol\theta}}$ is the fitted guide-minus-null discrepancy. Appendix~B gives the allocation of cells to the two classes when $L$ is odd.
\end{tablenotes}
\end{threeparttable}
\end{table}

\looseness=-1 The omnibus benchmarks are the tests of \citet{stute1997nonparametric} ($S$), \citet{stute2002model} ($SZ$), and \citet{escanciano2006consistent} ($E$), with wild-bootstrap critical values \citep{stute1998bootstrap}. Design~1 omits $SZ$ because the null fit is constant. At $n=12{,}000$, $E$ is too costly to compute, so we use the distance-kernel test $E^*$ of \citet{escanciano2024gaussian} with multiplier-bootstrap critical values.

\looseness=-1 The number of cells is a tuning parameter. For the main tables, we report configurations satisfying $L>p$ and allocating at least three cells to each preliminary sign class of the NP partitions. These are $L=7$ in Design~1, $L=7,7,13$ for $k=1,5,10$ in the linear--log design, and $L=6$ in the Hermite and nested designs. Tables~\ref{tab:trigonometric_selected}--\ref{tab:nested_selected} report size/power pairs in percent and the drift of each partition, which relates the simulations to Theorem~\ref{thm:new-local}. In every design, the alternative departs from the null model by $d\,\bar\delta$, up to terms absorbed by the null parameters, where $\bar\delta$ is given in the table notes and $d=0$ gives the null. The alternative of the linear--log design is a fixed function rather than a scaled departure, so $d=1$. We call $\bar{\lambda}$, the value of $\lambda(\boldsymbol\Gamma)$ with $\delta=\bar\delta$ at the limit cells, the drift of the partition. Because the noncentrality parameter is quadratic in the local departure, the local-power formula motivates approximating the power against the fixed alternative by $P\{\chi^{2}_{L-p}(nd^{2}\bar{\lambda})>\chi^{2}_{L-p,0.95}\}$. This approximation does not follow from Theorem~\ref{thm:new-local}. It is close in Design~1 (Figure~\ref{fig:powc}) and, in the nested design, for BC, KM, and their guided refinements, with larger errors for the direct FV partitions, and \citet{drost1989power} find it fair, but not always accurate, for the Pearson test of fit. In these cases, the limit cells do not change with the size of the departure. When the cells follow a fitted guide that contains the departure only under a fixed alternative, as with FV cells and non-nested guides, the drift need not predict power or the ranking of the partitions \citep[sec.~8]{kallenberg1985number}. Dashes mark partitions whose cells fail the rank condition, so the drift is undefined. KM drifts average repeated $K$-means fits to large simulated samples. Appendix~B gives further details of the designs and reports the same comparisons for other values of $n$, $d$, $L$, $k$, and $\rho$.

\paragraph{Design 1: mean independence.}

\looseness=-1 This design implements Example~1. Let $\boldsymbol X\sim\mathcal N(\boldsymbol0,\boldsymbol\Sigma_\rho)$ and $\sigma^2(\boldsymbol X)=\exp(0.75X_1-0.75^2/2)$. The null is $m_{\theta_0}(\boldsymbol x)=1$ and the alternative is $m^*_{\boldsymbol\alpha_0}(\boldsymbol x)=1+d\,g\{s(\boldsymbol x)\}$, where $s$ is the standardized index in~\eqref{eq:10}.  We use the trigonometric link $g_T(s)=\sin(\pi s)$ and the discontinuous link $g_D(s)=\sqrt{v_T}\{1_{(-\infty,c]}(|s|)-p_D\}/\sqrt{p_D(1-p_D)}$.
Here $p_{D}=1/7$ is the probability of the central region $|s|\leq c$,
with $c=\Phi^{-1}(4/7)$ and $v_T=(1-e^{-2\pi^2})/2$. Both links have $L^2(P_{\boldsymbol X})$ norm $\sqrt{v_T}$ and mean zero, so $\mathbb E(Y)=1$ under both. When $k=1$, $\rho$ plays no role.

\looseness=-1 BC and KM use no information about the departure. When $\rho=0$ and $k>1$, the KM cells need not have a fixed limit (Section~\ref{sec:partition_choice}), so KM is a benchmark outside Theorem~\ref{thm:ddp}. Index cuts
$s(\boldsymbol X)$ at its quantiles, so it uses the direction of the departure
but not its shape. The guides Trig and Disc are LS fits of $Y$ on an intercept
and $\sin\{\pi s(\boldsymbol X)\}$ or $1_{(-\infty,c]}(|s(\boldsymbol X)|)$. Trig is the matching guide, because its family contains the alternative, and Disc is misspecified.

\begin{sidewaystable}
\begingroup
\renewcommand{\baselinestretch}{1}\selectfont
\centering
\begin{threeparttable}
\caption{Mean independence, trigonometric alternative, $d=0.20$, $L=7$.}\label{tab:trigonometric_selected}
\footnotesize\setlength{\tabcolsep}{3.5pt}\renewcommand{\arraystretch}{0.92}
\centering
\begin{tabular}{rrr*{12}{r}}
\toprule
$n$ & $\rho$ & $k$ & BC & Index & KM & \shortstack{FV--\\Trig} & \shortstack{Trig--\\BC} & \shortstack{Trig--\\FV} & \shortstack{Trig--\\KM} & \shortstack{Disc--\\BC} & \shortstack{Disc--\\Index} & \shortstack{Disc--\\KM} & $S$ & $E$\\
\midrule
\multicolumn{15}{c}{\textit{Panel A: Size/power (percent)}}\\
300 & 0 & 1 & 3.2/40.6 & 3.2/40.6 & 4.0/38.3 & 3.8/39.0 & 4.0/51.3 & 3.9/39.5 & 2.5/47.1 & 4.3/32.0 & 4.3/32.0 & 3.9/39.3 & 4.2/8.8 & 4.9/6.9\\
 &  & 5 & 4.5/5.0 & 4.1/30.4 & 5.4/4.6 & 3.1/36.6 & 4.2/38.7 & 2.9/37.3 & 5.3/34.6 & 3.3/3.1 & 4.3/23.6 & 5.1/4.6 & 5.0/5.5 & 4.9/4.9\\
 &  & 10 & 4.1/4.3 & 5.1/30.4 & 3.5/4.3 & 4.7/39.9 & 4.0/38.2 & 5.6/41.3 & 3.4/34.2 & 4.3/4.3 & 4.1/24.3 & 2.9/4.3 & 2.2/2.4 & 4.7/5.0\\
\cmidrule(lr){2-15}
 & 0.5 & 1 & 4.1/39.8 & 4.1/39.8 & 4.7/37.8 & 4.5/39.6 & 4.6/51.3 & 4.0/38.0 & 3.8/48.2 & 4.0/30.4 & 4.0/30.4 & 3.9/40.4 & 6.2/11.7 & 6.0/8.9\\
 &  & 5 & 4.2/4.8 & 4.0/32.2 & 3.5/10.8 & 5.1/38.6 & 4.6/39.1 & 5.8/37.2 & 4.0/40.2 & 2.6/3.1 & 4.3/28.4 & 3.9/8.9 & 4.6/4.8 & 4.4/4.2\\
 &  & 10 & 4.0/4.1 & 4.8/30.7 & 4.7/5.9 & 5.4/37.1 & 4.1/36.8 & 4.6/39.9 & 4.1/34.9 & 3.9/4.0 & 3.8/24.3 & 3.3/6.4 & 4.0/5.1 & 4.6/4.6\\
\midrule
600 & 0 & 1 & 5.3/77.2 & 5.3/77.2 & 5.1/74.5 & 4.9/73.1 & 5.6/86.1 & 6.6/72.7 & 4.4/85.2 & 5.6/66.2 & 5.6/66.2 & 5.0/73.7 & 4.4/17.7 & 4.3/10.8\\
 &  & 5 & 5.0/4.9 & 5.2/63.3 & 6.9/5.5 & 5.3/72.0 & 5.6/70.1 & 5.4/71.5 & 5.0/71.0 & 4.7/4.9 & 4.4/54.2 & 5.1/5.8 & 5.1/5.2 & 5.1/4.4\\
 &  & 10 & 5.4/4.9 & 4.1/58.8 & 3.5/5.5 & 5.5/74.4 & 4.7/69.6 & 6.2/74.8 & 4.2/67.0 & 4.5/4.6 & 4.1/52.1 & 4.4/5.0 & 4.2/3.7 & 4.3/4.8\\
\cmidrule(lr){2-15}
 & 0.5 & 1 & 5.1/78.5 & 5.1/78.5 & 5.3/76.3 & 3.8/70.6 & 4.3/87.2 & 4.8/70.5 & 3.6/85.3 & 4.0/66.8 & 4.0/66.8 & 6.1/77.6 & 5.6/20.0 & 4.9/14.1\\
 &  & 5 & 4.4/6.4 & 4.4/62.7 & 3.8/18.0 & 5.1/72.8 & 4.7/72.5 & 5.0/72.1 & 4.9/76.2 & 4.5/4.9 & 3.6/55.1 & 4.2/15.3 & 5.3/6.3 & 4.2/6.1\\
 &  & 10 & 4.0/4.1 & 5.8/60.1 & 4.1/8.5 & 4.5/71.2 & 5.5/71.9 & 5.8/71.7 & 5.2/69.9 & 4.5/5.1 & 4.1/51.9 & 4.2/11.2 & 4.7/5.4 & 4.3/4.7\\
\midrule
1{,}200 & 0 & 1 & 4.7/98.9 & 4.7/98.9 & 5.2/97.5 & 4.4/97.4 & 5.1/99.8 & 4.9/97.2 & 4.4/99.3 & 5.0/96.1 & 5.0/96.1 & 4.9/98.7 & 4.7/35.9 & 4.5/25.3\\
 &  & 5 & 5.4/5.8 & 5.6/91.0 & 4.5/5.9 & 5.4/96.1 & 5.2/97.1 & 4.6/97.4 & 6.0/97.8 & 4.0/4.2 & 4.7/86.6 & 3.8/8.8 & 4.9/4.9 & 5.3/5.9\\
 &  & 10 & 5.0/5.4 & 6.3/90.5 & 5.2/4.5 & 4.5/96.9 & 4.2/96.3 & 4.6/97.4 & 4.6/94.9 & 3.6/3.7 & 4.7/86.0 & 3.9/7.2 & 4.6/4.3 & 4.8/5.3\\
\cmidrule(lr){2-15}
 & 0.5 & 1 & 4.7/98.4 & 4.7/98.4 & 5.8/97.6 & 5.2/97.1 & 3.4/99.6 & 5.7/96.5 & 5.7/99.3 & 4.4/94.7 & 4.4/94.7 & 4.9/97.8 & 5.7/39.1 & 4.9/26.3\\
 &  & 5 & 3.6/9.7 & 4.9/93.0 & 4.4/33.3 & 5.4/97.0 & 4.4/97.2 & 5.0/97.0 & 3.7/98.3 & 5.0/5.7 & 5.2/91.1 & 5.4/29.5 & 5.1/9.0 & 4.3/7.1\\
 &  & 10 & 5.0/4.8 & 5.7/90.2 & 4.9/11.5 & 3.8/97.2 & 4.4/96.4 & 4.8/96.8 & 4.0/96.2 & 4.0/4.3 & 3.6/85.7 & 4.6/24.4 & 5.7/6.9 & 3.8/5.9\\
\midrule
12{,}000 & 0 & 1 & 5.1/100 & 5.1/100 & 4.9/100 & 5.0/100 & 4.5/100 & 4.7/100 & 5.4/100 & 6.2/100 & 6.2/100 & 5.6/100 & 4.1/100.0 & 5.2/100$^{*}$\\
 &  & 5 & 5.7/12.3 & 4.1/100 & 5.2/17.7 & 4.2/100 & 4.6/100 & 5.0/100 & 5.2/100 & 4.5/4.8 & 5.1/100 & 5.6/66.0 & 3.9/12.6 & 4.4/11.9$^{*}$\\
 &  & 10 & 5.1/6.4 & 4.7/100 & 3.5/6.0 & 6.1/100 & 4.1/100 & 5.0/100 & 4.5/100 & 4.3/4.2 & 4.7/100 & 4.5/58.6 & 4.8/5.2 & 5.3/7.2$^{*}$\\
\cmidrule(lr){2-15}
 & 0.5 & 1 & 4.9/100 & 4.9/100 & 4.9/100 & 4.5/100 & 3.6/100 & 4.7/100 & 4.5/100 & 5.1/100 & 5.1/100 & 5.5/100 & 5.8/100.0 & 4.9/100$^{*}$\\
 &  & 5 & 6.0/66.9 & 5.7/100 & 5.1/97.2 & 4.9/100 & 4.8/100 & 4.7/100 & 6.3/100 & 4.0/13.6 & 4.0/100 & 4.6/100 & 5.0/47.1 & 4.7/56.7$^{*}$\\
 &  & 10 & 4.5/6.9 & 4.8/100 & 5.3/81.8 & 4.9/100 & 5.4/100 & 5.7/100 & 4.6/100 & 4.0/4.5 & 5.4/100 & 5.3/99.8 & 4.2/20.9 & 3.7/15.3$^{*}$\\
\midrule
\multicolumn{15}{c}{\textit{Panel B: Drift, $\bar{\lambda}$}}\\
 & 0 & 1 & 0.575 & 0.575 & 0.498 & 0.491 & 0.706 & 0.490 & 0.669 & 0.494 & 0.494 & 0.559 & -- & --\\
 &  & 5 & 0.003 & 0.402 & 0.004 & 0.494 & 0.480 & 0.493 & 0.525 & 0.000 & 0.377 & 0.025 & -- & --\\
 &  & 10 & 0.000 & 0.383 & 0.001 & 0.492 & 0.480 & 0.492 & 0.434 & 0.000 & 0.365 & 0.023 & -- & --\\
\cmidrule(lr){2-13}
 & 0.5 & 1 & 0.575 & 0.575 & 0.510 & 0.492 & 0.711 & 0.491 & 0.671 & 0.505 & 0.505 & 0.558 & -- & --\\
 &  & 5 & 0.022 & 0.424 & 0.050 & 0.490 & 0.491 & 0.489 & 0.517 & 0.004 & 0.390 & 0.087 & -- & --\\
 &  & 10 & 0.001 & 0.389 & 0.027 & 0.490 & 0.485 & 0.489 & 0.459 & 0.001 & 0.368 & 0.080 & -- & --\\
\bottomrule
\end{tabular}
\begin{tablenotes}[flushleft]\scriptsize
\item \textit{Notes:} 1,000 replications. $^{*}$At $n=12{,}000$, the distance-kernel test $E^{*}$ replaces $E$. The omnibus tests do not depend on $L$. Panel B uses $\bar\delta(\boldsymbol x)=\sin\{\pi s(\boldsymbol x)\}$.
\end{tablenotes}
\end{threeparttable}
\endgroup
\end{sidewaystable}

\begin{figure}[!htbp]
\centering \renewcommand{\baselinestretch}{1}\selectfont \captionsetup{font=small,skip=3pt}
\includegraphics[width=0.46\textwidth]{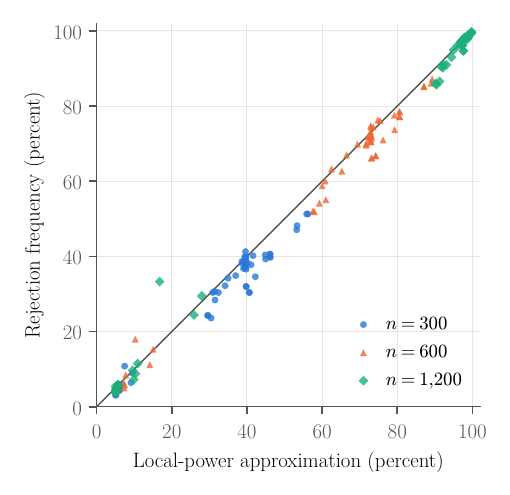}
\caption{Design~1 with $d=0.20$ and $L=7$. Rejection frequencies of Table~\ref{tab:trigonometric_selected} at $n=300$, 600, and $1{,}200$ against the approximation $P\{\chi^{2}_{6}(nd^{2}\bar{\lambda})>\chi^{2}_{6,0.95}\}$ motivated by Theorem~\ref{thm:new-local}. Each point corresponds to one partition, value of $k$, and value of $\rho$, and the marker identifies $n$.}\label{fig:powc}
\end{figure}

\looseness=-1 Table~\ref{tab:trigonometric_selected} reports the weakest trigonometric departure, $d=0.20$, with $L=7$, and includes $n=12{,}000$ to distinguish slow power gains from lack of power. Figure~\ref{fig:powc} plots, for $d=0.20$, the rejection frequencies of Panel~A against the power approximation implied by the drifts of Panel~B for the different partition methods. The points lie close to the 45-degree line, so the approximation is accurate and the drift accounts for most of the differences in power among partitions and sample sizes. FV--Trig and Trig--FV are within 3.2 points of the approximation in every case. The largest deviations are those of KM with $k=5$ and $\rho=0.5$, which rejects in 33.3 percent of the samples at $n=1{,}200$ against an approximation of 16.8 percent, and of Disc--BC and Disc--Index with $k=1$, which reject 7 to 10 points less often than the approximation at $n\leq600$. The sign split adds power to FV--Trig for $k=1$, where Trig--BC has drift 0.71 against 0.49 and power of 86 against 73 percent at $n=600$. It adds none for $k=10$, where the BC and KM cuts within each sign class mix values of the departure. When $k>1$ and $\rho=0$, BC, KM, and Disc--BC have drifts below 0.005 and power below 20 percent even at $n=12{,}000$. The omnibus tests have power below 10 percent at $n\leq1{,}200$ whenever $k>1$.

\begin{figure}[!t]
\centering \renewcommand{\baselinestretch}{1}\selectfont \captionsetup{font=small,skip=3pt}
\includegraphics[width=0.78\textwidth]{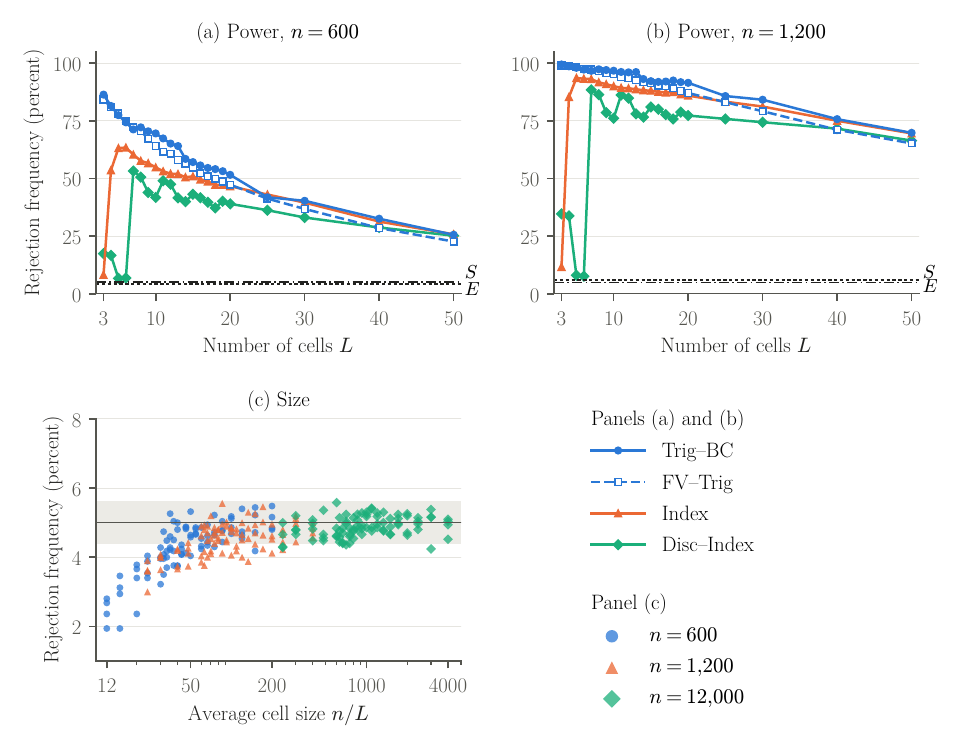}
\caption{Design~1 with $k=5$ and $\rho=0$, from 5,000 replications. Panels (a) and (b), rejection frequencies under the trigonometric alternative with $d=0.20$ against the number of cells $L$. Black lines mark the omnibus tests of Table~\ref{tab:trigonometric_selected}. Panel (c), null rejection frequencies of the same partitions for every $L$ against the average cell size $n/L$, with a band of two Monte Carlo standard errors around 5 percent.}\label{fig:L-design1}
\end{figure}

\looseness=-1 Figure~\ref{fig:L-design1} shows the role of $L$ for $k=5$ and $\rho=0$. The power of the Trig partitions tends to decrease as $L$ grows, because additional cells add degrees of freedom without a comparable increase in the drift, and at $n=600$ their advantage over Index nearly disappears for $L\geq25$. At $n=1{,}200$, Trig--BC still loses power as $L$ grows, from 97 percent at $L=7$ to 70 percent at $L=50$, but it remains far above the omnibus tests, which reject in about 5 percent of the samples. At $n=12{,}000$, not shown, the guided partitions reject in almost all samples for almost every $L$, whereas the omnibus tests reject in about 12 percent of them. BC, not shown, has power close to the nominal level for every $L$. Its cells are cut along one coordinate at a time, so within each cell the other coordinates move $s(\boldsymbol{X})$ across several periods of the sine, and positive and negative values of the departure cancel. Panel~(c) plots size in the Monte Carlo against the number of observations per cell, $n/L$. In this design, the tests are conservative below about 50 observations per cell, with a rejection frequency of 3.9 percent at 30, and close to the nominal level above. Accurate size thus depends on the number of observations per cell, whereas power depends, approximately, on $nd^{2}\bar{\lambda}$, which grows with $n$, and on the degrees of freedom $L-p$. These findings agree with \citet{kallenberg1985number}, who show for Pearson goodness-of-fit tests that local power falls to the level as the number of cells grows when the drift does not grow with it.

\paragraph{Design 2: non-nested alternatives.}

\looseness=-1 This design implements Example~2. Let $X_{1},\ldots,X_{k}$ be independent
$U[0,1]$ variables, with $\sigma^2(\boldsymbol X)=3\exp(3X_1)/(e^3-1)$. The null is linear in $\boldsymbol x$ and the alternative is linear in $\log\boldsymbol x$, with $m_{\boldsymbol\theta_0}(\boldsymbol x)=1+\sum_jx_j$ and $m^{*}_{\boldsymbol\alpha_0}(\boldsymbol x)=a_k+\sum_j\log x_j/2$, where $a_k=1+(k/2)\{1-\log(1/2)\}$.
The two functions coincide when $X_j=1/2$ for every $j$, and their
distance is not held fixed across $k$. The guide LL is the LS fit of $Y$ on
an intercept and $\log X_{1},\ldots,\log X_{k}$. Table~\ref{tab:linear_log_selected} reports $k=1,5,10$ with $L=7,7,13$ and $n=300$, 600, and $1{,}200$.

\begin{table}[!htbp]
\begingroup
\centering
\begin{threeparttable}
\caption{Linear--log design, $d=1$: rejection frequencies and drift.}\label{tab:linear_log_selected}
\footnotesize\renewcommand{\baselinestretch}{1}\selectfont\setlength{\tabcolsep}{3pt}
\renewcommand{\arraystretch}{0.93}\begin{tabular}{lrrrrrrrrr}
\toprule
& \multicolumn{3}{c}{$n=300$} & \multicolumn{3}{c}{$n=600$} & \multicolumn{3}{c}{$n=1{,}200$}\\
\cmidrule(lr){2-4}\cmidrule(lr){5-7}\cmidrule(lr){8-10}
Method & $k=1$ & $k=5$ & $k=10$ & $k=1$ & $k=5$ & $k=10$ & $k=1$ & $k=5$ & $k=10$\\
& $L=7$ & $L=7$ & $L=13$ & $L=7$ & $L=7$ & $L=13$ & $L=7$ & $L=7$ & $L=13$\\
\midrule

\looseness=-1 BC & 3.5/90.2 & 5.3/57.3 & 3.5/57.8 & 5.1/100.0 & 4.4/84.7 & 5.0/85.7 & 5.1/100.0 & 5.1/98.1 & 3.5/99.0\\
KM & 3.6/91.4 & 4.8/6.4 & 6.5/7.2 & 5.2/99.8 & 5.8/6.5 & 4.5/5.4 & 6.3/100.0 & 5.9/5.9 & 4.3/5.0\\
FV--LL & 3.5/90.2 & 4.6/37.5 & 5.3/33.4 & 5.1/100.0 & 5.8/52.0 & 5.5/52.2 & 5.1/100.0 & 5.4/64.3 & 4.5/72.7\\
LL--FV & 3.8/85.5 & 4.8/63.8 & 5.8/80.8 & 5.0/99.6 & 4.4/73.7 & 5.8/90.5 & 4.5/100.0 & 5.5/81.9 & 4.9/94.2\\
LL--BC & 3.8/85.5 & 5.1/99.9 & 6.0/100.0 & 5.0/99.6 & 6.4/100.0 & 4.5/100.0 & 4.5/100.0 & 4.2/100.0 & 3.9/100.0\\
LL--KM & 4.4/95.8 & 6.4/91.1 & 5.6/97.7 & 5.0/99.9 & 7.6/93.5 & 4.4/99.3 & 4.0/100.0 & 5.3/96.4 & 5.8/99.9\\
\midrule
$S$ & 5.0/46.9 & 5.4/61.8 & 1.6/8.9 & 4.9/88.9 & 5.0/96.3 & 4.9/30.1 & 4.9/99.8 & 5.2/99.9 & 4.7/62.2\\
$SZ$ & 4.8/46.5 & 5.0/19.6 & 6.7/18.9 & 4.6/88.5 & 5.4/34.6 & 6.6/25.9 & 4.9/99.8 & 5.4/64.8 & 5.0/50.3\\
$E$ & 4.5/83.7 & 3.2/99.3 & 0.1/99.8 & 4.7/99.1 & 3.7/100.0 & 0.9/100.0 & 4.6/100.0 & 3.8/100.0 & 2.0/100.0\\
\midrule
\multicolumn{10}{c}{\textit{Drift, $\bar{\lambda}$}}\\
 $k$ & BC & KM & FV--LL & LL--FV & LL--BC & LL--KM\\
1 & 0.080 & 0.080 & 0.080 & 0.037 & 0.037 & 0.148\\
5 & 0.020 & 0.000 & -- & -- & 0.000 & 0.000\\
10 & 0.021 & 0.000 & -- & -- & 0.001 & 0.005\\
\bottomrule
\end{tabular}
\begin{tablenotes}[flushleft]\scriptsize
\item \textit{Notes:} The upper panel reports size/power (percent). The drift uses $\bar\delta=m_{\rm log}-m_{\boldsymbol\theta_0}$, up to a constant absorbed by the intercept. KM entries, and LL--KM entries for $k=1$, come from a separate run with different samples.
\end{tablenotes}
\end{threeparttable}\par
\endgroup
\end{table}

\looseness=-1 For $k=1$, every partition rejects in at least 85 percent of the samples at $n=300$, against 47 percent for $S$ and $SZ$. For $k=5$ and 10, LL--BC and LL--KM reject in more than 90 percent of the samples at $n=300$. BC also detects this departure, with power of 57 percent, whereas KM has power close to the nominal level. Direct FV--LL has less power than BC, and LL--FV improves on it, with 90.5 against 52.2 percent at $n=600$ and $k=10$. When $k>1$, the FV--LL and LL--FV cells fail the rank condition, because the null has equal slopes and the covariates are i.i.d. Theorem~\ref{thm:ddp} does not cover them and we do not report their drift, although their size is close to the nominal level. With a non-nested guide, the limiting cells can differ under local and fixed alternatives, so the drift need not predict power against the fixed alternative. Among the omnibus tests, $E$ is powerful but has size 0.1 percent at $n=300$ and $k=10$, where $S$ and $SZ$ reject in fewer than a third of the samples.

\looseness=-1 \emph{Hermite departure.} Let $\log X_{1}$ and $\log X_{2}$
be independent $\mathcal{N}(0,1/4)$ variables, with $\sigma^2(\boldsymbol X)=\exp(1.5\log X_1-0.75^2/2)$. In the families~\eqref{eq:hermite_null_guide}, we use $m_{\boldsymbol\theta_0}(\boldsymbol x)=1+(x_1+2x_2)/\sqrt5$ and $m^*_{\boldsymbol\alpha_0}(\boldsymbol x)=1+dq_H(\boldsymbol x)$, with $q_H$ defined in~\eqref{eq:hermite_interaction}. Since $q_{H}$ is
orthogonal to the null regressors, the $L^2(P_{\boldsymbol X})$ distance of the alternative
from the linear class is $2d$. Table~\ref{tab:hermite_selected}
uses $d=0.20$ and $L=6$. The LS guides include an intercept
and either $q_{H}$ or $q_{\mathrm{LL}}=2(\log X_{1}+2\log X_{2})/\sqrt{5}$,
and each sign class is split into three FV, BC, or KM cells.

\begin{table}[!htbp]
\begingroup
\centering
\begin{threeparttable}
\caption{Hermite design, $d=0.20$, $L=6$: size/power and drift.}\label{tab:hermite_selected}
\footnotesize\renewcommand{\baselinestretch}{1}\selectfont\setlength{\tabcolsep}{7pt}
\renewcommand{\arraystretch}{0.93}\begin{tabular}{lrrrrr}
\toprule
Method & $n=300$ & $n=600$ & $n=1{,}200$ & $n=12{,}000$ & $\bar{\lambda}$\\\midrule
BC & 5.3/6.8 & 4.5/9.0 & 5.0/11.0 & 5.1/62.8 & 0.018\\
KM & 4.7/8.0 & 6.0/9.8 & 5.5/13.8 & 4.3/67.4 & 0.024\\
FV--H & 5.9/48.8 & 6.3/71.9 & 5.4/89.6 & 5.4/99.9 & --\\
H--FV & 4.5/64.7 & 4.8/84.5 & 4.7/96.6 & 5.0/100.0 & 0.686\\
H--BC & 4.7/79.5 & 5.5/98.3 & 5.1/100.0 & 4.6/100 & 1.248\\
H--KM & 5.2/77.7 & 5.1/97.7 & 4.7/99.9 & 4.6/100 & 1.141\\
FV--LL & 4.8/11.2 & 5.8/20.2 & 2.8/28.7 & 4.7/71.3 & 0.016\\
LL--FV & 5.4/14.2 & 3.5/30.3 & 4.2/49.0 & 5.2/99.8 & 0.016\\
LL--BC & 5.1/11.8 & 5.2/23.4 & 3.8/38.9 & 4.3/87.8 & 0.146\\
LL--KM & 4.5/12.6 & 5.1/17.5 & 4.6/25.0 & 5.5/80.5 & 0.112\\
\midrule
$S$ & 6.4/8.3 & 5.6/8.9 & 5.7/11.8 & 4.4/86.3 & --\\
$SZ$ & 4.8/13.0 & 5.8/20.0 & 5.2/28.1 & 4.6/78.7 & --\\
$E$ & 6.2/17.6 & 5.8/26.9 & 5.5/46.0 & 5.5/99.9$^{*}$ & --\\
\bottomrule\end{tabular}
\begin{tablenotes}[flushleft]\scriptsize
\item \textit{Notes:} Entries are size/power (percent); the final column uses $\bar\delta=q_H$.
\end{tablenotes}
\end{threeparttable}\par
\endgroup
\end{table}

\looseness=-1 At $n=300$, the partitions of the matching guide, H--BC and H--KM, reject in 78--80 percent of the samples and H--FV in 64.7 percent, against 48.8 percent for direct FV--H and at most 17.6 percent for the omnibus tests. The FV--H cells are quantile cells of $q_H$, which is symmetric in $\log X_1$ and $\log X_2$, so they fail the rank condition. Theorem~\ref{thm:ddp} does not cover them, and we do not report their drift. The misspecified log guide needs larger samples, but at $n=12{,}000$ its NP refinements reject more often than BC and KM. Its drifts do not order its partitions by power, as can happen with a non-nested guide, because LL--FV has the smallest drift and the highest power.

\paragraph{Design 3: nested guides.}

\looseness=-1 This design implements Example~3. We keep the covariates and scale of the Hermite design and set $m^*_{\boldsymbol\alpha_0}(\boldsymbol x)=1+\boldsymbol\alpha_0^{\prime}\boldsymbol x+d\,h_g(\boldsymbol x)$, with $\boldsymbol\alpha_0=(1,2)^{\prime}/\sqrt5$. For $g=\mathrm{LL}$ or $H$, $h_g$ is $q_g$ after removing its population linear projection on the null regressors, standardized to unit $L^2(P_{\boldsymbol X})$ norm, so $d$ is the $L^2(P_{\boldsymbol X})$ distance from the linear class. Each guide augments the linear null regression with $q_{\mathrm{LL}}$ or $q_H$, and both guides are applied under each alternative. Table~\ref{tab:nested_selected} reports $d=0.20$ and $L=6$.

\begin{table}[!ht]\centering
\begin{threeparttable}
\caption{Nested guides, $k=2$, $d=0.20$, $L=6$: rejection frequencies and drift.}\label{tab:nested_selected}
\renewcommand{\arraystretch}{0.93}\footnotesize\renewcommand{\baselinestretch}{1}\selectfont\setlength{\tabcolsep}{3pt}
\begin{tabular}{@{}lrrrr@{\hspace{10pt}}rrrr@{}}\toprule
 & \multicolumn{4}{c}{\textit{Log departure}} & \multicolumn{4}{c}{\textit{Hermite departure}}\\
\cmidrule(lr){2-5}\cmidrule(lr){6-9}
Method & $n=300$ & $n=600$ & $n=1{,}200$ & $n=12{,}000$ & $n=300$ & $n=600$ & $n=1{,}200$ & $n=12{,}000$\\
\midrule
BC & 5.9/18.8 & 4.8/28.3 & 4.8/57.8 & 5.1/100 & 5.9/5.5 & 4.8/4.8 & 4.8/5.8 & 5.1/18.8\\
KM & 5.2/34.8 & 4.3/59.7 & 4.8/90.6 & 4.3/100 & 5.1/5.8 & 4.2/5.4 & 4.5/7.1 & 4.3/25.3\\
FV--H & 5.8/20.4 & 4.8/30.8 & 4.2/38.1 & 5.0/68.8 & 5.8/6.9 & 4.8/11.6 & 4.2/18.6 & 5.0/95.5\\
H--BC & 4.5/14.2 & 4.7/24.3 & 4.8/47.4 & 5.3/100 & 4.5/28.8 & 4.7/57.3 & 4.8/89.0 & 5.3/100\\
H--KM & 5.4/28.9 & 4.2/50.4 & 4.6/81.5 & 5.3/100 & 5.3/30.0 & 4.1/55.5 & 4.6/86.9 & 5.3/100\\
FV--LL & 5.5/27.0 & 4.7/46.9 & 4.3/61.6 & 4.6/81.1 & 5.5/6.0 & 4.7/6.5 & 4.3/8.6 & 4.6/35.4\\
LL--BC & 5.0/51.9 & 5.4/86.9 & 5.2/99.5 & 4.9/100 & 5.0/7.3 & 5.4/9.7 & 5.2/11.9 & 4.9/76.0\\
LL--KM & 4.9/63.2 & 6.0/92.1 & 4.7/100.0 & 4.8/100 & 4.9/10.6 & 6.0/16.2 & 4.7/29.1 & 4.8/100\\
\midrule
$S$ & 4.8/35.9 & 5.5/70.4 & 4.8/96.0 & 4.9/100.0 & 4.8/4.9 & 5.5/5.6 & 4.8/5.9 & 4.9/28.1\\
$SZ$ & 5.4/38.2 & 4.4/73.1 & 5.0/95.5 & 4.8/100.0 & 5.4/6.1 & 4.4/10.1 & 5.0/10.3 & 4.8/86.8\\
$E$ & 5.8/57.9 & 6.3/89.1 & 4.7/99.2 & 6.1/100.0$^{*}$ & 5.8/8.2 & 6.3/9.6 & 4.7/13.4 & 6.1/98.4$^{*}$\\
\midrule
\multicolumn{9}{c}{\textit{Drift, $\bar{\lambda}$}}\\
Departure & BC & KM & FV--H & H--BC & H--KM & FV--LL & LL--BC & LL--KM\\
Log & 0.144 & 0.323 & 0.013 & 0.118 & 0.260 & 0.013 & 0.564 & 0.674\\
Hermite & 0.005 & 0.006 & 0.007 & 0.312 & 0.297 & 0.007 & 0.021 & 0.070\\
\bottomrule\end{tabular}
\begin{tablenotes}[flushleft]\scriptsize
\item \textit{Notes:} The upper panel reports size/power (percent). The drift uses $\bar\delta=h_g$, the unit-$L^2(P_{\boldsymbol X})$ residualized log or Hermite component.
\end{tablenotes}
\end{threeparttable}
\end{table}

\looseness=-1 At $n=1{,}200$, the NP refinements of the matching guide reject in 99.5--100 percent of the samples against the log departure and in 86.9--89.0 percent against the Hermite departure, compared with 61.6 and 18.6 percent for the direct FV partitions of the same guides. Their drifts are also the largest and order the generic partitions and the NP refinements by power under both departures. The direct FV partitions are the exception. Under local alternatives the fitted guide converges to the null regression function, so both FV partitions have the same limit cells and the same small drift. Under a fixed alternative their cells follow the fitted guide, which contains the departure, and their empirical power is higher than the drift suggests. The sign classes of the NP partitions do not depend on the size of the departure, so their cells are the same in both cases. Against the Hermite departure, BC, KM, and the omnibus tests reject in fewer than 14 percent of the samples at $n=1{,}200$. Against the log departure, the omnibus tests are powerful, and $E$ is close to LL--BC. A mismatched guide helps against the Hermite departure, where LL--KM rejects more often than KM, but not against the log departure, where H--BC and H--KM have slightly less power than BC and KM. Size is close to the nominal level. Appendix~B reports results for $L=4,6,\ldots,20$. The partitions of the matching guide have the highest power for every $L$, and larger $L$ helps the generic and mismatched partitions.

\section{General Conditional Moment Restrictions}

\label{sec:new-general}

 \looseness=-1 The Pearson construction extends to conditional moment restrictions
on generalized residuals. Following \citet{delgado2006consistent},
let $\boldsymbol W$ now denote the vector of all observables, with $\boldsymbol Z$ the subvector on which we condition. For a real-valued Borel-measurable generalized residual $\boldsymbol{w}\mapsto\psi_{\boldsymbol{\theta}}(\boldsymbol{w})$, indexed by $\boldsymbol{\theta}\in\Theta$, consider
\[
H_{0}:\quad\mathbb{E}\{\psi_{\boldsymbol{\theta}_{0}}(\boldsymbol{W})\mid\boldsymbol{Z}\}=0\quad\text{a.s.-}P_{\boldsymbol Z}\quad\text{for some }\boldsymbol{\theta}_{0}\in\Theta.
\]

 \looseness=-1 Given an i.i.d. sample $\{\boldsymbol{W}_{i}\}_{i=1}^{n}$, partitioning the support of $\boldsymbol{Z}$ into $\boldsymbol{\gamma}=(\gamma_{1},\ldots,\gamma_{L})$
gives the scaled sample cell moments 
\begin{equation}
\widehat{\boldsymbol{V}}_{\boldsymbol{\theta}}^{\psi}(\boldsymbol{\gamma})=\frac{1}{\sqrt{n}}\sum_{i=1}^{n}\psi_{\boldsymbol{\theta}}(\boldsymbol{W}_{i})\mathbf{1}_{\boldsymbol{\gamma}}(\boldsymbol{Z}_{i}).\label{eq:general-cells}
\end{equation}
Under the null, every population cell moment is zero. For a fixed
partition, disjointness of the cells gives the covariance matrix 
\[
\boldsymbol{\Sigma}_{0}^{\psi}(\boldsymbol{\gamma})=\operatorname{diag}\left\{ \mathbb{E}\left[\psi_{\boldsymbol{\theta}_{0}}^{2}(\boldsymbol{W})\mathbf{1}_{\boldsymbol\gamma}(\boldsymbol{Z})\right]\right\} .
\]
The cell-moment Jacobian is 
\[
\boldsymbol{J}_{0}^{\psi}(\boldsymbol{\gamma})=\left.\frac{\partial}{\partial\boldsymbol{\theta}'}\mathbb{E}\left[\psi_{\boldsymbol{\theta}}(\boldsymbol{W})\mathbf{1}_{\boldsymbol{\gamma}}(\boldsymbol{Z})\right]\right|_{\boldsymbol{\theta}=\boldsymbol{\theta}_{0}}.
\]
We replace the population second moments by sample estimates
and minimize the resulting weighted criterion over $\boldsymbol{\theta}$.
This gives the analogue of the regression statistic. Suppose that
the conditional moment versions of A1 or A2 hold; in particular, the
selected cell moments identify $\boldsymbol{\theta}$, the cell variances
are positive and the Jacobian has full column rank $p$. Then, for
$L>p$, the minimized statistic converges under the
null to a $\chi^2_{L-p}$ distribution.
The proof follows the regression case, with the same expansion of the cell moments, covariance estimation argument and projection. With multiple conditional restrictions, each cell contributes a vector of moments. The covariance matrix is then block diagonal across disjoint cells, with nonsingular within-cell blocks, and the degrees of freedom equal the total number of cell moments minus the number of estimated parameters. When the partition is estimated, the complexity and
stability conditions are stated under $P_{\boldsymbol{Z}}$, the distribution of $\boldsymbol Z$,
together with regularity conditions on the classes of generalized
residuals. The examples below show the constructions used in the empirical
applications.

\refstepcounter{example}\label{ex:instrumental-variables}
\paragraph{Example~\theexample\ (instrumental variables).}

 \looseness=-1 Consider the restriction 
\begin{equation*}
H_0:\quad
\mathbb E[Y-m_{\boldsymbol\theta_0}(\boldsymbol X)\mid\boldsymbol Z]=0
\quad\text{a.s.-}P_{\boldsymbol Z},\quad\text{for some }\boldsymbol\theta_0\in\Theta.
\end{equation*}
Here $\boldsymbol{X}$ may contain endogenous regressors and $\boldsymbol{Z}$
is a vector of instruments. Set $U=Y-m_{\boldsymbol{\theta}_{0}}(\boldsymbol{X})$.
The cells are constructed in instrument space, and \eqref{eq:general-cells}
is 
\begin{equation*}
\widehat{\boldsymbol V}_{\boldsymbol\theta}^{\mathrm{IV}}(\boldsymbol\gamma)
=\frac1{\sqrt n}\sum_{i=1}^n
\{Y_i-m_{\boldsymbol\theta}(\boldsymbol X_i)\}\mathbf1_{\boldsymbol\gamma}(\boldsymbol Z_i).
\end{equation*}
The analog of $\boldsymbol{\Sigma}_{0}$ is $\boldsymbol{\Sigma}_{0}^{\mathrm{IV}}(\boldsymbol{\gamma})=\operatorname{diag}\{\mathbb{E}[U^{2}\mathbf{1}_{\boldsymbol\gamma}(\boldsymbol{Z})]\}$,
and the Jacobian is $\boldsymbol{J}_{0}^{\mathrm{IV}}(\boldsymbol{\gamma})=-\mathbb{E}[\mathbf{1}_{\boldsymbol{\gamma}}(\boldsymbol{Z})\boldsymbol{\dot{m}}(\boldsymbol{X})']$,
where $\boldsymbol{\dot{m}}$ is the derivative of $m_{\boldsymbol{\theta}}$
at $\boldsymbol{\theta}_{0}$. The selected instrument cells must retain identification
of the structural parameter. This construction applies to linear and
nonlinear specifications of $m$.

\refstepcounter{example}\label{ex:expected-elasticities}
\paragraph{Example~\theexample\ (expected output elasticities).}

We apply the mean-independence test of Example~\ref{ex:mean-independence} to an exponential residual identified by unconditional normal equations.

\looseness=-1 Let $Y>0$ be output, $\boldsymbol X$ a vector of positive inputs, and \begin{equation}\label{eq:cd-mean}
m_{\boldsymbol\vartheta}(\boldsymbol x)
=A\prod_{j=1}^k x_j^{\beta_j},
\qquad \boldsymbol\vartheta=(A,\boldsymbol\beta')',\quad A>0,
\end{equation}
the Cobb--Douglas technology. If $\mathbb E(Y\mid\boldsymbol X)=m_{\boldsymbol\vartheta_0}(\boldsymbol X)$,
with $\boldsymbol\vartheta_0=(A_0,\boldsymbol\beta_0')'$, the elasticities
of expected output are the components of $\boldsymbol\beta_0$. These elasticities are often estimated by the slopes of the LS projection of $\ln Y$ on $\boldsymbol z(\boldsymbol X)=(1,\ln X_1,\ldots,\ln X_k)'$. The projection parameter $\boldsymbol\eta_0=(\eta_{00},\eta_{01},\ldots,\eta_{0k})'$ solves the unconditional moment restriction $\mathbb E\{\nu(\boldsymbol\eta_0)\boldsymbol z(\boldsymbol X)\}=\boldsymbol0$, where $\nu(\boldsymbol\eta):=\ln Y-\boldsymbol z'(\boldsymbol X)\boldsymbol\eta$.
Under the maintained Cobb--Douglas specification, $\eta_{0j}=\beta_{0j}$, $j=1,\ldots,k$, if and only if
\[
\mathbb E\{e^{\nu(\boldsymbol\eta_0)}-\varpi_0\mid\boldsymbol X\}=0,
\qquad \varpi_0:=\mathbb E\{e^{\nu(\boldsymbol\eta_0)}\}.
\]
Under this restriction, the LS slopes $\eta_{0j}$ equal the elasticities of expected output $\beta_{0j}$ \citep{santos2006log}. We call this the mean-independence (MI) test of the retransformation factor. To account for estimation of $\boldsymbol\eta_0$, we augment the cell moments with the unconditional normal equations.

\looseness=-1 For implementation, first estimate $\boldsymbol\eta_0$ by LS and $\varpi_0$ by the sample mean of the resulting exponential residuals. For a partition $\boldsymbol\Gamma$, combine the $k+1$ LS normal equations with the $L$ sample cell moments $n^{-1}\sum_{i=1}^n1_{\Gamma_\ell}(\boldsymbol X_i)\{e^{\nu_i(\boldsymbol\eta)}-\varpi\}$, $\ell=1,\ldots,L$, where $\nu_i(\boldsymbol\eta)=\ln Y_i-\boldsymbol z'(\boldsymbol X_i)\boldsymbol\eta$.
Use the preliminary estimates to estimate the full covariance matrix of these stacked moments, including the cross-covariances between the two blocks. Holding this weight matrix and the cells fixed, minimize the quadratic criterion jointly over $\boldsymbol\eta$ and $\varpi$. Thus the initial LS slopes are updated in computing the test, and their estimation is accounted for. The $L+k+1$ moment restrictions and $k+2$ parameters leave $L-1$ overidentifying restrictions. Standard GMM arguments \citep{NeweyMcFadden1994}, with the treatment of estimated cells in Section~\ref{sec:ddp}, give a $\chi^2_{L-1}$ null limit. As in Section~\ref{sec:ddp}, the statistic computed with the data-dependent cells has the same limiting distribution as the statistic computed with their fixed limits. Additional functions of the log regressors can be included in $\boldsymbol z$, as in the restricted-translog application below.

\section{Tests for the Specification of Production Functions}
 
\label{sec:applications} \looseness=-1 We apply the proposed tests to production-function specifications for 411 Colombian food-products plants observed in 1991. Output $Y$ is real gross output, and the three inputs are labor ($X_1$), capital ($X_2$), and real intermediate inputs ($X_3$). These data, studied by \citet{gandhi2020identification}, are available as \texttt{colombian} in the \texttt{gnrprod} package \citep{jin2023gnrprod}. Appendix~B gives implementation details and applies the tests to cost functions of U.S. electric utilities.
 
\looseness=-1 We test the Cobb--Douglas (CD) specification in~\eqref{eq:cd-mean} and the constant-elasticity-of-substitution (CES) specification,
$m_{\mathrm{CES}}(\boldsymbol x)=e^{\xi}(\sum_{j=1}^3 \varphi_jx_j^\rho)^{\kappa/\rho}$,
with $\sum_{j=1}^3 \varphi_j=1$, which are estimated by nonlinear least squares (NLS). CES has elasticity of substitution $1/(1-\rho)$ and approaches CD as $\rho\to0$. We also implement the test in Example~5 for mean independence of $e^{\nu(\boldsymbol\eta_0)}$ and the inputs, with $\nu(\boldsymbol\eta_0)$ defined in Example~\ref{ex:expected-elasticities}, denoted MI--CD. The MI--RT test replaces the log-log projection by the restricted-translog (RT) projection with index
\[
r_{\mathrm{RT}}(\boldsymbol x)=\theta_0+\textstyle\sum_{j=1}^3\theta_j\ln x_j
+\theta_4\cdot\{s(\boldsymbol x)-\bar s\}^2/2,
\]
where $s(\boldsymbol x)=\sum_{j=1}^3\ln x_j$ and $\bar s=n^{-1}\sum_{i=1}^{n}s(\boldsymbol X_i)$. Under the maintained CD specification, MI--CD tests the condition under which the log-projection slopes equal the elasticities of expected output. For the regression specification tests and partition construction, we also use the exponentiated RT specification
$m_{\mathrm{exp\text{-}RT}}(\boldsymbol x)=\exp\{r_{\mathrm{RT}}(\boldsymbol x)\}$,
denoted exp-RT and fitted by NLS.

\paragraph{Specification tests.}
\looseness=-1 The regression specification tests use the $\chi^2_{L-p}$ approximation, with $p=4$ for CD ($A$ and
the three elasticities) and $p=5$ for CES ($\xi$, $\kappa$, $\rho$, and two
free weights). After accounting for the log-projection coefficients, the
MI--CD and MI--RT tests estimate $\varpi_0=\mathbb E(e^{\nu(\boldsymbol\eta_0)})$ and use the $\chi^2_{L-1}$ approximation.
 
CD is tested with partitions guided by CES and exp-RT, and CES with partitions guided by CD and exp-RT. MI--CD and MI--RT use the same partitions. Their preliminary sign splits compare the CES or exp-RT guide with the NLS fit of CD, so the guided partitions identify the model used to construct the cells, not the restriction under test.

\begin{sidewaystable}
\begingroup
\centering
\begin{threeparttable}
\setcounter{table}{5}\caption{Production-function specification tests: Colombian plants, $n=411$. $p$-values.}\label{Table:Applications}
\footnotesize\renewcommand{\baselinestretch}{1}\selectfont
\setlength{\tabcolsep}{3.2pt}\renewcommand{\arraystretch}{0.92}
\smallskip\noindent
\begin{minipage}[t]{.485\linewidth}\centering
\textit{Conditional-mean specifications}\par\smallskip
\begin{tabular}{@{}lrrrrrrr@{}}\toprule
Partition & \multicolumn{7}{c}{$L$}\\
\cmidrule(lr){2-8}
 & 6 & 7 & 8 & 9 & 10 & 11 & 12\\\midrule
\multicolumn{8}{l}{\textit{CD null}}\\
BC & \textbf{.550} & \textbf{.170} & .004 & .004 & \textbf{.259} & .024 & .021\\
KM & .001 & $<\!.001$ & $<\!.001$ & $<\!.001$ & $<\!.001$ & .002 & $<\!.001$\\
FV--CES & \textbf{.264} & .010 & \textbf{.129} & .003 & $<\!.001$ & .003 & .013\\
CES--FV & \textbf{.596} & .036 & .010 & .019 & .004 & $<\!.001$ & .022\\
CES--BC & $<\!.001$ & $<\!.001$ & $<\!.001$ & $<\!.001$ & $<\!.001$ & $<\!.001$ & $<\!.001$\\
CES--KM & $<\!.001$ & $<\!.001$ & $<\!.001$ & $<\!.001$ & $<\!.001$ & $<\!.001$ & $<\!.001$\\
FV--exp-RT & \textbf{.370} & \textbf{.302} & .048 & \textbf{.169} & .022 & .010 & .016\\
exp-RT--FV & .033 & $<\!.001$ & .002 & $<\!.001$ & $<\!.001$ & .007 & .024\\
exp-RT--BC & \textbf{.092} & .002 & .018 & .034 & .002 & .004 & \textbf{.061}\\
exp-RT--KM & .049 & .020 & .036 & .013 & .006 & .001 & $<\!.001$\\
\midrule
\multicolumn{8}{l}{\textit{CES null}}\\
BC & \textbf{.927} & \textbf{.244} & \textbf{.079} & \textbf{.337} & \textbf{.533} & \textbf{.426} & \textbf{.302}\\
KM & .001\textsuperscript{$\dagger$} & $<\!.001$ & .032 & .013 & \textbf{.091} & \textbf{.141} & \textbf{.187}\\
FV--CD & \textbf{.334} & \textbf{.300} & \textbf{.965} & \textbf{.668} & \textbf{.826} & \textbf{.830} & \textbf{.177}\\
CD--FV & \textbf{.362} & \textbf{.589} & \textbf{.422} & \textbf{.102} & \textbf{.248} & \textbf{.852} & \textbf{.201}\\
CD--BC & \textbf{.506} & \textbf{.382} & .038 & .003 & \textbf{.097} & \textbf{.453} & \textbf{.086}\\
CD--KM & \textbf{.158} & \textbf{.150} & \textbf{.052} & .017 & .027 & .036 & .036\\
FV--exp-RT & \textbf{.769} & \textbf{.581} & \textbf{.480} & \textbf{.776} & \textbf{.544} & \textbf{.586} & \textbf{.332}\\
exp-RT--FV & \textbf{.146} & \textbf{.134} & \textbf{.413} & \textbf{.479} & \textbf{.342} & \textbf{.175} & \textbf{.141}\\
exp-RT--BC & \textbf{.288} & .009 & .020 & \textbf{.128} & \textbf{.209} & .010 & \textbf{.205}\\
exp-RT--KM & \textbf{.748} & \textbf{.924} & \textbf{.518} & \textbf{.111} & \textbf{.166} & .038 & .017\\
\bottomrule\end{tabular}
\par\smallskip\textit{Omnibus comparisons}\par
\setlength{\tabcolsep}{6pt}
\begin{tabular}{@{}lrrrr@{}}\toprule
 & CD & CES & MI--CD & MI--RT\\\midrule
$S$ & \textbf{.685} & \textbf{.557} & \textbf{.060} & \textbf{.152}\\
$SZ$ & \textbf{.481} & \textbf{.620} & \textbf{.215} & \textbf{.110}\\
$E$ & \textbf{.518} & \textbf{.753} & \textbf{.287} & \textbf{.094}\\\bottomrule
\end{tabular}
\end{minipage}\hfill\begin{minipage}[t]{.485\linewidth}\centering
\textit{Mean-independence restrictions}\par\smallskip
\begin{tabular}{@{}lrrrrrrr@{}}\toprule
Partition & \multicolumn{7}{c}{$L$}\\
\cmidrule(lr){2-8}
 & 6 & 7 & 8 & 9 & 10 & 11 & 12\\\midrule
\multicolumn{8}{l}{\textit{MI--CD null}}\\
BC & \textbf{.236} & $<\!.001$ & $<\!.001$ & $<\!.001$ & \textbf{.123} & $<\!.001$ & $<\!.001$\\
KM & $<\!.001$ & $<\!.001$ & $<\!.001$ & $<\!.001$ & $<\!.001$ & $<\!.001$ & $<\!.001$\\
FV--CES & $<\!.001$ & $<\!.001$ & .025 & $<\!.001$ & $<\!.001$ & $<\!.001$ & $<\!.001$\\
CES--FV & $<\!.001$ & $<\!.001$ & $<\!.001$ & $<\!.001$ & $<\!.001$ & $<\!.001$ & $<\!.001$\\
CES--BC & $<\!.001$ & $<\!.001$ & $<\!.001$ & $<\!.001$ & $<\!.001$ & $<\!.001$ & $<\!.001$\\
CES--KM & $<\!.001$ & $<\!.001$ & $<\!.001$ & $<\!.001$ & $<\!.001$ & $<\!.001$ & $<\!.001$\\
FV--exp-RT & $<\!.001$ & .009 & .006 & \textbf{.060} & .005 & .002 & $<\!.001$\\
exp-RT--FV & \textbf{.067} & .024 & .031 & .004 & .004 & .004 & .009\\
exp-RT--BC & \textbf{.144} & \textbf{.065} & \textbf{.344} & \textbf{.258} & $<\!.001$ & .029 & .022\\
exp-RT--KM & \textbf{.238} & \textbf{.356} & \textbf{.312} & .008 & .013 & $<\!.001$ & $<\!.001$\\
FV--CD & $<\!.001$ & .010 & \textbf{.219} & .009 & .004 & .013 & .001\\
\midrule
\multicolumn{8}{l}{\textit{MI--RT null}}\\
BC & \textbf{.637} & .019 & .017 & .004 & \textbf{.227} & .003 & .017\\
KM & $<\!.001$ & .002 & $<\!.001$ & $<\!.001$ & $<\!.001$ & $<\!.001$ & $<\!.001$\\
FV--CES & .005 & .019 & \textbf{.316} & .005 & .002 & $<\!.001$ & $<\!.001$\\
CES--FV & $<\!.001$ & $<\!.001$ & $<\!.001$ & $<\!.001$ & $<\!.001$ & $<\!.001$ & $<\!.001$\\
CES--BC & $<\!.001$ & $<\!.001$ & $<\!.001$ & $<\!.001$ & $<\!.001$ & $<\!.001$ & $<\!.001$\\
CES--KM & $<\!.001$ & $<\!.001$ & $<\!.001$ & $<\!.001$ & $<\!.001$ & $<\!.001$ & $<\!.001$\\
FV--exp-RT & .005 & \textbf{.127} & \textbf{.067} & \textbf{.264} & .049 & .022 & .007\\
exp-RT--FV & \textbf{.079} & \textbf{.104} & .039 & .017 & .014 & .009 & .028\\
exp-RT--BC & \textbf{.377} & \textbf{.459} & \textbf{.148} & \textbf{.162} & .003 & \textbf{.065} & \textbf{.088}\\
exp-RT--KM & \textbf{.522} & \textbf{.425} & \textbf{.225} & .002 & .003 & .001 & $<\!.001$\\
FV--CD & .006 & \textbf{.159} & \textbf{.724} & \textbf{.107} & .048 & \textbf{.121} & .009\\
\bottomrule\end{tabular}\end{minipage}
\begin{tablenotes}[flushleft]\scriptsize
\item \textit{Notes:} Entries are $p$-values; values above 0.05 are in bold. The dagger marks the CES-null KM cell at $L=6$, whose unconstrained criterion fit has a negative CES weight. Omnibus comparisons use the same four nulls.
\end{tablenotes}
\end{threeparttable}\par
\endgroup\end{sidewaystable}

\looseness=-1 Table~\ref{Table:Applications} reports $p$-values for $L=6,\ldots,12$, with values above 0.05 in bold. The entries are dependent and do not form a combined test. Pairing CES-guided tests of CD with CD-guided tests of CES gives 28 comparisons at 5 percent. Only CD is rejected in 19, both in six, neither in three, and only CES is rejected in none. Overall, these comparisons favor CES over CD. The CES estimate $\hat\rho=0.60$ (robust standard error 0.20) implies an elasticity of substitution of 2.5, versus one imposed under CD.

\looseness=-1 The three CES-guided refinements, and KM, reject both MI--CD and MI--RT for every $L$, and the exp-RT-guided partitions give more mixed results, especially for small $L$. Thus, as discussed in Example~\ref{ex:expected-elasticities}, the log-projection slopes cannot generally be interpreted as elasticities of expected output. The RT restriction differs from the CD one only by the scale term, and when both restrictions are rejected this term does not restore mean independence. These findings do not rule out every richer log specification. No omnibus test rejects either level specification or either mean-independence restriction at the 5 percent level.

\label{sec:iv-application}
 
\begin{table}[!tbp]\centering
\begingroup
\begin{threeparttable}
\setcounter{table}{6}\caption{Instrumental-variable production-function specification tests: Colombian plants, $n=406$. $p$-values.}\label{tab:iv-main}
\footnotesize\renewcommand{\baselinestretch}{1}\selectfont\setlength{\tabcolsep}{4pt}\renewcommand{\arraystretch}{1}
\begin{tabular}{llrrrrrrr}
\toprule
Null & Partition & \multicolumn{7}{c}{$L$}\\
\cmidrule(lr){3-9}
 & & 6 & 7 & 8 & 9 & 10 & 11 & 12\\
\midrule
 
\looseness=-1 CD & BC & \textbf{0.057} & \textbf{0.294} & $<$0.001 & \textbf{0.056} & \textbf{0.195} & 0.007 & 0.003\\
CD & KM & 0.016 & 0.005 & $<$0.001 & $<$0.001 & $<$0.001 & $<$0.001 & $<$0.001\\
CD & FV--CES & \textbf{0.724} & 0.002 & \textbf{0.086} & 0.004 & 0.022 & 0.002 & 0.004\\
CD & FV--exp-RT & 0.035 & 0.015 & 0.022 & $<$0.001 & 0.001 & 0.001 & 0.001\\
CD & CES--FV & 0.050 & \textbf{0.644} & \textbf{0.775} & 0.011 & 0.012 & \textbf{0.072} & \textbf{0.124}\\
\midrule
CES & BC & \textbf{0.234} & \textbf{0.584} & \textbf{0.708}$^{\dagger}$ & \textbf{0.714} & \textbf{0.966} & \textbf{0.202} & \textbf{0.205}\\
CES & KM & \textbf{0.270} & 0.005 & 0.004 & \textbf{0.123} & \textbf{0.282} & \textbf{0.269} & \textbf{0.250}\\
CES & FV--exp-RT & \textbf{0.769} & \textbf{0.179} & \textbf{0.683}$^{\dagger}$ & \textbf{0.061} & \textbf{0.093} & 0.034$^{\dagger}$ & \textbf{0.243}\\
CES & exp-RT--FV & \textbf{0.787}$^{\dagger}$ & \textbf{0.888}$^{\dagger}$ & \textbf{0.910}$^{\dagger}$ & \textbf{0.774} & \textbf{0.851}$^{\dagger}$ & \textbf{0.935} & \textbf{0.691}\\
\midrule
\multicolumn{9}{l}{\textit{Omnibus tests}}\\
Null & $S$ & $SZ$ & $E$ & & & & &\\
CD & \textbf{0.052} & \textbf{0.327} & 0.008 & & & & &\\
CES & \textbf{0.287} & \textbf{0.485} & \textbf{0.536} & & & & &\\
\bottomrule
\end{tabular}
\begin{tablenotes}[flushleft]\scriptsize
\item \textit{Notes:} Entries are $p$-values, and values above 0.05 are bold, with boldface determined before rounding. Daggers mark negative CES weights.
\end{tablenotes}
\end{threeparttable}\par
\endgroup
\end{table}

\paragraph{Instrumental-variable application.}
\looseness=-1 To allow endogenous inputs \citep{griliches1999production,blundell2000gmm}, we apply Example~4 to 406 plants observed in 1990--1991. The instruments are the 1990 log inputs, their squares, and their cross products. The cells are formed with the lagged log inputs, the guides use the first-stage fitted log inputs, and the structural residuals use the observed inputs. The smallest conditional first-stage $F$ statistic \citep{sanderson2016weak} is 540.

\looseness=-1 The classical $J$ test \citep{hansen1982large} rejects the CD and exp-RT IV specifications ($p<0.001$), but not CES ($p=0.23$). With the cell indicators as instruments, Table~\ref{tab:iv-main} also gives stronger evidence against CD than CES. Appendix~B reports the IV tests for all partitions of Table~\ref{Table:Applications}. The CES-guided BC and KM refinements reject CD for every reported $L$, whereas the CD-guided refinements reject CES only for some $L\geq8$. Only $E$ among the omnibus tests rejects CD, and none rejects CES. These tests jointly assess specification and instrument validity.

\section{Conclusions}
 
\looseness=-1 We propose a Pearson-type test based on cell moments whose asymptotic null
distribution is chi-square, also when the cells are chosen from the data. Local
power depends on the cell drift left after projecting out the parameter
directions, so guiding models, which need not be correctly specified, can
direct power toward relevant departures. In our simulations, guided partitions
give large power gains with size accuracy comparable to that of bootstrap
omnibus tests. For the Colombian plants, the cell tests often reject
Cobb--Douglas and the mean-independence restrictions but rarely CES, whereas in the regression and mean-independence comparisons the omnibus tests reject none of them. The IV tests point in the same direction.

{\small \paragraph{Acknowledgments:} The authors used Codex (OpenAI) for most of the code and the graphs, and Claude (Anthropic) to revise the text and check it against the tables and code. The authors verified all output and take full responsibility for the paper.\par}

{\small \paragraph{Disclosure statement:} The authors report there are no competing interests to declare.\par}

\begingroup
\normalsize
\setlength{\bibsep}{0pt}\enlargethispage{\baselineskip}\interlinepenalty=10000
\bibliographystyle{asa_jbes}
\bibliography{References_revised}

\begin{thebibliography}{43}
\newcommand{\enquote}[1]{``#1''}
\expandafter\ifx\csname natexlab\endcsname\relax\def\natexlab#1{#1}\fi

\bibitem[{Andrews(1988)}]{andrews1988chi}
Andrews, D. W.~K. (1988), \enquote{Chi-Square Diagnostic Tests for Econometric
  Models: Theory,} \textit{Econometrica}, 56, 1419--1453.

\bibitem[{Andrews(1997)}]{andrews1997conditional}
--- (1997), \enquote{A Conditional {K}olmogorov Test,} \textit{Econometrica},
  65, 1097--1128.

\bibitem[{Bhattacharya(1974)}]{bhattacharya1974}
Bhattacharya, P.~K. (1974), \enquote{Convergence of Sample Paths of Normalized
  Sums of Induced Order Statistics,} \textit{The Annals of Statistics}, 2,
  1034--1039.

\bibitem[{Bhattacharya(1976)}]{bhattacharya1976}
--- (1976), \enquote{An Invariance Principle in Regression Analysis,}
  \textit{The Annals of Statistics}, 4, 621--624.

\bibitem[{Bickel and Rosenblatt(1973)}]{bickel1973some}
Bickel, P.~J. and Rosenblatt, M. (1973), \enquote{On Some Global Measures of
  the Deviations of Density Function Estimates,} \textit{The Annals of
  Statistics}, 1, 1071--1095.

\bibitem[{Bierens(1982)}]{bierens1982consistent}
Bierens, H.~J. (1982), \enquote{Consistent Model Specification Tests,}
  \textit{Journal of Econometrics}, 20, 105--134.

\bibitem[{Bierens and Ploberger(1997)}]{bierens1997asymptotic}
Bierens, H.~J. and Ploberger, W. (1997), \enquote{Asymptotic Theory of
  Integrated Conditional Moment Tests,} \textit{Econometrica}, 65, 1129--1151.

\bibitem[{Birch(1964)}]{birch1964detection}
Birch, M.~W. (1964), \enquote{A New Proof of the {Pearson--Fisher} Theorem,}
  \textit{The Annals of Mathematical Statistics}, 35, 817--824.

\bibitem[{Blundell and Bond(2000)}]{blundell2000gmm}
Blundell, R. and Bond, S. (2000), \enquote{{GMM} Estimation with Persistent
  Panel Data: An Application to Production Functions,} \textit{Econometric
  Reviews}, 19, 321--340.

\bibitem[{Delgado et~al.(2006)Delgado, Dom{\'i}nguez, and
  Lavergne}]{delgado2006consistent}
Delgado, M.~A., Dom{\'i}nguez, M.~A., and Lavergne, P. (2006),
  \enquote{Consistent Tests of Conditional Moment Restrictions,}
  \textit{Annales d'{\'E}conomie et de Statistique}, 81, 33--67.

\bibitem[{Delgado and Stute(2008)}]{delgado2008distribution}
Delgado, M.~A. and Stute, W. (2008), \enquote{Distribution-Free Specification
  Tests of Conditional Models,} \textit{Journal of Econometrics}, 143, 37--55.

\bibitem[{Delgado and Vainora(2022)}]{delgado2022conditional}
Delgado, M.~A. and Vainora, J. (2022), \enquote{Conditional Distribution
  Specification Testing Based on Data-Dependent Partitions,} arXiv:2210.00624.

\bibitem[{Drost et~al.(1989)Drost, Kallenberg, Moore, and
  Oosterhoff}]{drost1989power}
Drost, F.~C., Kallenberg, W. C.~M., Moore, D.~S., and Oosterhoff, J. (1989),
  \enquote{Power Approximations to Multinomial Tests of Fit,} \textit{Journal
  of the American Statistical Association}, 84, 130--141.

\bibitem[{Escanciano(2006)}]{escanciano2006consistent}
Escanciano, J.~C. (2006), \enquote{A Consistent Diagnostic Test for Regression
  Models Using Projections,} \textit{Econometric Theory}, 22, 1030--1051.

\bibitem[{Escanciano(2024)}]{escanciano2024gaussian}
--- (2024), \enquote{A Gaussian Process Approach to Model Checks,} \textit{The
  Annals of Statistics}, 52, 2456--2481.

\bibitem[{Gandhi et~al.(2020)Gandhi, Navarro, and
  Rivers}]{gandhi2020identification}
Gandhi, A., Navarro, S., and Rivers, D.~A. (2020), \enquote{On the
  Identification of Gross Output Production Functions,} \textit{Journal of
  Political Economy}, 128, 2973--3016.

\bibitem[{Greenwood and Nikulin(1996)}]{greenwood1996guide}
Greenwood, P.~E. and Nikulin, M.~S. (1996), \textit{A Guide to Chi-Squared
  Testing}, New York: Wiley.

\bibitem[{Griliches and Mairesse(1999)}]{griliches1999production}
Griliches, Z. and Mairesse, J. (1999), \enquote{Production Functions: The
  Search for Identification,} in \textit{Econometrics and Economic Theory in
  the 20th Century: The Ragnar Frisch Centennial Symposium}, ed. Str{\o}m, S.,
  Cambridge: Cambridge University Press, pp. 169--203.

\bibitem[{Hansen(1982)}]{hansen1982large}
Hansen, L.~P. (1982), \enquote{Large Sample Properties of Generalized Method of
  Moments Estimators,} \textit{Econometrica}, 50, 1029--1054.

\bibitem[{H{\"a}rdle and Mammen(1993)}]{hardle1993comparing}
H{\"a}rdle, W. and Mammen, E. (1993), \enquote{Comparing Nonparametric Versus
  Parametric Regression Fits,} \textit{The Annals of Statistics}, 21,
  1926--1947.

\bibitem[{Heckman(1984)}]{heckman1984chi}
Heckman, J.~J. (1984), \enquote{The {$\chi^2$} Goodness-of-Fit Statistic for
  Models With Parameters Estimated From Microdata,} \textit{Econometrica}, 52,
  1543--1547.

\bibitem[{Hosmer and Lemeshow(1980)}]{hosmer1980goodness}
Hosmer, D.~W. and Lemeshow, S. (1980), \enquote{Goodness of Fit Tests for the
  Multiple Logistic Regression Model,} \textit{Communications in
  Statistics---Theory and Methods}, 9, 1043--1069.

\bibitem[{Jin(2023)}]{jin2023gnrprod}
Jin, D.~J. (2023), \textit{{gnrprod}: Estimates Gross Output Functions}, r
  package version 1.1.2, Available at
  \url{https://CRAN.R-project.org/package=gnrprod}.

\bibitem[{Kallenberg et~al.(1985)Kallenberg, Oosterhoff, and
  Schriever}]{kallenberg1985number}
Kallenberg, W. C.~M., Oosterhoff, J., and Schriever, B. (1985), \enquote{The
  Number of Classes in Chi-Squared Goodness-of-Fit Tests,} \textit{Journal of
  the American Statistical Association}, 80, 959--968.

\bibitem[{Khmaladze(1982)}]{khmaladze1982martingale}
Khmaladze, E.~V. (1982), \enquote{Martingale Approach in the Theory of
  Goodness-of-Fit Tests,} \textit{Theory of Probability \& Its Applications},
  26, 240--257.

\bibitem[{Khmaladze and Koul(2004)}]{10.1214/Khmaladze2004}
Khmaladze, E.~V. and Koul, H.~L. (2004), \enquote{{Martingale Transforms
  Goodness-of-Fit Tests in Regression Models},} \textit{The Annals of
  Statistics}, 32, 995--1034.

\bibitem[{Kosorok(2008)}]{Kosorok2008}
Kosorok, M.~R. (2008), \textit{Introduction to Empirical Processes and
  Semiparametric Inference}, Springer Series in Statistics, New York: Springer.

\bibitem[{Newey(1985)}]{newey1985conditional}
Newey, W.~K. (1985), \enquote{Maximum Likelihood Specification Testing and
  Conditional Moment Tests,} \textit{Econometrica}, 53, 1047--1070.

\bibitem[{Newey and McFadden(1994)}]{NeweyMcFadden1994}
Newey, W.~K. and McFadden, D. (1994), \enquote{Large Sample Estimation and
  Hypothesis Testing,} in \textit{Handbook of Econometrics}, eds. Engle, R.~F.
  and McFadden, D.~L., Amsterdam: Elsevier, vol.~4, chap.~36, pp. 2111--2245.

\bibitem[{Pearson(1900)}]{pearson1900x}
Pearson, K. (1900), \enquote{On the Criterion That a Given System of Deviations
  From the Probable in the Case of a Correlated System of Variables Is Such
  That It Can Be Reasonably Supposed to Have Arisen From Random Sampling,}
  \textit{The London, Edinburgh, and Dublin Philosophical Magazine and Journal
  of Science}, 50, 157--175.

\bibitem[{Pollard(1979)}]{pollard1979general}
Pollard, D. (1979), \enquote{General Chi-Square Goodness-of-Fit Tests With
  Data-Dependent Cells,} \textit{Zeitschrift f{\"u}r Wahrscheinlichkeitstheorie
  und verwandte Gebiete}, 50, 317--331.

\bibitem[{Pollard(1980)}]{pollard1980minimum}
--- (1980), \enquote{The Minimum Distance Method of Testing,} \textit{Metrika},
  27, 43--70.

\bibitem[{Pollard(1981)}]{pollard1981strong}
--- (1981), \enquote{Strong Consistency of {$K$}-Means Clustering,} \textit{The
  Annals of Statistics}, 9, 135--140.

\bibitem[{Sanderson and Windmeijer(2016)}]{sanderson2016weak}
Sanderson, E. and Windmeijer, F. (2016), \enquote{A Weak Instrument {F}-Test in
  Linear {IV} Models With Multiple Endogenous Variables,} \textit{Journal of
  Econometrics}, 190, 212--221.

\bibitem[{Santos~Silva and Tenreyro(2006)}]{santos2006log}
Santos~Silva, J. M.~C. and Tenreyro, S. (2006), \enquote{The Log of Gravity,}
  \textit{The Review of Economics and Statistics}, 88, 641--658.

\bibitem[{Stute(1997)}]{stute1997nonparametric}
Stute, W. (1997), \enquote{Nonparametric Model Checks for Regression,}
  \textit{The Annals of Statistics}, 25, 613--641.

\bibitem[{Stute et~al.(1998)Stute, {Gonz{\'a}lez Manteiga}, and {Presedo
  Quindimil}}]{stute1998bootstrap}
Stute, W., {Gonz{\'a}lez Manteiga}, W., and {Presedo Quindimil}, M. (1998),
  \enquote{Bootstrap Approximations in Model Checks for Regression,}
  \textit{Journal of the American Statistical Association}, 93, 141--149.

\bibitem[{Stute and Zhu(2002)}]{stute2002model}
Stute, W. and Zhu, L.-X. (2002), \enquote{Model Checks for Generalized Linear
  Models,} \textit{Scandinavian Journal of Statistics}, 29, 535--545.

\bibitem[{Surjanovic et~al.(2024)Surjanovic, Lockhart, and
  Loughin}]{surjanovic2024generalized}
Surjanovic, N., Lockhart, R.~A., and Loughin, T.~M. (2024), \enquote{A
  Generalized {Hosmer--Lemeshow} Goodness-of-Fit Test for a Family of
  Generalized Linear Models,} \textit{TEST}, 33, 589--608.

\bibitem[{Tauchen(1985)}]{tauchen1985diagnostic}
Tauchen, G. (1985), \enquote{Diagnostic Testing and Evaluation of Maximum
  Likelihood Models,} \textit{Journal of Econometrics}, 30, 415--443.

\bibitem[{van~der Vaart(1998)}]{van2000asymptotic}
van~der Vaart, A.~W. (1998), \textit{Asymptotic Statistics}, Cambridge:
  Cambridge University Press.

\bibitem[{van~der Vaart and Wellner(1996)}]{van1996weak}
van~der Vaart, A.~W. and Wellner, J.~A. (1996), \textit{Weak Convergence and
  Empirical Processes: With Applications to Statistics}, New York: Springer.

\bibitem[{van~der Vaart and Wellner(2000)}]{vdvw2000preservation}
--- (2000), \enquote{Preservation Theorems for {G}livenko--{C}antelli and
  Uniform {G}livenko--{C}antelli Classes,} in \textit{High Dimensional
  Probability {II}}, eds. Gin{\'e}, E., Mason, D.~M., and Wellner, J.~A.,
  Boston: Birkh{\"a}user, vol.~47 of \textit{Progress in Probability}, pp.
  115--133.

\end{thebibliography}
\endgroup

\bigskip
\setcounter{section}{0}\renewcommand{\thesection}{A.\arabic{section}}
\setcounter{equation}{0}\renewcommand{\theequation}{A.\arabic{equation}}
\setcounter{table}{0}\renewcommand{\thetable}{A.\arabic{table}}
\setcounter{figure}{0}\renewcommand{\thefigure}{A.\arabic{figure}}
\begin{center}
{\large\bf APPENDIX A: PROOFS OF THE THEOREMS}
\end{center}
\medskip
\spacingset{1.45} 
\label{app:S.1}

The proofs of Theorems~1 and~2 adapt the argument of \citet{pollard1979general} for the Pearson
statistic using data-dependent cells, with the IRF in the role of the
distribution function. Thus $M(A)$ and $M_{\bth}(A)$ are values
of signed measures rather than cell probabilities (Section~\ref{sec:test}). Notation is that of Sections~\ref{sec:test} and~\ref{sec:ddp} of the
paper. In addition, $\|\cdot\|$ also denotes the induced operator norm
of a matrix. We write
$\Theta_{\epsilon}\coloneqq\{\bth\in\Theta:\|\bth-\bth_0\|<\epsilon\}$, so that $\mathcal{M}_{\epsilon}=\{m_{\bth}:\bth\in\Theta_{\epsilon}\}$,
$r_{\bth}\coloneqq m_{\bth}-m_{\bth_0}-\dm'(\bth-\bth_0)$ is the
remainder of the $L^2(\PX)$ expansion in A1(iv) and A2(v), and
$\sigma^2_{\bth}(A)\coloneqq\mathbb{E}[U_{\bth}^2(\bW)1_A(\bX)]$, so
that $\sigma^2_{\bth_0}(A)=\sigma_0^2(A)$ under $H_0$. CLT and CMT
abbreviate the multivariate central limit theorem and the continuous
mapping theorem.

The proofs of Theorems~1 and~2 have four steps. Consistency and a local identifiability
bound give $\sqrt n$-consistency of $\hat{\bth}$ (Step~1). On sets where $\sqrt n(\bth-\bth_0)$ is bounded, the criterion is uniformly approximated by a quadratic form (Step~2) whose minimum is explicit (Step~3), and the
CLT gives the limit (Step~4). For Theorem~2 the approximations are made
uniform over partitions, and A2(i) with stochastic equicontinuity
reduces the problem to the fixed limiting partition $\bGam$, where
Steps~3 and~4 of Theorem~1 apply without change.

A bound used in both proofs follows from
$U_{\bth}^2-U_0^2=-(m_{\bth}-m_{\bth_0})(U_{\bth}+U_0)$ under $H_0$ and
the Cauchy--Schwarz inequality: for every Borel set $A$,
\begin{equation}
\bigl|\sigma^2_{\bth}(A)-\sigma^2_0(A)\bigr|
\leq
\|m_{\bth}-m_{\bth_0}\|_{\PX,2}
\bigl(2\|U_0\|_{P,2}+\|m_{\bth}-m_{\bth_0}\|_{\PX,2}\bigr).
\label{S1:var}
\end{equation}
The right-hand side does not depend on $A$ and tends to zero as
$\bth\to\bth_0$, because $L^2(\PX)$ differentiability implies
$L^2(\PX)$ continuity and $\mathbb{E}Y^2<\infty$.

\section{Proof of Theorem 1}\label{app:S.1.1}

\begin{lemma}\label{S1:lem-ident}
Under A1(iv)--(v), there exist $\delta>0$ and $\kappa>0$ such that
$\kappa\|\bth-\bth_0\|\leq\|\bM_{\bth}-\bM_{\bth_0}\|$ whenever
$\|\bth-\bth_0\|<\delta$.
\end{lemma}

\begin{proof}
Following \citet[Lemma~2]{pollard1979general}, put
$2\kappa\coloneqq\inf_{\|\boldsymbol t\|=1}\|\dM\boldsymbol t\|>0$.
Positivity follows from A1(v), continuity and compactness of the unit sphere.
By \eqref{eq:03}, there is a $\delta>0$ such that
$\|\bM_{\bth}-\bM_{\bth_0}-\dM(\bth-\bth_0)\|\leq \kappa\|\bth-\bth_0\|$
for $\|\bth-\bth_0\|<\delta$. By the reverse triangle inequality,
$\|\bM_{\bth}-\bM_{\bth_0}\|\geq\|\dM(\bth-\bth_0)\|-\kappa\|\bth-\bth_0\|
\geq \kappa\|\bth-\bth_0\|$.
\end{proof}

\begin{lemma}\label{S1:lem-donsker}
Under A1(iii) and A1(vi), for $\epsilon$ as in A1(vi) and each
$\ell=1,\ldots,L$, the classes
$\mathcal{M}_\epsilon\cdot1_{C_\ell}\coloneqq\{\bx\mapsto m_{\bth}(\bx)1_{C_\ell}(\bx):\bth\in\Theta_{\epsilon}\}$
and
$\mathcal{U}_{\epsilon\ell}\coloneqq\{(y,\bx)\mapsto
(y-m_{\bth}(\bx))1_{C_\ell}(\bx):\bth\in\Theta_{\epsilon}\}$ are
$P$-Donsker with envelopes in $L^2(P)$, and the class of squares of
$\mathcal{U}_{\epsilon\ell}$ is $P$-Glivenko--Cantelli.
\end{lemma}

\begin{proof}
The singleton $\{(y,\bx)\mapsto y\}$ is $P$-Donsker with envelope
$|Y|\in L^2(P)$ by A1(iii), and $\mathcal{M}_{\epsilon}$ is
$\PX$-Donsker by A1(vi), with the $L^2(\PX)$ envelope $F$ of
$\mathcal{M}$. The Donsker property is preserved by sums and by
multiplication by the fixed bounded function $1_{C_\ell}$ \citep[Examples~2.10.7 and~2.10.10]{van1996weak}, so
$\mathcal{M}_\epsilon\cdot1_{C_\ell}$ is $P$-Donsker with envelope $F(\bx)$ and
$\mathcal{U}_{\epsilon\ell}$ is $P$-Donsker with envelope
$|y|+F(\bx)$. Being Donsker, it is Glivenko--Cantelli; squaring is continuous and $(|y|+F(\bx))^2$ is integrable. Its squares are Glivenko--Cantelli by \citet[Theorem~3]{vdvw2000preservation}.
\end{proof}

\begin{lemma}\label{S1:lem-var}
Under A1(iii), A1(iv) and A1(vi), if
$\tilde{\bth}\to_P\bth_0$, then
$\hat\sigma^2_{\tilde{\bth}\ell}\to_P\sigma^2_{0\ell}>0$ for
$\ell=1,\ldots,L$; hence
$\hat{\bSig}_{\tilde{\bth}}^{-1}=\bSig_0^{-1}+o_P(1)$.
\end{lemma}

\begin{proof}
Write
$\hat\sigma^2_{\tilde{\bth}\ell}-\sigma^2_{0\ell}
=\{\hat\sigma^2_{\tilde{\bth}\ell}-\sigma^2_{\tilde{\bth}}(C_\ell)\}
+\{\sigma^2_{\tilde{\bth}}(C_\ell)-\sigma^2_0(C_\ell)\}$.
Since $P(\tilde{\bth}\in\Theta_{\epsilon})\to1$, the first term is
bounded with probability tending to one by
$\sup_{\bth\in\Theta_{\epsilon}}
|\hat\sigma^2_{\bth\ell}-\sigma^2_{\bth}(C_\ell)|=o_P(1)$, by
Lemma~\ref{S1:lem-donsker}. The second term is $o_P(1)$ by
(\ref{S1:var}) and the CMT. Positivity is part of A1(iii), and the CMT
gives the result for the inverse.
\end{proof}

\paragraph{Step 1: $\sqrt n$-consistency.}
The approximate-minimization inequality, Lemma~\ref{S1:lem-var} and the CLT give
\[
\|\hat{\bV}_{\hat{\bth}}\|^2
\leq\|\hat{\bSig}_{\tilde{\bth}}\|
\bigl\{\hat{\mathsf{X}}_{\tilde{\bth}}^2(\bth_0)+o_P(1)\bigr\}
\leq\|\hat{\bSig}_{\tilde{\bth}}\|
\bigl\{\|\hat{\bSig}_{\tilde{\bth}}^{-1}\|\,\|\hat{\bV}_0\|^2+o_P(1)\bigr\}
=O_P(1).
\]
Under $H_0$, $\bM=\bM_{\bth_0}$ and
$\hat{\bV}_{\hat{\bth}}=\sqrt n(\hat{\bM}-\hat{\bM}_{\hat{\bth}})$.
The triangle inequality therefore yields
\begin{equation}
\|\bM_{\hat{\bth}}-\bM_{\bth_0}\|
\leq\|\bM_{\hat{\bth}}-\hat{\bM}_{\hat{\bth}}\|
+n^{-1/2}\|\hat{\bV}_{\hat{\bth}}\|
+\|\hat{\bM}-\bM\|.
\label{S1:four}
\end{equation}
By A1(vi), multiplication by a fixed bounded indicator preserves the
Glivenko--Cantelli property \citep[Theorem~3]{vdvw2000preservation}.
The first term is thus bounded by
$\sup_{\bth\in\Theta}\|\hat{\bM}_{\bth}-\bM_{\bth}\|=o_P(1)$;
the last is $O_P(n^{-1/2})$ by the CLT. Hence
$\|\bM_{\hat{\bth}}-\bM_{\bth_0}\|=o_P(1)$, and A1(ii) implies
$\hat{\bth}\to_P\bth_0$.

We may now restrict the first term of \eqref{S1:four} to
$\Theta_\epsilon$. The local Donsker assumption gives
$\sup_{\bth\in\Theta_\epsilon}\sqrt n\|\hat{\bM}_{\bth}-\bM_{\bth}\|=O_P(1)$,
so all three terms are $O_P(n^{-1/2})$.
Lemma~\ref{S1:lem-ident} then gives
$\hat{\bth}-\bth_0=O_P(n^{-1/2})$.

\paragraph{Step 2: Quadratic approximation.}
Let $\hat B\coloneqq\{\bth\in\Theta:\|\bth-\bth_0\|\leq\hat\rho\}$,
where $\hat\rho=O_P(n^{-1/2})$ satisfies
$\hat\rho\geq\|\hat{\bth}-\bth_0\|$ and is fixed in Step~3. With
$\boldsymbol{R}_{\bth}\coloneqq\bM_{\bth}-\bM_{\bth_0}-\dM(\bth-\bth_0)$,
\begin{equation}
\hat{\bV}_{\bth}
=\bigl[\hat{\bV}_0-\dM\sqrt n(\bth-\bth_0)\bigr]
-\sqrt n\,\boldsymbol{R}_{\bth}
-\sqrt n\bigl\{(\hat{\bM}_{\bth}-\bM_{\bth})
-(\hat{\bM}_{\bth_0}-\bM_{\bth_0})\bigr\}.
\label{S1:decomp}
\end{equation}
By \eqref{eq:03},
$\sup_{\bth\in\hat B}\sqrt n\|\boldsymbol{R}_{\bth}\|
\leq\sqrt n\hat\rho\,
\sup_{0<\|\bth-\bth_0\|\leq\hat\rho}
\|\boldsymbol{R}_{\bth}\|/\|\bth-\bth_0\|=o_P(1)$.
The last term of (\ref{S1:decomp}) is $o_P(1)$ uniformly on $\hat B$ by
stochastic equicontinuity of the empirical processes indexed by the
Donsker classes $\mathcal{M}_\epsilon\cdot1_{C_\ell}$, $\ell=1,\ldots,L$ (Lemma~\ref{S1:lem-donsker}).
Their $L^2(\PX)$ continuity at $\bth_0$ follows from
$\|(m_{\bth}-m_{\bth_0})1_{C_\ell}\|_{\PX,2}
\leq\|m_{\bth}-m_{\bth_0}\|_{\PX,2}\to0$ \citep[sec.~2.1.2]{van1996weak}; the random radius is handled by noting that
$P(\hat\rho\leq n^{-1/4})\to1$. Hence
$\sup_{\bth\in\hat B}\|\hat{\bV}_{\bth}-[\hat{\bV}_0-\dM\sqrt
n(\bth-\bth_0)]\|=o_P(1)$. The bracket is $O_P(1)$ uniformly on
$\hat B$, and $\hat{\bSig}_{\tilde{\bth}}^{-1}=\bSig_0^{-1}+o_P(1)$ by
Lemma~\ref{S1:lem-var}; therefore
\begin{equation}
\sup_{\bth\in\hat B}\bigl|\hat{\mathsf{X}}_{\tilde{\bth}}^2(\bth)
-\hat Q(\bth)\bigr|=o_P(1),
\qquad\text{with}\qquad
\hat Q(\bth)\coloneqq\bigl\|\bSig_0^{-1/2}
\bigl[\hat{\bV}_0-\dM\sqrt n(\bth-\bth_0)\bigr]\bigr\|^2 .
\label{S1:quad}
\end{equation}

\paragraph{Step 3: Approximation of the minimum.}
Over $\mathbb{R}^p$, $\hat Q$ is minimized at $\hat{\bth}^{*}$, with
$\sqrt n(\hat{\bth}^{*}-\bth_0)=(\dM'\bSig_0^{-1}\dM)^{-1}\dM'\bSig_0^{-1}\hat{\bV}_0
=O_P(1)$ and
\begin{equation*}
\hat Q(\hat{\bth}^{*})=\bigl\|\bPi\,\bSig_0^{-1/2}\hat{\bV}_0\bigr\|^2,
\qquad\text{with}\qquad
\bPi\coloneqq\boldsymbol{I}_L-\bSig_0^{-1/2}\dM(\dM'\bSig_0^{-1}\dM)^{-1}\dM'\bSig_0^{-1/2},
\end{equation*}
the orthogonal projection onto the orthogonal complement of the column
space of $\bSig_0^{-1/2}\dM$. By A1(i), $\hat{\bth}^{*}\in\Theta$ with
probability tending to one. Take
$\hat\rho=\|\hat{\bth}-\bth_0\|+\|\hat{\bth}^{*}-\bth_0\|$, so that
$\hat{\bth},\hat{\bth}^{*}\in\hat B$. By (\ref{S1:quad}) and the
approximate-minimizing property of $\hat{\bth}$,
$\hat Q(\hat{\bth}^{*})\leq\hat Q(\hat{\bth})
=\hat{\mathsf{X}}_{\tilde{\bth}}^2(\hat{\bth})+o_P(1)
\leq\hat{\mathsf{X}}_{\tilde{\bth}}^2(\hat{\bth}^{*})+o_P(1)
=\hat Q(\hat{\bth}^{*})+o_P(1)$, so that
\begin{equation}
\hat{\mathsf{X}}_{\tilde{\bth}}^2(\hat{\bth})
=\bigl\|\bPi\,\bSig_0^{-1/2}\hat{\bV}_0\bigr\|^2+o_P(1).
\label{S1:repr}
\end{equation}

\paragraph{Step 4: Limiting distribution.}
By the CLT, $\bSig_0^{-1/2}\hat{\bV}_0\to_d \mathcal{N}_L(\boldsymbol{0},\boldsymbol{I}_L)$.
Since $\bPi$ is symmetric and idempotent with rank $L-p$, (\ref{S1:repr})
and the CMT give
$\hat{\mathsf{X}}_{\tilde{\bth}}^2(\hat{\bth})\to_d\chi^2_{L-p}$.
\hfill$\blacksquare$

\section{Proof of Theorem 2}\label{app:S.1.2}

Functions of a partition are evaluated cellwise, as in Section~\ref{sec:ddp} of the
paper; $\mathcal{I}_{\mathcal{C}}=\{1_C:C\in\mathcal{C}\}$ and $F$ is
the $L^2(\PX)$ envelope of $\mathcal{M}$ in A2(vii). Only the
modifications of the proof of Theorem~1 are given. Population expectations indexed by $\hat{\bGam}$ are evaluated at the realized partition, treating its cells as fixed sets.

\paragraph{Continuity in the partition.}
For $A,B\in\mathcal{C}$, the Cauchy--Schwarz inequality gives
\begin{align}
\sup_{\bth\in\Theta}|M_{\bth}(A)-M_{\bth}(B)|&\leq\|F\|_{\PX,2}\,d_{\PX}(A,B),
\quad
\|\dM(A)-\dM(B)\|\leq\|\dm\|_{\PX,2}\,d_{\PX}(A,B),\nonumber\\
|\sigma^2_0(A)-\sigma^2_0(B)|&\leq w\bigl(d^2_{\PX}(A,B)\bigr),
\label{S1:cont}
\end{align}
where
\[
w(t)\coloneqq\sup_{\{D\in\mathcal B^k:\,\PX(D)\leq t\}}\mathbb{E}[U_0^2(\bW)1_D(\bX)]
\]
tends to zero as $t\downarrow0$ by the absolute continuity of the
integral, because $\mathbb{E}U_0^2\leq2\mathbb{E}Y^2+2\mathbb{E}m_{\bth_0}^2(\bX)<\infty$
by A2(iv) and A2(vii).

\begin{lemma}\label{S1:lem-part}
Under A2(i), A2(iv)--(vii):
(a) $\sup_{\bth\in\Theta}\|\bM_{\bth}(\hat{\bGam})-\bM_{\bth}(\bGam)\|=o_P(1)$;
(b) $\dM(\hat{\bGam})=\dM(\bGam)+o_P(1)$;
(c) $\bSig_0(\hat{\bGam})=\bSig_0(\bGam)+o_P(1)$.
Moreover, define
\[
c(\bgam)\coloneqq\inf_{\|\boldsymbol t\|=1}\|\dM(\bgam)\boldsymbol t\|
\]
and $\kappa=c(\bGam)/4>0$. There is a neighborhood $\mathcal{G}_{\bGam}$ of
$\bGam$ in $(\mathcal{G},d_{\PX})$ such that $c(\bgam)\geq2\kappa$
for all $\bgam\in\mathcal{G}_{\bGam}$, and
$P(\hat{\bGam}\in\mathcal{G}_{\bGam})\to1$.
\end{lemma}

\begin{proof}
Parts (a)--(c) follow from (\ref{S1:cont}), A2(i) and the fact that
$L$ is fixed. As in \citet[Lemma~3]{pollard1979general}, the bound
$|c(\bgam)-c(\bGam)|\leq\|\dM(\bgam)-\dM(\bGam)\|$
shows that $c$ is continuous at $\bGam$, and A2(vi) gives $c(\bGam)>0$.
This gives $\mathcal{G}_{\bGam}$, and A2(i) gives
$P(\hat{\bGam}\in\mathcal{G}_{\bGam})\to1$.
\end{proof}

\begin{lemma}\label{S1:lem-ident2}
Under A2(v) and A2(vi), with $\kappa$ and $\mathcal{G}_{\bGam}$ as in
Lemma~\ref{S1:lem-part}, there is a $\delta>0$ such that
$\kappa\|\bth-\bth_0\|\leq\|\bM_{\bth}(\bgam)-\bM_{\bth_0}(\bgam)\|$ for all
$\bgam\in\mathcal{G}_{\bGam}$ and $\|\bth-\bth_0\|<\delta$.
\end{lemma}

\begin{proof}
By the same argument as in \citet[Lemma~3]{pollard1979general},
$\|\dM(\bgam)(\bth-\bth_0)\|\geq2\kappa\|\bth-\bth_0\|$ on
$\mathcal{G}_{\bGam}$. The remainder bound of Section~\ref{sec:ddp} gives
\[
\sup_{\bgam\in\mathcal{G}}\|\bM_{\bth}(\bgam)-\bM_{\bth_0}(\bgam)-\dM(\bgam)(\bth-\bth_0)\|
\leq\|r_{\bth}\|_{\PX,2}.
\]
Choose $\delta>0$ so that $\|r_{\bth}\|_{\PX,2}\leq \kappa\|\bth-\bth_0\|$
whenever $\|\bth-\bth_0\|<\delta$. Subtracting this bound from the
preceding lower bound proves the claim uniformly in $\bgam$.
\end{proof}

\begin{lemma}\label{S1:lem-donsker2}
Under A2(iv), A2(vii) and A2(viii), the classes
$\{(y,\bx)\mapsto y1_C(\bx):C\in\mathcal{C}\}$,
$\mathcal{M}_\epsilon\cdot\mathcal{I}_{\mathcal{C}}$ and
$\mathcal{U}_\epsilon\cdot\mathcal{I}_{\mathcal{C}}
\coloneqq\{(y,\bx)\mapsto(y-m_{\bth}(\bx))1_C(\bx):\bth\in\Theta_{\epsilon},C\in\mathcal{C}\}$
are $P$-Donsker with envelopes in $L^2(P)$; the squares of
$\mathcal{U}_\epsilon\cdot\mathcal{I}_{\mathcal{C}}$ form a
$P$-Glivenko--Cantelli class; and
$\mathcal{M}\cdot\mathcal{I}_{\mathcal{C}}$ is
$\PX$-Glivenko--Cantelli.
\end{lemma}

\begin{proof}
The singleton $\{y\}$ is BUEI with envelope $|Y|\in L^2(P)$, since its
covering numbers equal one; $\mathcal{M}_\epsilon$ is BUEI with an
$L^2(\PX)$ envelope by A2(vii); and $\mathcal{I}_{\mathcal{C}}$ is BUEI
with envelope one, because $\mathcal{C}$ is VC by A2(viii) \citep[Lemma~9.8 and Theorem~9.3]{Kosorok2008}. BUEI and pointwise measurability are preserved by sign changes, sums and products
\citep[Lemma~9.17(iv)--(vi)]{Kosorok2008}, so the three
classes are BUEI with $L^2(P)$ envelopes and hence $P$-Donsker \citep[Theorem~8.19]{Kosorok2008}. Squaring is continuous with an integrable envelope, so the squares are $P$-Glivenko--Cantelli by \citet[Theorem~3]{vdvw2000preservation}. Finally, $\mathcal{M}$
is $\PX$-Glivenko--Cantelli with an integrable envelope and
$\mathcal{I}_{\mathcal{C}}$ is uniformly bounded and
Glivenko--Cantelli, so their product is $\PX$-Glivenko--Cantelli \citep[Theorem~3]{vdvw2000preservation}.
\end{proof}

\begin{lemma}\label{S1:lem-var2}
Under A2(i), A2(iv), A2(v), A2(vii) and A2(viii), if
$\tilde{\bth}\to_P\bth_0$, then
$\hat{\bSig}_{\tilde{\bth}}(\hat{\bGam})=\bSig_0(\bGam)+o_P(1)$ and
$\hat{\bSig}_{\tilde{\bth}}^{-1}(\hat{\bGam})=\bSig_0^{-1}(\bGam)+o_P(1)$.
\end{lemma}

\begin{proof}
For each $\ell$,
$\hat\sigma^2_{\tilde{\bth}}(\hat\Gamma_\ell)-\sigma^2_0(\Gamma_\ell)
=\{\hat\sigma^2_{\tilde{\bth}}-\sigma^2_{\tilde{\bth}}\}(\hat\Gamma_\ell)
+\{\sigma^2_{\tilde{\bth}}-\sigma^2_0\}(\hat\Gamma_\ell)
+\{\sigma^2_0(\hat\Gamma_\ell)-\sigma^2_0(\Gamma_\ell)\}$.
The first term is bounded, with probability tending to one, by
$\sup_{\bth\in\Theta_{\epsilon},C\in\mathcal{C}}
|\hat\sigma^2_{\bth}(C)-\sigma^2_{\bth}(C)|=o_P(1)$
(Lemma~\ref{S1:lem-donsker2}); the second is $o_P(1)$ by
(\ref{S1:var}), which holds uniformly over cells; and the third is
$o_P(1)$ by Lemma~\ref{S1:lem-part}(c). Positive definiteness of $\bSig_0(\bGam)$ (A2(iv)) and the CMT give the result for the inverse.
\end{proof}

\paragraph{Step 1: $\sqrt n$-consistency.}
By Lemma~\ref{S1:lem-donsker2},
$\sup_{\bgam\in\mathcal G}\|\hat{\bV}_0(\bgam)\|=O_P(1)$.
The approximate-minimization inequality and Lemma~\ref{S1:lem-var2}
therefore give $\|\hat{\bV}_{\hat{\bth}}(\hat{\bGam})\|=O_P(1)$,
by the same bound as in Step~1 of Theorem~1.
Apply \eqref{S1:four} with all vectors evaluated at $\hat{\bGam}$.
Its first term is $o_P(1)$ uniformly over $\Theta\times\mathcal G$
by the global Glivenko--Cantelli conclusion of Lemma~\ref{S1:lem-donsker2},
and its last term is $O_P(n^{-1/2})$ uniformly over $\mathcal G$.
Thus $\|(\bM_{\hat{\bth}}-\bM_{\bth_0})(\hat{\bGam})\|=o_P(1)$.
Lemma~\ref{S1:lem-part}(a) permits replacement of $\hat{\bGam}$ by
$\bGam$ in this last statement, and A2(iii) gives $\hat{\bth}\to_P\bth_0$.

With $\hat{\bth}\in\Theta_\epsilon$ with probability tending to one,
the local Donsker conclusion of Lemma~\ref{S1:lem-donsker2} bounds
all three terms in \eqref{S1:four} by $O_P(n^{-1/2})$, uniformly over
partitions. Consequently
$\|(\bM_{\hat{\bth}}-\bM_{\bth_0})(\hat{\bGam})\|=O_P(n^{-1/2})$.
Since $P(\hat{\bGam}\in\mathcal{G}_{\bGam})\to1$,
Lemma~\ref{S1:lem-ident2} gives $\hat{\bth}-\bth_0=O_P(n^{-1/2})$.

\paragraph{Step 2: Reduction to the limiting partition.}
Let $\hat B$ be as in Theorem~1. The decomposition (\ref{S1:decomp})
holds at every $\bgam$, and we show that
\begin{equation}
\sup_{\bth\in\hat B}\bigl\|\hat{\bV}_{\bth}(\hat{\bGam})
-\bigl[\hat{\bV}_0(\bGam)-\dM(\bGam)\sqrt n(\bth-\bth_0)\bigr]\bigr\|=o_P(1).
\label{S1:red}
\end{equation}
The difference is bounded by four terms:
(i) $\sup_{\bth\in\hat B,\bgam\in\mathcal{G}}\sqrt n\|\boldsymbol{R}_{\bth}(\bgam)\|=o_P(1)$,
by the uniform remainder bound quoted in the proof of
Lemma~\ref{S1:lem-ident2}, as in Step~2 of Theorem~1.
(ii) The centered increments of the empirical process are
$o_P(1)$ uniformly over $\bth\in\hat B$ and $\bgam\in\mathcal{G}$, by
stochastic equicontinuity of the Donsker class
$\mathcal{M}_\epsilon\cdot\mathcal{I}_{\mathcal{C}}$ (Lemma~\ref{S1:lem-donsker2})
and
$\sup_{C\in\mathcal{C}}\|(m_{\bth}-m_{\bth_0})1_C\|_{\PX,2}\leq\|m_{\bth}-m_{\bth_0}\|_{\PX,2}\to0$;
this is the uniform expansion stated in Section~\ref{sec:ddp} for
deterministic $\delta_n\downarrow0$, and it applies on $\hat B$ because
$P(\hat\rho\leq n^{-1/4})\to1$.
(iii) $\|\dM(\hat{\bGam})-\dM(\bGam)\|\sqrt n\hat\rho=o_P(1)$ by
Lemma~\ref{S1:lem-part}(b).
(iv) $\hat{\bV}_0(\hat{\bGam})-\hat{\bV}_0(\bGam)=o_P(1)$: the class
$\{U_01_C:C\in\mathcal{C}\}\subset\mathcal{U}_\epsilon\cdot\mathcal{I}_{\mathcal{C}}$
is Donsker, its intrinsic semimetric is bounded by
$w(d^2_{\PX})^{1/2}$ by (\ref{S1:cont}), and
$d_{\PX}(\hat{\bGam},\bGam)\to_P0$ by A2(i); asymptotic uniform
equicontinuity gives the claim \citep[Lemma~19.24]{van2000asymptotic}, as noted in Section~\ref{sec:ddp}. No
independence between $\hat{\bGam}$ and the sample is needed.

By (\ref{S1:red}) and Lemma~\ref{S1:lem-var2},
$\sup_{\bth\in\hat B}|\hat{\mathsf{X}}_{\tilde{\bth}}^2(\hat{\bGam},\bth)-\hat Q_{\bGam}(\bth)|=o_P(1)$,
where $\hat Q_{\bGam}$ is the quadratic form $\hat Q$ of
(\ref{S1:quad}) for the fixed cells $\bC=\bGam$.

\paragraph{Steps 3 and 4.}
The approximating quadratic form is that of Theorem~1 for the fixed
partition $\bGam$, which satisfies the conditions used there: $\bth_0$
is interior (A2(ii)), $\dM(\bGam)$ has rank $p$ (A2(vi)) and
$\bSig_0(\bGam)$ is positive definite (A2(iv)). Steps~3 and~4 of the
proof of Theorem~1 therefore apply without change, with $\hat\rho$ chosen as
there, and give
$\hat{\mathsf{X}}_{\tilde{\bth}}^2(\hat{\bGam},\hat{\bth})
=\|\bPi(\bGam)\bSig_0^{-1/2}(\bGam)\hat{\bV}_0(\bGam)\|^2+o_P(1)
\to_d\chi^2_{L-p}$.
The fixed-partition expansion established in the proof of Theorem~1 gives the same representation for $\hat{\mathsf{X}}_{\tilde{\bth}}^2(\bGam,\hat{\bth}(\bGam))$. Comparing the two representations proves equation~\eqref{eq:partition-equivalence}.
\hfill$\blacksquare$

\section{Proof of Theorem 3}\label{app:S.1.3}
\label{S1:local-proof-local}

The regression perturbation follows \citet[sec.~3]{stute1997nonparametric}.
The projected-drift calculation is also related to the minimum-distance
argument of \citet[sec.~6]{pollard1980minimum}. We work throughout on
the fixed probability space carrying the i.i.d. pairs $(\varepsilon_i,\bX_i)$
in Section~\ref{sec:new-contiguous}, with $Y_i=m_{\bth_0}(\bX_i)+\varepsilon_i+n^{-1/2}\delta(\bX_i)$. Under $H_{1n}$ the true regression function is $m_{\bth_0}+n^{-1/2}\delta$, so the regression error is $\varepsilon_i$, whereas the residual at the null parameter is $U_{\bth_0}(\boldsymbol W_i)=\varepsilon_i+n^{-1/2}\delta(\bX_i)$.
No change-of-measure or contiguity argument is used.
As in Section~\ref{sec:new-contiguous}, we write $\boldsymbol D=\boldsymbol D(\bGam)$, $\bSig_0=\bSig_0(\bGam)$,
$\dM=\dM(\bGam)$, $\boldsymbol\Pi=\boldsymbol\Pi(\bGam)$, and $\boldsymbol d=\boldsymbol d(\bGam)$ for the cell drift, the covariance matrix, the Jacobian, the projection \eqref{eq:Pi}, and the standardized drift defined in Section~\ref{sec:new-contiguous}, all evaluated at the limit partition $\bGam$, and $\hat\bth=\hat\bth(\hat\bGam)$. For $\bgam\in\mathcal G$, let
$\hat{\boldsymbol D}(\bgam)=n^{-1}\sum_{i=1}^{n}\delta(\bX_i)\mathbf1_{\bgam}(\bX_i)$.
By A3, $\dM$ has column rank $p$ and $\bSig_0$ is positive definite.

\paragraph{Preliminary decomposition.}
We use the empirical processes $\hat{\bV}_{\bth}(\bgam)$ and $\hat{\bV}_0(\bgam)$
of Section~\ref{sec:ddp}. Under $H_{1n}$ the regression error
is $U_0(\bW_i)=Y_i-m_{\bth_0}(\bX_i)-n^{-1/2}\delta(\bX_i)=\varepsilon_i$, so
$\hat{\bV}_0(\bgam)=n^{-1/2}\sum_i\varepsilon_i\mathbf1_{\bgam}(\bX_i)$, and the
residual at $\bth_0$ is $U_{\bth_0}(\bW_i)=\varepsilon_i+n^{-1/2}\delta(\bX_i)$.
Hence, for every $\bgam\in\mathcal G$,
\[
\hat{\bV}_{\bth_0}(\bgam)
=\hat{\bV}_0(\bgam)+\hat{\boldsymbol D}(\bgam).
\]
At $\bgam=\bGam$, the multivariate CLT gives
$\hat{\bV}_0(\bGam)\to_d\bSig_0^{1/2}\boldsymbol G$, with $\boldsymbol G\sim\mathcal N_L(\boldsymbol0,\boldsymbol I_L)$, because the
cells are disjoint, $\mathbb E(\varepsilon\mid\bX)=0$ and $\mathbb E\varepsilon^2<\infty$,
and the law of large numbers gives $\hat{\boldsymbol D}(\bGam)\to_P\boldsymbol D$.
Therefore
\begin{equation}
\hat{\boldsymbol v}:=\hat{\bV}_{\bth_0}(\bGam)
=\hat{\bV}_0(\bGam)+\boldsymbol D+o_P(1)
\ \longrightarrow_d\ \boldsymbol D+\bSig_0^{1/2}\boldsymbol G.
\label{S1:local-cell-limit}
\end{equation}

The data-dependent cells change neither term. The errors $\varepsilon_i$ have a fixed
law, so the argument of part (iv) of Step~2 in the proof of Theorem~2 applies verbatim and gives
$\hat{\bV}_0(\hat\bGam)-\hat{\bV}_0(\bGam)=o_P(1)$. For the drift term, the VC
property and the integrable envelope $|\delta(\bX)|$ give a uniform law of large
numbers over $\mathcal C$, and, for each cell,
\[
\bigl|\mathbb E[\delta(\bX)\{1_{\hat\Gamma_\ell}(\bX)-1_{\Gamma_\ell}(\bX)\}]\bigr|
\leq\|\delta\|_{P_{\bX},2}
P_{\bX}(\hat\Gamma_\ell\triangle\Gamma_\ell)^{1/2}=o_P(1).
\]
Consequently $\hat{\boldsymbol D}(\hat\bGam)=\boldsymbol D+o_P(1)$ and
\[
\hat{\bV}_{\bth_0}(\hat\bGam)=\hat{\bV}_0(\bGam)+\boldsymbol D+o_P(1)=\hat{\boldsymbol v}+o_P(1).
\]
The cell drift is therefore $\boldsymbol D=\mathbb E[\delta(\bX)\mathbf1_{\bGam}(\bX)]$,
the drift of the limit cells, whether or not the cells are estimated. The partitions
may depend on the local responses: these are uniform statements, so no independence
between observations and cells is required.

\paragraph{Step 1: $\sqrt n$-consistency.}
We first establish consistency of the estimated covariance matrix under the local
alternative. Define the auxiliary response
\[
Y_i^{(0)}=m_{\bth_0}(\bX_i)+\varepsilon_i,\qquad
Y_i=Y_i^{(0)}+\frac{\delta(\bX_i)}{\sqrt n}.
\]
The pairs $(Y_i^{(0)},\bX_i)$ have a fixed distribution satisfying the null model.
They are introduced only for the proof. Throughout this comparison, $\tilde\bth$ and
$\hat\bGam$ are the estimates constructed from the actual observations $(Y_i,\bX_i)$.
In particular, we do not refit the model to the auxiliary responses. Write
\[
\tilde a_i=Y_i^{(0)}-m_{\tilde\bth}(\bX_i),\qquad
\hat\sigma_\ell^{2,(0)}=\frac1n\sum_{i=1}^{n}\tilde a_i^2 1_{\hat\Gamma_\ell}(\bX_i).
\]
Apply the uniform laws of large numbers used in Lemma~\ref{S1:lem-var2} to the
observations $(Y_i^{(0)},\bX_i)$. The bounds hold uniformly over parameters in a
neighborhood of $\bth_0$ and over cells in $\mathcal C$. They therefore also hold at
the sample-dependent choices $\tilde\bth$ and $\hat\Gamma_\ell$, and no independence
between these estimates and the observations is required. The assumed consistency
$\tilde\bth\to_P\bth_0$ under $H_{1n}$ and partition convergence in A3 then give, for
each $\ell$, $\hat\sigma_\ell^{2,(0)}\to_P\sigma_{0\ell}^2$.

The residual based on the actual response is $Y_i-m_{\tilde\bth}(\bX_i)=\tilde a_i+n^{-1/2}\delta(\bX_i)$.
Expanding its square shows that the actual cell variance estimate differs from the
auxiliary one by
\[
\hat\sigma^2_{\tilde\bth}(\hat\Gamma_\ell)-\hat\sigma_\ell^{2,(0)}
=\frac{2}{n^{3/2}}\sum_{i=1}^{n}\tilde a_i\delta(\bX_i)1_{\hat\Gamma_\ell}(\bX_i)
+\frac1{n^2}\sum_{i=1}^{n}\delta^2(\bX_i)1_{\hat\Gamma_\ell}(\bX_i).
\]
Taking absolute values, bounding the indicators by one, and applying Cauchy--Schwarz
bounds this difference by
\[
\frac2{\sqrt n}\Bigl(\frac1n\sum_{i=1}^{n}\tilde a_i^2\Bigr)^{1/2}
\Bigl(\frac1n\sum_{i=1}^{n}\delta^2(\bX_i)\Bigr)^{1/2}
+\frac1n\Bigl(\frac1n\sum_{i=1}^{n}\delta^2(\bX_i)\Bigr)=O_P(n^{-1/2}).
\]
The first sample average is $O_P(1)$ by consistency of $\tilde\bth$ and the local
square-integrable envelope, and the second converges to $\mathbb E\delta^2(\bX)<\infty$.
Since $L$ is fixed, covariance consistency holds for the whole diagonal matrix.
Positive definiteness of $\bSig_0$ also gives consistency of its inverse,
\[
\hat{\bSig}_{\tilde\bth}(\hat\bGam)=\bSig_0+o_P(1),\qquad
\hat{\bSig}_{\tilde\bth}^{-1}(\hat\bGam)=\bSig_0^{-1}+o_P(1).
\]

The approximate-minimization inequality and the covariance bounds above give \[ \|\hat{\bV}_{\hat\bth}(\hat\bGam)\|=O_P(1). \] For $\bth\in\Theta$ and
$\bgam\in\mathcal G$, let
\begin{equation}
\boldsymbol R_{\bth}(\bgam)=\bM_{\bth}(\bgam)-\bM_{\bth_0}(\bgam)
-\dM(\bgam)(\bth-\bth_0)
\label{S1:local-R}
\end{equation}
and
\[
\hat{\boldsymbol E}_{\bth}(\bgam)=\frac{1}{\sqrt n}\sum_{i=1}^{n}\bigl\{(m_{\bth}-m_{\bth_0})(\bX_i)\mathbf1_{\bgam}(\bX_i)
-\mathbb E[(m_{\bth}-m_{\bth_0})(\bX)\mathbf1_{\bgam}(\bX)]\bigr\}.
\]
Since $Y_i-m_{\bth}(\bX_i)=\varepsilon_i+n^{-1/2}\delta(\bX_i)-(m_{\bth}-m_{\bth_0})(\bX_i)$,
the definitions of $\hat{\bV}_{\bth}$, $\hat{\bV}_0$, $\hat{\boldsymbol D}$, and
$\hat{\boldsymbol E}_{\bth}$ give the exact identity
\begin{equation}
(\bM_{\bth}-\bM_{\bth_0})(\bgam)
=n^{-1/2}\bigl\{\hat{\bV}_0(\bgam)+\hat{\boldsymbol D}(\bgam)
-\hat{\bV}_{\bth}(\bgam)-\hat{\boldsymbol E}_{\bth}(\bgam)\bigr\}
\label{S1:local-identity}
\end{equation}
for all $\bth\in\Theta$ and $\bgam\in\mathcal G$. Each of the $L$ entries of $\hat{\boldsymbol D}(\bgam)$ is bounded in absolute
value by $n^{-1}\sum_{i=1}^{n}|\delta(\bX_i)|$, so
\[
n^{-1/2}\sup_{\bgam\in\mathcal G}\|\hat{\boldsymbol D}(\bgam)\|
\leq n^{-1/2}\,\sqrt L\,\frac1n\sum_{i=1}^{n}|\delta(\bX_i)|=O_P(n^{-1/2}),
\]
because $n^{-1}\sum_{i=1}^{n}|\delta(\bX_i)|\to_P\mathbb E|\delta(\bX)|<\infty$ by the
law of large numbers. Thus the local alternative moves the cell moments by
$O_P(n^{-1/2})$, uniformly over partitions.

Evaluate \eqref{S1:local-identity} at $\hat\bth$ and $\hat\bGam$. By
Lemma~\ref{S1:lem-donsker2}, $\sup_{\bgam\in\mathcal G}\|\hat{\bV}_0(\bgam)\|=O_P(1)$,
and $\|\hat{\bV}_{\hat\bth}(\hat\bGam)\|=O_P(1)$ as shown above. The global
Glivenko--Cantelli conclusion of Lemma~\ref{S1:lem-donsker2} gives
$n^{-1/2}\sup_{\bth\in\Theta,\,\bgam\in\mathcal G}\|\hat{\boldsymbol E}_{\bth}(\bgam)\|=o_P(1)$.
Hence $\|(\bM_{\hat\bth}-\bM_{\bth_0})(\hat\bGam)\|=o_P(1)$.
Lemma~\ref{S1:lem-part}(a) permits replacement of $\hat\bGam$ by $\bGam$ in this
statement, and identification in A3 and A2(iii) gives $\hat\bth\to_P\bth_0$.

As $\hat\bth\in\Theta_\epsilon$ with probability approaching one, the local Donsker conclusion of Lemma~\ref{S1:lem-donsker2} gives
$\sup_{\bth\in\Theta_\epsilon,\,\bgam\in\mathcal G}\|\hat{\boldsymbol E}_{\bth}(\bgam)\|=O_P(1)$.
Every term inside the braces of \eqref{S1:local-identity} is then $O_P(1)$ at
$\hat\bth$ and $\hat\bGam$, so
$\|(\bM_{\hat\bth}-\bM_{\bth_0})(\hat\bGam)\|=O_P(n^{-1/2})$. Since
$P(\hat\bGam\in\mathcal G_{\bGam})\to1$, Lemma~\ref{S1:lem-ident2} gives
$\sqrt n(\hat\bth-\bth_0)=O_P(1)$, exactly as in Step~1 of the proof of Theorem~2.

\paragraph{Step 2: Quadratic approximation.}
Combining \eqref{S1:local-identity} with \eqref{S1:local-R} gives the exact expansion
\begin{align*}
\hat{\bV}_{\bth}(\hat\bGam)
={}&\hat{\bV}_{\bth_0}(\hat\bGam)
-\dM\sqrt n(\bth-\bth_0)
-\hat{\boldsymbol E}_{\bth}(\hat\bGam)\\
&-\sqrt n\,\boldsymbol R_{\bth}(\hat\bGam)
-\{\dM(\hat\bGam)-\dM\}\sqrt n(\bth-\bth_0).
\end{align*}
Fix $K_0<\infty$. We show that the last three terms are $o_P(1)$ as $n\to\infty$, uniformly over $\|\bth-\bth_0\|\leq K_0/\sqrt n$.

(i) \emph{Empirical-process term.} Over this neighborhood, $\sup_{C\in\mathcal C}\|(m_{\bth}-m_{\bth_0})1_C\|_{\PX,2}\leq\|m_{\bth}-m_{\bth_0}\|_{\PX,2}\to0$, so asymptotic equicontinuity of the Donsker class $\mathcal M_\epsilon\cdot\mathcal I_{\mathcal C}$ (Lemma~\ref{S1:lem-donsker2}) gives $\hat{\boldsymbol E}_{\bth}(\hat\bGam)=o_P(1)$ \citep[sec.~2.1.2]{van1996weak}, as in part~(ii) of Step~2 of the proof of Theorem~2.

(ii) \emph{Linearization remainder.} Let $r_{\bth}=m_{\bth}-m_{\bth_0}-\dm'(\bth-\bth_0)$ be the remainder of the $L^2(\PX)$ expansion in A2(v), so that $\boldsymbol R_{\bth}(\bgam)=\mathbb E[r_{\bth}(\bX)\mathbf 1_{\bgam}(\bX)]$. Since the cells are disjoint, $\|\boldsymbol R_{\bth}(\hat\bGam)\|\leq\|r_{\bth}\|_{\PX,2}$, so for $0<\|\bth-\bth_0\|\leq K_0/\sqrt n$, $\sqrt n\|\boldsymbol R_{\bth}(\hat\bGam)\|\leq K_0\|r_{\bth}\|_{\PX,2}/\|\bth-\bth_0\|\to0$ by $L^2$ differentiability, while $\boldsymbol R_{\bth_0}(\hat\bGam)=\boldsymbol 0$.

(iii) \emph{Jacobian term.} $\|\{\dM(\hat\bGam)-\dM\}\sqrt n(\bth-\bth_0)\|\leq K_0\|\dM(\hat\bGam)-\dM\|=o_P(1)$ by Lemma~\ref{S1:lem-part}(b).

Together with $\hat{\bV}_{\bth_0}(\hat\bGam)=\hat{\boldsymbol v}+o_P(1)$, these bounds yield
\begin{equation*}
\sup_{\|\boldsymbol t\|\leq K_0}
\|\hat{\bV}_{\bth_0+\boldsymbol t/\sqrt n}(\hat\bGam)
-\hat{\boldsymbol v}+\dM\boldsymbol t\|=o_P(1).
\end{equation*}
Since $\bth_0$ is interior, $\bth_0+\boldsymbol t/\sqrt n\in\Theta$ for all $\|\boldsymbol t\|\leq K_0$ when $n$ is large.
Define
\[
\hat Q(\boldsymbol t)=(\hat{\boldsymbol v}-\dM\boldsymbol t)'
\bSig_0^{-1}(\hat{\boldsymbol v}-\dM\boldsymbol t).
\]
Since $\hat\bSig^{-1}_{\tilde\bth}(\hat\bGam)=\bSig_0^{-1}+o_P(1)$ by Step~1, the covariance and local-expansion bounds give, for each fixed $K_0<\infty$,
\[
\sup_{\|\boldsymbol t\|\leq K_0}\bigl|\hat{\mathsf X}^2_{\tilde\bth}(\hat\bGam,\bth_0+\boldsymbol t/\sqrt n)-\hat Q(\boldsymbol t)\bigr|=o_P(1),
\]
the counterpart of \eqref{S1:quad} in the proof of Theorem~1.

\paragraph{Step 3: Approximation of the minimum.}
Its minimizer satisfies
\[
\hat{\boldsymbol t}^{*}=(\dM'\bSig_0^{-1}\dM)^{-1}\dM'\bSig_0^{-1}\hat{\boldsymbol v}=O_P(1),
\]
and
\[
\hat Q(\boldsymbol t)-\hat Q(\hat{\boldsymbol t}^{*})
=(\boldsymbol t-\hat{\boldsymbol t}^{*})'\dM'\bSig_0^{-1}
\dM(\boldsymbol t-\hat{\boldsymbol t}^{*}).
\]
Root-$n$ consistency, the positive quadratic lower bound, and the fact that $\hat\bth$ minimizes $\hat{\mathsf X}^2_{\tilde\bth}(\hat\bGam,\cdot)$ up to an $o_P(1)$ error (A3)
therefore give
\[
\sqrt n(\hat\bth-\bth_0)=(\dM'\bSig_0^{-1}\dM)^{-1}\dM'\bSig_0^{-1}\hat{\boldsymbol v}+o_P(1),
\]
which is the expansion in~\eqref{eq:theta-local}. Formally, apply the uniform
approximation on a ball containing $\sqrt n(\hat\bth-\bth_0)$ and $\hat{\boldsymbol t}^{*}$ with arbitrarily high probability and then let its radius
increase. Substitution into the local expansion gives
\[
\hat{\bSig}_{\tilde\bth}^{-1/2}(\hat\bGam)\hat{\bV}_{\hat\bth}(\hat\bGam)
=\boldsymbol\Pi\bSig_0^{-1/2}\hat{\boldsymbol v}+o_P(1),
\]
where $\boldsymbol\Pi$ is the $L\times L$ orthogonal projection defined in \eqref{eq:Pi} that annihilates the column space of $\bSig_0^{-1/2}\dM$. It is symmetric and idempotent, with $\operatorname{rank}(\boldsymbol\Pi)=\operatorname{trace}(\boldsymbol\Pi)=L-p$.

\paragraph{Step 4: Limiting distribution.}
By (\ref{S1:local-cell-limit}), $\bSig_0^{-1/2}\hat{\boldsymbol v}\to_d\boldsymbol G+\boldsymbol d$,
with $\boldsymbol G\sim\mathcal N_L(\boldsymbol 0,\boldsymbol I_L)$. Since $\boldsymbol\Pi$
is symmetric and idempotent with rank $L-p$, the CMT gives
\[
\hat{\mathsf X}^2_{\tilde\bth}(\hat\bGam,\hat\bth)\to_d
\|\boldsymbol\Pi(\boldsymbol G+\boldsymbol d)\|^2\sim\chi^2_{L-p}\{\lambda(\bGam)\},
\qquad \lambda(\bGam)=\boldsymbol d'\boldsymbol\Pi\boldsymbol d.
\]
The decomposition $\lambda=\lambda_1-\lambda_2$ and its interpretation are given in
Section~\ref{sec:new-contiguous}. For fixed cells the same proof uses A1
and omits partition replacement. This proves both cases of Theorem~3.\hfill$\blacksquare$

\end{document}